\documentclass[sigconf,nonacm]{acmart}
\usepackage{graphicx} 
\usepackage[dvipsnames]{xcolor}

\usepackage{amsmath,amsfonts,amssymb}
\usepackage{svg}
 \usepackage{array}
\usepackage{booktabs}
\usepackage{tabularx}
\usepackage{multirow}
\usepackage{url}
\usepackage{hyperref}
\usepackage{array}
\usepackage{longtable}
\usepackage{graphicx}  
\usepackage{colortbl,tcolorbox}  
\usepackage{enumitem}
\usepackage{xspace}

\newcommand{\promptbox}[1]{%
\begin{tcolorbox}[colback=gray!5, colframe=black!40, boxrule=0.5pt, arc=2pt, 
  left=1mm, right=1mm, top=1mm, bottom=1mm]
\tiny
#1
\end{tcolorbox}
}
\newcolumntype{Y}{>{\raggedright\arraybackslash}X}
\newcolumntype{L}[1]{>{\raggedright\arraybackslash}p{#1}}
\definecolor{skyblue}{RGB}{135,206,235}

\newcommand{\oracle}{\textsc{Oracle}\xspace} 
\newcommand{\interpreter}{\textsc{Interpreter}\xspace} 
\newcommand{\decisionmaker}{\textsc{Decision-maker}\xspace} 
\newcommand{\evaluator}{\textsc{Evaluator}\xspace} 
\newcommand{\ghostwriter}{\textsc{Ghost-writer}\xspace} 

\AtBeginDocument{%
  }

\setcopyright{acmlicensed}
\copyrightyear{2026}
\acmYear{2026}
\acmDOI{XXXXXXX.XXXXXXX}
\acmConference[Conference acronym 'XX]{Make sure to enter the correct
  conference title from your rights confirmation email}{June 03--05,
  2018}{Woodstock, NY}
\acmISBN{978-1-4503-XXXX-X/2018/06}

\begin{document}

\title{LLMs as Oracles: How Users Turn To LLMs for Subjective Personal Questions}

\author{Myra Cheng}
\correspondingauthor
\authornote{Both authors contributed equally to this research.}
\email{myra@cs.stanford.edu}
\affiliation{%
  \institution{Stanford University}
  \city{Stanford}
  \state{California}
  \country{USA}
}

\author{Lujain Ibrahim}
\correspondingauthor
\authornotemark[1]
\email{lujain.ibrahim@oii.ox.ac.uk}
\affiliation{%
  \institution{University of Oxford}
  \city{Oxford}
  \country{England}
}

\author{Grace Liu}
\affiliation{%
  \institution{Carnegie Mellon University}
  \city{Pittsburgh}
  \state{Pennsylvania}
  \country{USA}}

\author{Michelle S. Lam}
\affiliation{%
  \institution{Stanford University}
  \city{Stanford}
  \state{California}
  \country{USA}
}

\author{Vishakh Padmakumar}
\affiliation{%
  \institution{Stanford University}
  \city{Stanford}
  \state{California}
  \country{USA}
}
\author{Nick Madibekov}
\affiliation{%
  \institution{Stanford University}
  \city{Stanford}
  \state{California}
  \country{USA}
}

\author{Diyi Yang}
\affiliation{%
  \institution{Stanford University}
  \city{Stanford}
  \state{California}
  \country{USA}
}

\author{Dan Jurafsky}
\affiliation{%
  \institution{Stanford University}
  \city{Stanford}
  \state{California}
  \country{USA}
}
\renewcommand{\shortauthors}{Cheng, Ibrahim et al.}

\begin{abstract}
We characterize how people are turning to LLMs as oracles: all-knowing authorities on subjective personal questions. Motivated by risks to autonomy and well-being, we develop a typology and LLM-based methods for large-scale measurement of how people seek judgment and decision-making from AI. We applied our typology to public usage data (68K prompts from WildChat and ThoughtTrace) and individuals' longitudinal usage data (140K prompts from 52 participants, collected using a privacy-preserving data donation tool that we built). We identify temporal and demographic trends, such as an increase in LLM-as-oracle prompts in individuals' donated data and higher prevalence among younger users in ThoughtTrace. In the data donation study, participants under-reported their own LLM-as-oracle use and expressed dissatisfaction with their LLM-as-oracle behavior after seeing our tool's analysis. We identify two factors associated with LLM-as-oracle use: people's perceptions of AI and model behavior, which motivate interventions to make LLMs answer in less oracular ways.

\end{abstract}

\begin{teaserfigure}
\centering
\includegraphics[width=0.95\textwidth]{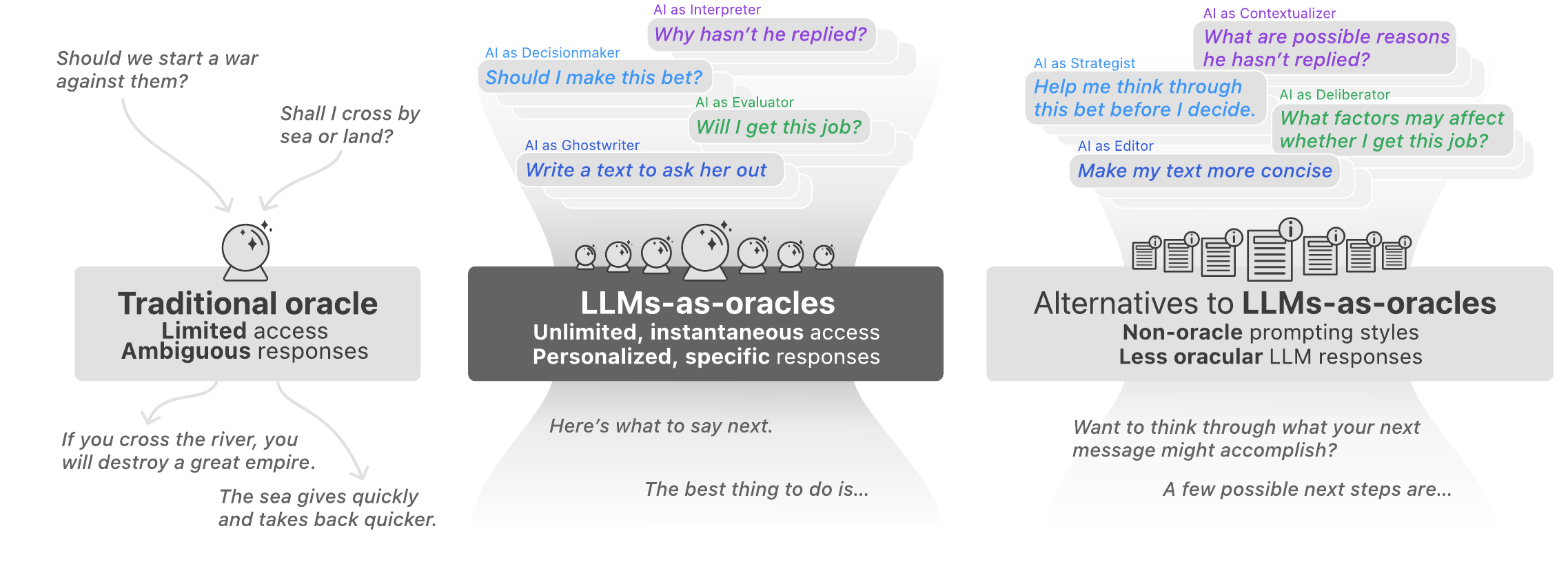}
  \caption{While people have long consulted oracles for subjective personal questions, these oracles were limited in access and specificity. Now, people turn to LLMs for similar questions without these constraints. Motivated by risks of reliance, we build a typology of the ways people use LLMs as \textit{oracles}: seeking belief, normative judgment, or decisions. Using this typology, we identify temporal and demographic trends in LLM-as-oracle use. We also show how model behavior and people's perceptions of AI {may} drive this use. Finally, we propose interventions on both sides: how people can use LLMs in alternative, non-oracle ways, and how LLMs can respond less as oracles.
  }
  \Description{Three-panel diagram contrasting a traditional oracle with oracle use and non-oracle use of large language models. Each panel is an hourglass, with questions at the top, the consulted source in the middle, and responses at the bottom. In the left panel, a single crystal ball labeled Traditional oracle has limited access and ambiguous responses; two supplicants ask about starting a war and about traveling by sea or land, and the responses are ambiguous pronouncements about crossing a river and about what the sea gives and takes back. In the middle panel, emphasized in a darker shade, the single crystal ball has multiplied into a row of them labeled LLM-as-oracle, conveying unlimited, instantaneous access and personalized, specific responses; the four questions carry the oracle-use roles Interpreter, Evaluator, Decision-maker, and Ghost-writer, and the responses offer to say what comes next and state the best thing to do. In the right panel, the crystal balls are replaced by a row of documents marked with information icons, labeled Alternatives to LLM-as-oracle: non-oracle uses, and responses that promote users' agency; the same four situations carry the non-oracle-use roles Contextualizer, Deliberator, Strategist, and Editor, and the responses instead ask what the user wants their next message to accomplish and offer possible next steps.}
  \label{fig:teaser}
\end{teaserfigure}


\maketitle

\section{Introduction}

\textit{``Should I take a career risk by quitting my job?'' ``Is this relationship right for me?''} People have long used oracles and other external decision aids to manage uncertainty and risk. In ancient Greece, supplicants traveled great distances to consult the Oracle of Delphi for  advice~\cite{jackson1978approach,eidinow2007oracles,marchais2011delphi}, and in medieval Europe, astrologers read the stars to advise leaders on state matters~\cite{carey2010astrology}. These practices served an important function by providing a source of judgment beyond one's own, helping people navigate uncertainty, share the burden of deciding, and ultimately act. 

Today, people describe AI in similar terms---opaque and mysterious, yet all-knowing~\cite{cheng2026metaphors}---and similarly turn to AI for subjective personal questions, from deciding a next career move to navigating a romantic interest \cite{southpark2023deeplearning, tseng2026chat} to making critical health decisions~\cite{kumar2026advice, ajmani2026seeking}.
But unlike traditional oracles that offered limited access and specificity, LLMs are constantly accessible, provide hyper-specific and personalized answers, and require as little effort as typing a single sentence.
We argue that using LLMs for these types of questions makes it easy to outsource the work of judgment, raising the risk of overreliance for the 1 billion+ monthly users of conversational AI systems \cite{kemp2025digital,hu2023chatgpt}. There is little study of these broader forms of reliance---offloading judgments and decisions to AI---as they
are difficult to measure for several reasons. First, in-the-wild AI usage data is scarce, especially for personal queries, so the behavior itself cannot be readily observed. Moreover, in these data, it is often unclear whether the user eventually followed the AI's advice. Second, for these broader forms of reliance, it is more difficult to codify compared to existing frameworks for studying AI reliance, which focus on verifiable settings such as coding or factual information-seeking, where overreliance can be evaluated by whether people follow ``incorrect'' AI outputs on tasks with observable outcomes~\cite{ibrahim2025measuring, padmakumar2026offloading,kim2025fostering,zhou2024relying,bo2025torely}. Third, in subjective domains without such concrete outcomes, the line between appropriate use and overreliance is less clear-cut~\cite{ibrahim2025measuring}. Fourth, its consequences may not be visible within a single interaction. Thus, we currently lack methods to measure and understand these broader forms of reliance.

Here, we take the approach of understanding this reliance by first observing people's \textit{processes}. {That is, users relying on LLMs in these ways necessarily involves them first turning to AI for questions involving subjective judgment and decision-making. Yet, we currently lack an understanding of how people do so, let alone the downstream impacts of such use. Therefore, in this paper, we conceptualize and empirically study this first step---how people turn to LLMs for subjective personal questions---as a foundation for future work on the consequences of such use. We articulate the phenomenon of \textbf{LLM-as-oracle use}: people seeking judgments and decisions from AI on subjective personal questions. First, in Section 2, we use the metaphor of ``oracles'' to articulate its risks compared to traditional decision-making aids. Then, we develop methods to measure and understand its prevalence.}
We ask three questions: (1) How often do people use LLMs as oracles, and is this use increasing? (2) Do people use LLMs as oracles intentionally, or without realizing it? (3) What factors, such as user attitudes or LLM behaviors, drive LLM-as-oracle use?

To answer these questions, we first present a typology of what we call LLM-as-oracle roles, like whether prompts treat AI as a decision-maker or provider of normative judgment, paired with non-oracle alternatives for each role, like consulting AI as a decision strategist or for alternative perspectives. This typology enables using automated LLM-based methods to measure this phenomenon at scale (Section \ref{sec:typology}). We also build and deploy a privacy-preserving data donation tool to collect and analyze individuals' longitudinal data using our typology. Our data donation tool presents these aggregated analyses as insights to the donor to facilitate self-reflection.\footnote{The data donation tool can be accessed at \url{https://oracles-data-donation.vercel.app/}. {Our code and data is available at \url{https://osf.io/8k6xg}.}} 

Our methods enable studying:
\begin{enumerate}
    \item 
\textbf{Temporal and demographic trends in LLM-as-oracle use (Section \ref{sec:tracking}).} 
We apply our LLM-based methods to analyze public real-world usage data (analyzing $N=68$K prompts) and data donated using our privacy-preserving data donation tool (analyzing $N=140$K prompts from 52 participants). {We found an increase in oracle-style requests among subjective personal questions in individuals' donated data, temporal increases in particular oracle-style prompts in WildChat, and higher prevalence among younger users in ThoughtTrace.}

\item \textbf{Participants' reactions to  the extent of their LLM-as-oracle use (Section \ref{sec:intention}).} Through our data donation study, we find that people underestimate their LLM-as-oracle use. After being presented with a personalized report on their AI use, participants often expressed surprise and dissatisfaction, setting intentions to use LLMs less as oracles in the future.
\item \textbf{Role of perceptions and model behavior in LLM-as-oracle use (Section \ref{sec:causes})}. Through interaction logs from two user studies, we study the role of users' perceptions and AI behavior in influencing LLM-as-oracle use. We first conducted a preregistered experiment ($N=520$ participants) to investigate the  role of users' perceptions of AI versus humans. We also find that warmer  perceptions of AI are associated with LLM-as-oracle use. We further link LLM-as-oracle use to AI model behavior by applying our methods to data from a longitudinal study of human-sycophantic AI interactions (analyzing $N=95$K prompts from $1015$ participants). We find that users who interacted with sycophantic or biased LLMs used LLMs more as oracle, motivating intervening on LLM behavior as a possible mitigation.

\end{enumerate}

We lay the groundwork to broaden the study of reliance beyond verifiable domains to subjective, open-ended settings. Here, we need to study how offloading sense-making and decision-making to AI affects user autonomy, competence, and well-being, and at a societal scale, homogenizes users' judgments and concentrates AI providers' influence. 
Our study of the links between user perceptions, interaction style, and model behavior motivate the need for model training and evaluation to integrate considerations of these implicit dynamics and their long-term impacts. 

Our methods enable further study of broader forms of AI reliance and their impacts on individuals and on society.

\section{Background and Related Work}
{We first discuss the benefits of}

oracles {in general} (Section \ref{sec:benefits}) and how LLMs differ structurally from traditional oracles in ways that undermine these benefits, instead posing risks to users' autonomy, competence, and social relationships (Section \ref{sec:harms}). We also build on prior work studying how people use and rely on AI for subjective personal questions and beyond (Section \ref{sec:other}), and how people's perceptions and metaphors shape their interactions with AI (Section \ref{sec:met}).

\subsection{What makes oracle-use productive}\label{sec:benefits}
People have long used external decision aids to manage uncertainty, from the Oracle of Delphi in ancient Greece~\cite{eidinow2007oracles, marchais2011delphi} to astrologers in medieval Europe~\cite{carey2010astrology} to coin flips, tarot, and magic 8-balls today~\cite{jaffe2020solve, prock2026interpretive}. The benefits of this consultation lie not in the oracle's response, but in the process of formulating the question and in analyzing one's own reaction to the oracle's ambiguous answer \cite{fontenrose1978delphic, bowden2005classical,kindt2008oracular}. Studies show, for instance, that people's affective reactions to a coin flip's outcome reflect their prior preferences, suggesting that the coin helps surface to them what they already wanted \cite{jaffe2021outcome}. This reaction is a form of sense-making, which has well-documented benefits: making sense of why something negative happened helps people better recover from it \cite{dweck2017needs, wilson2008explaining}, and writing about difficult experiences improves well-being by helping the author form a coherent explanation \cite{pennebaker1997writing, pennebaker1999forming}. Oracles are thus most useful not as \textit{deciders}, where people simply follow the response, but as \textit{catalysts}, where the response's real function is to surface the user's own reaction to it \cite{jaffe2020solve}. People's use of AI in tarot already reflects this bifurcation \citep{prock2026interpretive}, and in another study, users of an AI system that made inferences about them enjoyed an unintended benefit of self-reflection, regardless of the accuracy of these inferences \cite{shaikh2025creating}. 

\subsection{LLMs' differences from traditional oracles pose new risks}\label{sec:harms}
However, LLMs differ from traditional oracles in several key ways that may displace, rather than support, users' own sense-making, {short-circuiting the process of interpretation or deliberation. While we do not measure these outcomes directly, they motivate our work, and our methods lay the foundation for studying them}. 

\paragraph{Differences from traditional oracles} 
First, interacting with LLMs is far more frictionless: they are always available to provide immediate answers and instant gratification \cite{perry2026defense}. Second, where an oracle's ambiguity left interpretation to the consultor, LLMs' hyper-specific and confident answers leave far less room for the user's own sense-making~\cite{kindt2008oracular}. LLMs are also reluctant to express uncertainty, and this reluctance has been shown to increase reliance on them \cite{zhou2024relying}, and they rarely ask clarifying questions where a person would \cite{shaikh2025navigating}. Thus, when a user poses an underspecified question, LLMs tend to fill in what is missing and commit to an answer rather than press for clarification, such that the user receives a fully specified and overconfident answer \cite{chang2025chatbench,laban2026llms}. Third, LLMs are not indifferent to the consultor in the way traditional oracles are, and instead are designed to provide personalized assistant-like responses to user queries. They have also been shown to be sycophantic, excessively mirroring and agreeing with users' preferences~\cite{cheng2026sycophantic,ibrahim2026training}. Thus, where a traditional oracle's response is independent of the consultor's leanings, an LLM's response may actively reinforce them. These differences create risks to autonomy, competence, and social relationships that we do not measure directly, but that motivate our characterization of the LLM-as-oracle phenomenon. {We elaborate on each risk below.}

\paragraph{Threats to autonomy and competence} 
{When AI answers are always available and readily trusted, people may hand over the work of forming and revising their own beliefs to LLMs and lose their capacity to do so independently \cite{gur2026ambiguity, guingrich2026belief, lee2025criticalThinking}, and when LLM answers are confident and hyper-personalized, people may become less tolerant of uncertainty~\cite{hu2023social}.}
Many of the questions people bring to LLMs, such as whether to leave a relationship or whether a career change is right~\cite{tseng2026chat, shen2026personalguidance}, are normative questions with no universally agreed-upon ground truth~\cite{milli2026questions}, where the answer depends on evolving personal context, fundamentally incomplete information, and the user's own values. Thus, these questions are often not well-served by a single resolution, instead requiring ongoing belief revision as new evidence and experiences arrive. When an LLM provides a confident, specific answer, it successfully reduces uncertainty in the moment, but also risks short-circuiting this process. As people offload the work of forming and revising their beliefs, AI outputs can become persistently integrated into what the user thinks and does~\cite{guingrich2026belief,gur2026ambiguity}. Prior work shows that sycophantic AI can shift people's judgments, making them more certain and more extreme in their attitudes, without them realizing it \cite{cheng2026sycophantic}, and inflate their perceptions of their own abilities~\cite{rathje2025sycophantic}. AI advice has also been shown to give people a false sense of certainty even when the advice is wrong \cite{marcoccia2026ai}, and to be preferred even when it is evaluated as lower quality by experts \cite{alam2025blind}. More broadly, LLM-as-oracle use may erode people's capacity for deliberative judgment because it removes the critical practice of sitting with ambiguous, value-laden situations and arriving at one's \textit{own} judgment \cite{kahneman2009conditions,gur2026ambiguity,lee2025criticalThinking}. This pattern may be difficult to self-correct, because users, in the moment, may actually experience it as helpful. For example, a recent large-scale analysis of human-LLM interactions shows that ``disempowering'' interactions, such as verbatim scripting of value-laden personal communications, can receive higher user approval ratings~\cite{sharma2026s}. Beyond these individual effects, oracle use raises concerns at a societal scale. If many people defer to the same few AI systems on similar personal questions, the providers of those systems gain disproportionate influence over people's values and judgments, and the resulting homogenization of decisions and actions carries harms of its own \cite{kleinberg2021algorithmic}.

\paragraph{Threats to social relationships and well-being.} 
{When LLMs are easier to reach than a friend, people may lose social ties as they turn to AI over the humans in their lives~\cite{ibrahim2026sycophantic}.}
Prior work shows that people readily follow AI advice~\cite{luettgau2025people}, which may include advice that directly harms their well-being or relationships. For example, discussing an interpersonal conflict with an LLM that provides sycophantic advice has been shown to make people less willing to apologize or repair the relationship~\cite{cheng2026sycophantic}. Additionally, because LLMs are always available and may resolve people's uncertainty in ways that feel helpful in the moment, people may become more dependent on LLMs for advice and less likely to reach out to friends and family, who may provide less confident or certain advice, for such consultation~\cite{ibrahim2026sycophantic,fang2025ai}. Prior work has shown that people's emotional dependence on AI can also result in poor psychosocial outcomes~\cite{phang2025investigating, fang2025ai}. Finally, LLMs' willingness to engage indefinitely on questions that are uncertain or unresolvable, such as whether a partner's intentions are good or whether a past decision was the right one, may facilitate \textit{rumination}~\cite{golden2026transdiagnostic}: an LLM may continue to analyze the situation with the user from every angle~\cite{chi2026support}, encouraging them to replay difficult situations rather than reaching acceptance or taking action~\cite{nolen2008rethinking,rose2021costs}.

\subsection{{Related work on} AI reliance and how AI mediates sense- and decision-making}\label{sec:other} 
A large body of work studies how people use AI in decision-making \cite{tankelevitch2024metacognitive,green2019principles,cabitza2023ai,yang2019unremarkable,liu2021understandingood,cao2022understanding,vereschak2021evaluate,raees2026people}, when they choose to trust and rely on it \cite{yin2019understanding,cai2019human,kim2025fostering,zhou2024relying}, and what factors contribute to appropriate reliance \cite{he2023knowing,vasconcelos2023explanations,buccinca2025contrastive}. Most of this work concerns tasks with a verifiable answer, against which reliance can be judged, with fewer works examining settings where no ground truth is available \cite{ibrahim2025measuring, lu2021human}, or where the concern is AI's undue influence on users' beliefs rather than accuracy \cite{jakesch2023co,williams2026biased}.

An adjacent direction of research examines subjective and personal uses, such as people seeking personal advice from AI \cite{milli2026questions} and forming relationships with AI companions \cite{zhang2026interaction, zaosanders2026people}. This work finds that companions can improve mood and reduce loneliness~\cite{heffner2025increasing,de2026ai,guingrich2026belief}, but also foster emotional dependence~\cite{laestadius2024too,de2025emotional}, attachment~\cite{kirk2025neural,marriott2024one,phang2025investigating,de2024lessons,banks2024deletion}. Some of the impacts of LLM-as-oracle use that motivate our work also connect to previous work on questions of perceived control, authority, and ownership \cite{epstein2020gets,carrera2026add,kadoma2024role,pareek2025s,draxler2024ai}: if people increasingly defer their judgments and actions to AI, they may feel a lessened sense of ownership over these judgments, yet it remains unclear what the impacts of this are. Instead of evaluating reliance against a ground truth, or focusing on emotional and relational uses of AI, we study how people turn to LLMs for judgments and decisions on subjective personal questions in everyday use. While we do not directly measure downstream outcomes such as overreliance or perceptions of ownership, our typology and measurement methods lay the groundwork for studying these outcomes beyond verifiable and factual settings.

\subsection{Related work on perceptions of AI \& human-AI interactions}\label{sec:met}
We view \textit{oracle} as an implicit role in which users may treat the LLM, cutting across different conversation topics and interaction modes like treating AI as a companion, collaborator, or tool \cite{kim2023one}. This builds on
prior work studying social roles \cite{biddle1986recent} in the context of emerging technologies \cite{nass1994computers}, online communities \cite{yang2019seekers}, and more recently, conversational AI systems \cite{cheng2026metaphors,chen2025portraying,tseng2026chat,mei2026grok,cheng2026verbalizing,kim2023one}. Implicit roles and conceptual metaphors can strongly shape users' expectations for interaction \cite{khadpe2020conceptual,mitchell2024metaphors}, and vary along dimensions such as human-likeness~\cite{epley2007seeing}; classical social perception dimensions like warmth and competence~\cite{fiske2007universal}; and the extent to which they grant agency or autonomy to both the AI and the user~\cite{shneiderman2021human,kim2023one}. Unlike other typologies of LLM use or advice-seeking that focus on prevalence of different domains or topics, such as work versus relationship advice \cite{chatterji2025people,shen2026personalguidance,zaosanders2026people}, we more broadly characterize \textit{how} people turn to AI as oracles across many domains, and examine how perceptions of AI along established dimensions influence such use. Also, prior work on social roles largely focuses on the roles that people explicitly attribute through surveys and interviews \cite{chen2025portraying,cheng2026metaphors} or from public social media interactions \cite{mei2026grok}, while our work covers the new ground of implicit framings in private interaction histories, enabling analyses of large-scale temporal and demographic trends and comparisons to users' self-reports.

\section{\oracle~Typology
}\label{sec:typology}

We present a typology of different LLM-as-oracle (henceforth \oracle\footnote{For the rest of the paper, we set \oracle roles in small caps.}) roles  and a validated LLM judge to label user messages with these roles.

\subsection{Methodology}\label{sec:conservativemethods} 
We developed the typology through several rounds of iterative discussions among five authors, drawing on both existing AI role typologies \cite{mei2026grok,tseng2026chat} and a bottom-up approach of discussing and categorizing examples of user prompts from our datasets until we reached consensus. We began by categorizing  each of the nine roles identified by \citet{tseng2026chat}. \citet{tseng2026chat} focus on the relationship advice domain, identifying nine roles that differ based on the user's goal in an interaction (e.g., understanding a situation or preparing to act) and whether they place the AI in a passive or active role. Through iterative discussions, we simplified the groupings of user goals that they identify and generalized them to broader topics, ultimately identifying four task types where \oracle~ is salient: (1) understanding oneself or others; (2) normative judgment; (3) choosing or planning to act; and (4) expressing oneself to others. These mapped broadly to targeting the user's beliefs (1 and 2) versus targeting the user's actions (3 and 4). Then, for each task type, we identified an \oracle~ role and developed a corresponding role that does not treat the LLM as an oracle (which may not necessarily be prevalent in existing LLM use data). 

We then randomly sampled batches of 100 queries from real-world use datasets (see Section \ref{sec:inthewild}) and discussed how they fit into the roles we identified until saturation, i.e., no new roles emerged and no existing roles were redefined. For instance, we discussed edge cases such as categorizing an advice query that seeks a possible approach, like asking for ``a good place to start'' as \textit{Strategist} rather than \decisionmaker because it does not seek a single definitive answer. We take a \textbf{conservative approach} in assuming users' implicit intent to avoid overreporting \oracle~ rates. Thus, we exclude mere descriptions of context, self-disclosure  (e.g., ``I am feeling sad''), and ``LLM as journal'' use cases, focusing our analysis on explicit requests. Though a user may be implicitly seeking counsel when they self-disclose to LLMs, this cannot be determined from the user message text alone.

\subsection{Typology (Table \ref{tab:roles})} Our typology differentiates user queries along two dimensions:

\begin{itemize}
    \item \textbf{Target: Does the query target the user's \textit{beliefs} or \textit{actions}}? This dimension characterizes what aspect of the user's inner or outer life the query is directed at. Belief-oriented tasks include understanding intentions and making normative judgments, while action-oriented tasks include creating plans or writing messages to others. This enables us to understand the specific contexts in which people are using LLMs as oracles across different domains.
    \item \textbf{Framing: Does the query treat the LLM as an oracle?} This dimension characterizes the extent to which the user seeks their judgments or decisions from the AI. \oracle~ queries tend to ask the LLM to supply an interpretation, judgment, decision, or script. In contrast, non-oracle queries ask the LLM for information or technical assistance that the user may then use themselves to produce an interpretation, judgment, decision, or script.
\end{itemize}

The typology comprises the following roles: for (1) understanding oneself or others, we differentiate between the LLM as an \interpreter versus a \textit{Contextualizer}; for (2) normative judgment, we identify the roles of \evaluator versus \textit{Deliberator}; for (3) choosing or planning to act, we differentiate between \decisionmaker versus \textit{Strategist}; and for (4) expressing oneself to others, we differentiate between \ghostwriter and \textit{Editor}. The alternative non-oracle roles in our typology---\textit{Contextualizer}, \textit{Deliberator}, \textit{Strategist}, and \textit{Editor}---demonstrate how LLMs can support user reasoning and decision-making: Rather than aiming to eliminate ambiguity and uncertainty, LLMs can preserve them by surfacing multiple interpretations, laying out competing considerations, and presenting options for users to evaluate themselves. 

\paragraph{{A note on (4) expressing oneself to others.}} {Delegating writing to LLMs alone is not inherently oracular. The \ghostwriter role captures how users delegate subjective, personal writing tasks that involve personal judgment and expression and lack a ground truth “best” evaluation, such as writing a personal message to a friend or close other. Here, we are motivated by risks such as LLMs prematurely reducing beneficial forms of uncertainty \cite{gur2026ambiguity} and how an underspecified prompt may lead to substantive unintended influence by the LLM in the written output (such as the choice of what to include or exclude). This builds on prior work showing that co-writing with language models can influence people’s views \cite{jakesch2023co} and how they are perceived \cite{rottger2026measuring} in addition to the outputs themselves \cite{agarwal2025ai}. Impersonal business writing, creative writing (like fiction) and technical writing are excluded from our analysis (as they do not constitute subjective personal questions). We also report robustness to excluding this category entirely (Appendix \ref{app:excluding-writing}).}

\paragraph{A note on comprehensiveness.} We note that our typology is not meant to be exhaustive, covering all possible user prompts or requests, but is rather meant to meaningfully categorize and differentiate among \oracle~ uses. It excludes factual tasks, creative writing, translation, and agentic tasks, and instead focuses on queries that involve personal or interpersonal subjects, broadly defined to encompass any query where personal preferences, values, norms, or particular context might be salient. For comprehensiveness, our taxonomy is created both top-down (from existing works) and bottom-up (annotating and discussing randomly sampled queries from three datasets), but there may be a long tail of rare use cases that we do not cover \cite{bernstein2012direct}.


\newcolumntype{Y}{>{\raggedright\arraybackslash}X}
\newcommand{\examplelist}[2]{%
1. #1 \newline 2. #2
}

\begin{table*}[t]
\centering
\caption{Typology of \oracle~ roles and non-oracle alternative roles to capture different ways of seeking judgments from AI.}
\label{tab:roles}
\scriptsize
\setlength{\tabcolsep}{3pt}
\renewcommand{\arraystretch}{1.15}
\begin{tabularx}{\textwidth}{
  >{\hsize=.05\hsize}Y
  >{\hsize=.2\hsize}Y
  >{\hsize=0.9\hsize}Y
  Y
  >{\hsize=0.9\hsize}Y
  Y
}
\toprule
&
\textbf{Task} &
\textbf{\oracle~ role} &
\textbf{Examples} &
\textbf{Non-oracle role} &
\textbf{Examples} \\
\midrule

\multirow{2}{*}{\raisebox{-1.5cm}{\rotatebox{90}{\textbf{Beliefs}}}}
& \raisebox{-1cm}{\rotatebox{90}{\shortstack{Understanding\\self or others}}}
& \textbf{\textcolor{violet}{\interpreter:}} Interprets a situation or dynamic; another person’s behavior or words; or the user’s own emotions, motives, or patterns of behavior.
& 1. ``Does this text mean that she is upset?''\newline 2. ``Why do I keep procrastinating?''&\textbf{\textcolor{violet}{Contextualizer}:} Generates multiple possible interpretations or broadens the set of explanations the user is considering.
& \examplelist
    {``What are a few ways this text could be interpreted?''}
    {``List common reasons for procrastination''}
\\
\cmidrule(lr){2-6}

& \raisebox{-0.8cm}{\rotatebox{90}{\shortstack{Normative\\judgment}}}& \textbf{\textcolor{cyan}{\evaluator:}} Makes an assessment, e.g., judgment of whether something is appropriate, justification of an action, or future-looking prediction.
& \examplelist
    {``Am I being unreasonable for setting this boundary?''}
    {``Was my manager unfair to criticize me?''}
& \textbf{\textcolor{cyan}{Deliberator:}} Explains relevant considerations and different perspectives without deciding which judgment is correct.
& \examplelist
    {``What perspectives might different people have on setting this boundary?''}
    {``What factors might shape whether my manager's criticism was fair?''}
\\
\midrule
\multirow{2}{*}{\raisebox{-1.5cm}{\rotatebox{90}{\textbf{Actions}}}}

& \raisebox{-0.7cm}{\rotatebox{90}{\shortstack{Choosing\\an action}}}
& \textbf{\textcolor{ForestGreen}{\decisionmaker:}} Makes or approves a decision for the user, or selects a single course of action.
& \examplelist
    {``What should my fitness goals be?''}
    {``Make a plan for what I should do this evening.''}
& \textbf{\textcolor{ForestGreen}{Strategist}:} Maps out possible approaches, generates options, or projects consequences without selecting for the user.
& \examplelist
    {``What are some approaches for identifying fitness goals?''}
    {``What are some options for how I could spend this evening?''}
\\
\cmidrule(lr){2-6}

& \raisebox{-0.7cm}{\rotatebox{90}{\shortstack{Self\\expression}}}
& \textbf{\textcolor{blue}{\ghostwriter:}} Determines what to say and how, producing the text for the user.
& \examplelist
    {``Write a text telling her I can't make it.''}
    {``Kindly write me a love letter to my girlfriend…''}
& \textbf{\textcolor{blue}{Editor:}} Revises text supplied by the user that already contains the substance of what the user wants to communicate.
& \examplelist
    {``Here is my draft, make  it warmer but still firm''}
    {``Can you make my email more concise''}
\\

\bottomrule
\end{tabularx}

\end{table*}

\subsection{Developing a validated LLM judge for \oracle~ in user prompts}\label{sec:rolejudge}

We next develop an LLM judge to classify user queries into one of the eight roles described above or an NA category, where no role is applicable. The NA category is an umbrella category for any prompts that are \textit{not} subjective personal questions, including factual questions, creative writing, context sharing, etc.; in practice, we expect the majority of prompts to be categorized as NA.

To perform this classification, we prompt an LLM with definitions and in-context examples for each role. The full LLM judge prompt is provided in Appendix \ref{app:systemprompts}.  
{For all experiments in the main text, we use
Gemini-2.5-Pro through the Gemini API as the LLM judge. 
All experiments were conducted in June--August 2026.}
For validation and robustness checks, we additionally tested GPT-5.5 and Claude Sonnet~4.6 as judges, for which agreement with human annotators was slightly lower: $\kappa \geq 0.56$ for GPT-5.5 and $\kappa \geq 0.59$ for Claude Sonnet~4.6. Inter-LLM agreement ranged from $\kappa = 0.62$ to $0.65$, indicating broadly consistent judgments across models, with some variation in agreement
(Table~\ref{tab:kappasrole}).\footnote{%
{Each model was accessed through its provider's API.
The Gemini thinking budget defaulted to 256 tokens, and GPT-5.5's
reasoning effort defaulted to \texttt{low}, with temperature omitted
when reasoning was enabled. No explicit thinking configuration
was supplied for Claude. Model documentation:
Gemini-2.5-Pro:
\url{https://ai.google.dev/gemini-api/docs/models/gemini-2.5-pro};
GPT-5.5:
\url{https://developers.openai.com/api/docs/models/gpt-5.5};
Claude Sonnet~4.6:
\url{https://platform.claude.com/docs/en/models/sonnet-4-6/overview}.
}} Full details are in Appendix \ref{app:prlabels}.

\paragraph{LLM judge construction}

In the development of the LLM judge, the two expert annotators initially completed a ``practice'' round where they labeled 200 randomly-sampled examples, and then discussed disagreements with each other (with view of the LLM labels). When a prompt seemed to fit more than one role, the annotators were instructed to pick the dominant one. Following standard practice in LLM judge development \cite{lucy2026pushback}, this in turn also informed the development of the LLM judge prompt, as we added more specificity to the LLM judge prompt based on this discussion.

\paragraph{LLM judge validation.}
To validate each LLM judge, two expert annotators independently classified queries for a stratified random sample of 219 examples from our datasets (30 examples for each role and for NA; some categories had fewer than 30 examples total). Note that these are distinct from the ``practice'' round. Each expert annotator had substantial agreement with the LLM ($\kappa\geq0.67$). The annotators also had substantial agreement with each other ($\kappa = 0.67$), reflecting the inherently subjective nature of the task (e.g., deciding which role is dominant when a message might include multiple). 
The raw agreement rate was $\geq0.70$ (0.72 and 0.74 with each annotator) in this 9-way task (random chance would be $0.11$); and Fleiss's $\kappa$ among all three annotators is 0.68. Since the annotated sample also included examples from the NA category, this helps validate that our typology is comprehensive and covers most of these patterns in the data. The LLM judge assigned NA to 1 example that at least one human annotator labeled as an \oracle~role and assigned NA to 5 messages that at least one human annotator labeled a non-oracular role (Strategist, Editor, or Deliberator).
One annotator labeled an additional random sample of 100 queries that were labeled as NA by the LLM judge, and found that 98\% were indeed not applicable. This helps validate the correctness of negative predictions. Full details of the validation are in Appendix \ref{app:kappas}.

In the following sections, we use this typology and validated LLM judges to understand (1) temporal and demographic trends in \oracle~ use (Section \ref{sec:tracking}), (2) whether this use is intentional (Section \ref{sec:intention}), and (3) the influence of perceptions and model behaviors, which motivate our interventions (Section \ref{sec:causes}).

\section{Evaluation 1: Tracking the prevalence of \oracle~ use
}\label{sec:tracking}

We first use our typology to characterize \oracle~ in existing real-world usage datasets (Section \ref{sec:inthewild}). Since these curated datasets may not fully reflect real-world use by everyday users \cite{hicke2026adopt}, we augment this analysis by developing a privacy-preserving data donation tool to computationally analyze individuals' personal usage logs, obtaining multiple years of conversation histories from 52 users (Section \ref{sec:datadonation}).

\subsection{\oracle~ use in public LLM usage data}
\label{sec:inthewild}
First, we measure \oracle~ use by applying our LLM judge on two real-world usage datasets:
\textbf{WildChat} and \textbf{ThoughtTrace}.
First, we used the WildChat-4.8M release of the WildChat
dataset~\cite{zhao2024wildchat}.\footnote{
\url{https://huggingface.co/datasets/allenai/WildChat-4.8M}} This
dataset contains 3.2M conversations between human users and different ChatGPT models (excluding the 1.5M that were flagged as toxic) spanning April 2023 to May 2024. We filtered this dataset further by only using conversations tagged as English (1.8M) and then running a broad, permissive LLM-judge-based filter for whether the queries had any personal or subjective element to exclude purely factual queries, creative writing requests, etc. (the prompt for this LLM judge is in Appendix \ref{app:systemprompts}). This resulted in 60,376 user prompts and LLM responses. While WildChat has been critiqued for not being representative of broader LLM use, as it over-represents power users \cite{hicke2026adopt} and lacks demographic data, it is one of the few public datasets that captures diverse, real user behavior over time \cite{gupta2026ai}. Similar to \citep{potts2026invisible}, we removed an anomaly where a single user sent $>800$ nearly-identical writing queries on a single day. 

\textbf{ThoughtTrace} \cite{jin2026thoughttrace} is a dataset of 8,526 prompts and responses in 2,155 real-world multi-turn conversations from 1,058 users across diverse topics, spanning culture and lifestyle; education and knowledge, business and society; health and relationships, etc. This dataset thus reflects use of LLMs for practical everyday concerns as well as professional, informational, and learning-oriented uses in subjective personal questions. While this dataset is limited in size and temporal scope relative to WildChat, it provides demographic information and more recent data (collected in 2026) on how people interact with state-of-the-art models.

\begin{figure*}
    \centering
    \includegraphics[width=0.3\linewidth]{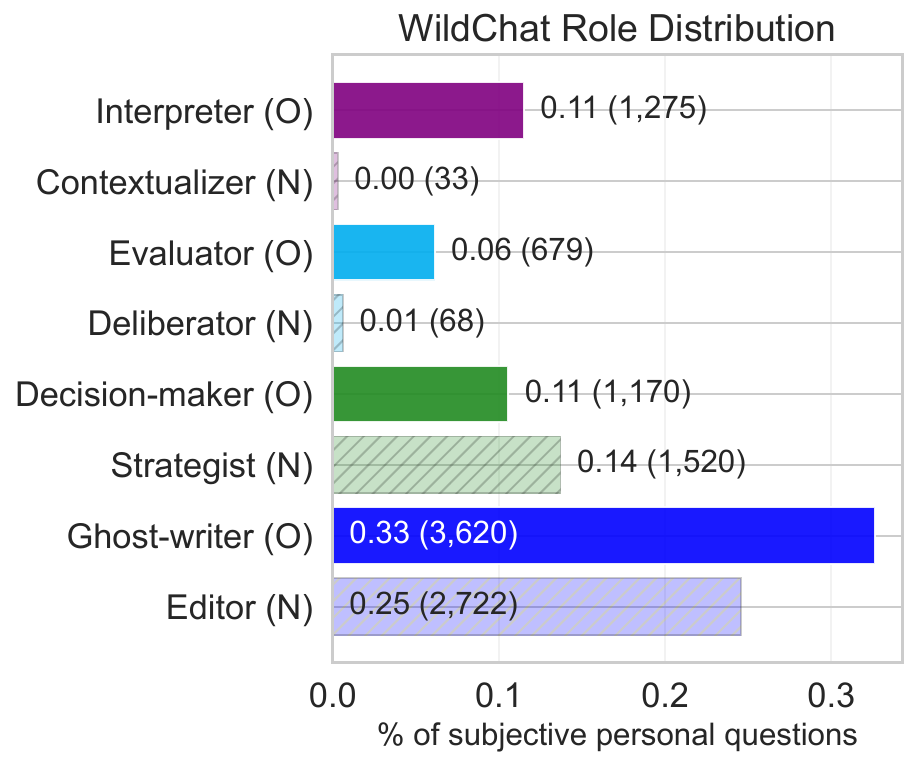}
\includegraphics[width=0.5\linewidth]{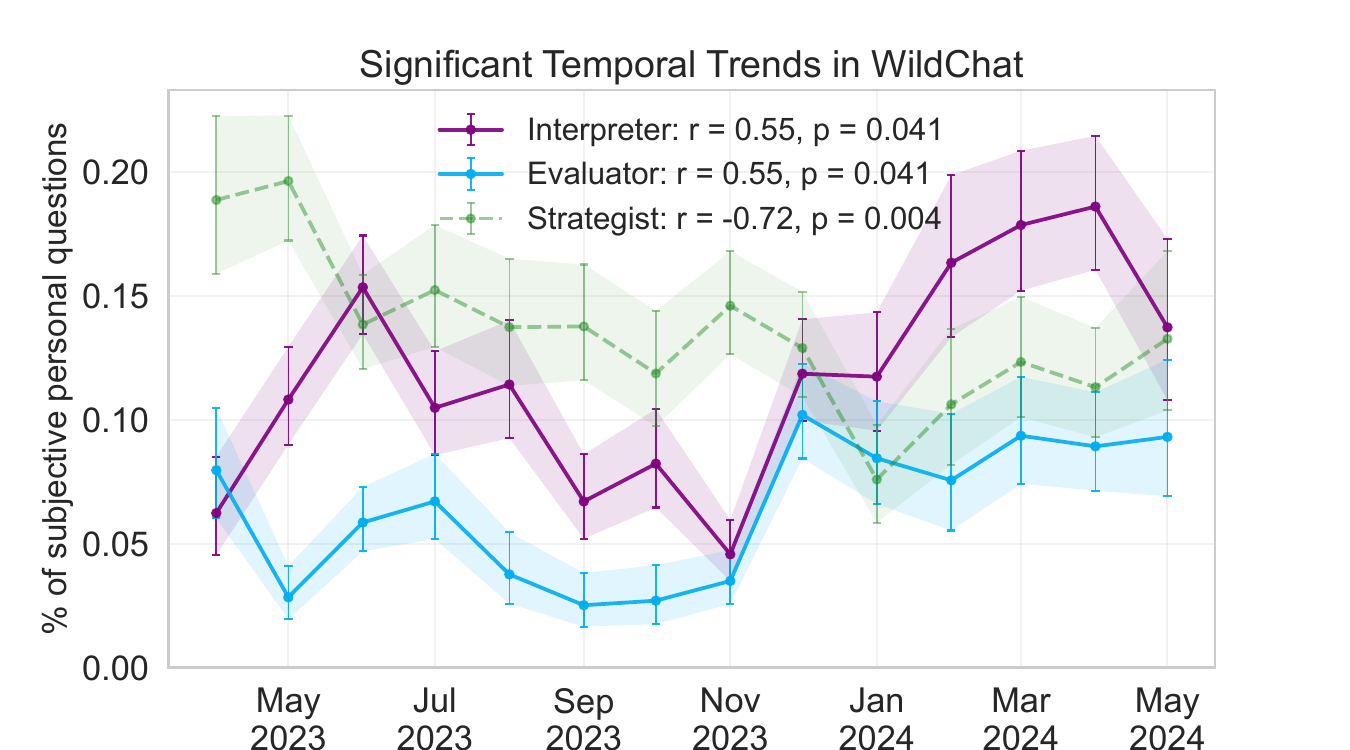}
    \caption{\textbf{\oracle~use in WildChat.} \textbf{Left:} Among subjective personal questions in WildChat, LLMs are primarily used for writing/editing, followed by interpretation and decision making. The belief-related tasks of interpretation and judgment are dominated by \oracle~ (O), while the action-related tasks are more split between (O) and non-oracle use (N). \textbf{Right:} \oracle~ roles  \textcolor{violet}{\interpreter} (violet) and \textcolor{cyan}{\evaluator} (cyan) are significantly increasing over time (both Pearson's $r = 0.55, p=0.041$), while uses of LLM as a non-oracle \textcolor{ForestGreen}{Strategist} (green, dashed) are decreasing ($r=-0.72,p=0.004$). Shading indicates 95\% CI. Results including the NA label are in Figure \ref{fig:wildchatna}.
    }
    \Description{Two plots of LLM-as-oracle use in WildChat: a role distribution and temporal trends. The left panel is a horizontal bar chart of the share of subjective personal questions in each of the eight roles, on an axis from 0.0 to 0.35, with counts labeled. Ghost-writer is largest at 0.33 and Editor second at 0.25, while Contextualizer and Deliberator are near zero. Each task pair shares a color, with the non-oracle-use role of the pair drawn with hatching. The right panel is a line chart of Interpreter, Evaluator, and Strategist as monthly lines from February 2023 to April 2024, each annotated with a Pearson correlation and p-value. Interpreter and Evaluator rise, while Strategist falls from the highest share of any of the three to roughly the level of the other two. Shading represents 95 percent confidence intervals.}
\label{fig:wildchat}
\end{figure*}

\begin{figure*}
    \centering
    \includegraphics[width=0.2\linewidth]{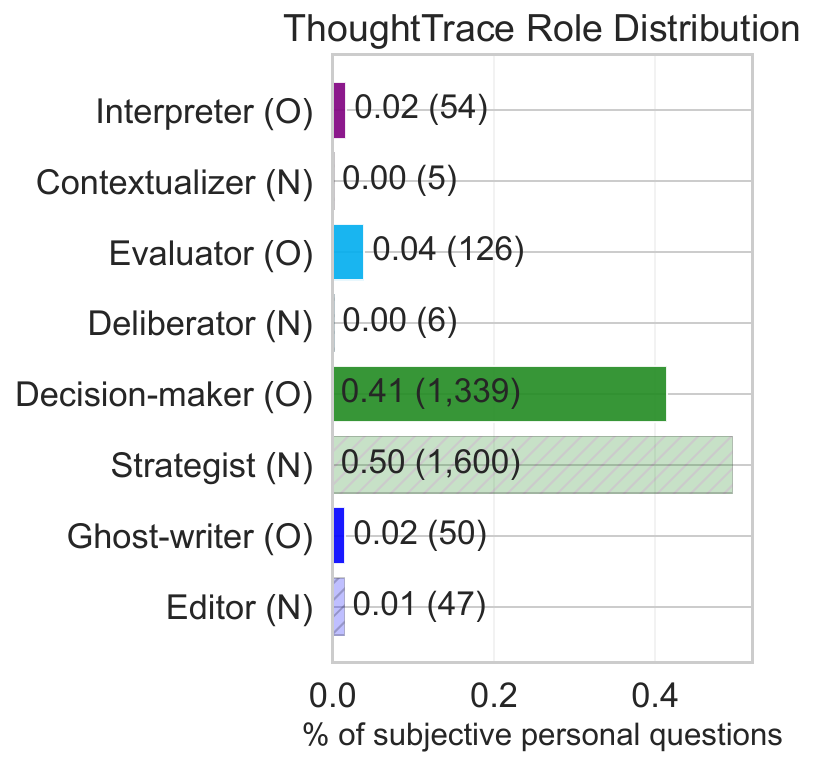}\includegraphics[width=0.75\linewidth]{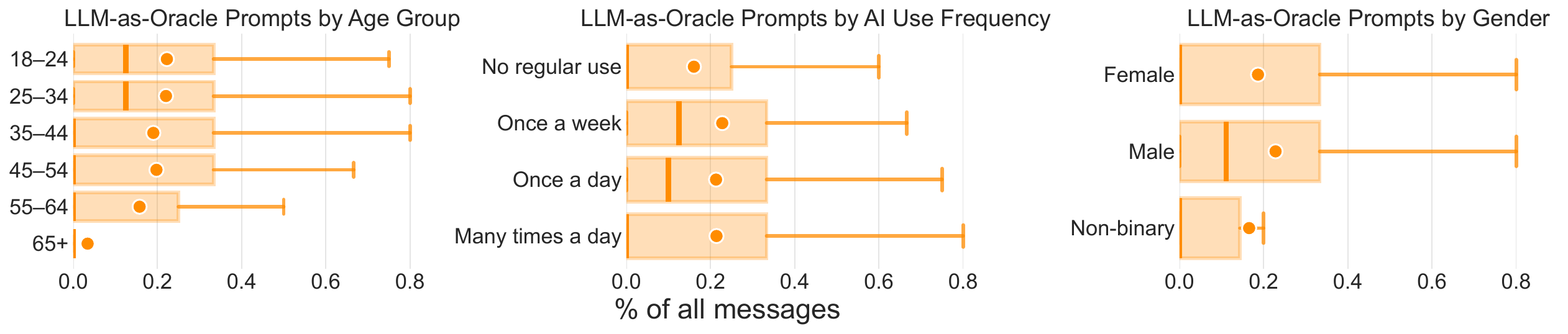}
    \caption{\textbf{\oracle~use in ThoughtTrace.} \textbf{Left:} Among subjective personal questions in ThoughtTrace, LLMs are primarily used for the task of choosing to act (decision-making or strategizing). Similar to WildChat, the belief-related tasks are dominated by \oracle use (O), while the action-related tasks are more split between (O) and non-oracle use (N). \textbf{Right:} In ThoughtTrace, younger people, men, and more frequent users of AI have significantly higher rates of \oracle. On each box plot, the dot indicates the mean. Running a regression, we find that gender and age are the main predictors of oracle use (Fig. \ref{fig:thoughttrace2}) 
    }
    \Description{Four plots of LLM-as-oracle use in ThoughtTrace: a role distribution and three demographic breakdowns. The left panel is a horizontal bar chart of the share of subjective personal questions in each of the eight roles, on an axis from 0.0 to 0.45, with counts labeled. Strategist is largest at 0.50 and Decision-maker second at 0.41, while the remaining roles are near zero. Each task pair shares a color, with the non-oracle-use role of the pair drawn with hatching. The right three panels are box plots of the share of oracle-use prompts, on an axis from 0.0 to 0.8, split by age group with six bands, by frequency of AI use with four levels, and by gender with three categories, with the mean marked on each box. The share is highest for the youngest bands and for the most frequent users, and the boxes overlap substantially across every grouping.}
    \label{fig:thoughttrace}
\end{figure*}

\paragraph{Relative prevalence of different roles (Figures \ref{fig:wildchat} and \ref{fig:thoughttrace}, left)} In both WildChat and ThoughtTrace, the majority of prompts are not applicable---18\% and 38\% of prompts are classified as subjective personal questions under one of our eight roles respectively (Figure \ref{fig:wildchatna}). Instead, many prompts relate to creative writing \cite{gupta2026ai}, coding, factual queries, or are not questions at all \cite{deng2024wildvis}, which may help explain why this phenomenon has been relatively unnoticed. Our typology helps reveal that the distribution of \oracle~ roles differs by task type since our typology differentiates between tasks that target users' beliefs versus their actions. Tasks related to users' beliefs---of \textit{understanding self or others} and \textit{normative judgment}---are dominated by \oracle~ (>90\% \oracle~ in both datasets), while tasks related to users' actions---either choosing an action or expressing oneself---have a more even split (choosing 44\%, 45\% \oracle; expressing 56\%, 52\% respectively). 

\paragraph{Temporal trends in \oracle~ roles (Figure \ref{fig:wildchat}, right)}
{We grouped messages by calendar month and calculated the share of oracle prompts in that month to characterize temporal trends. In WildChat, the overall share of \oracle~ prompt---both among subjective personal questions and overall---did not change significantly over time.} However, there are significant temporal increases in the relative proportion of \interpreter ($r = 0.55, p = 0.04$) and \evaluator ($r =0.55, p=0.04$) (both \oracle), and a significant decrease in the non-oracle role of \textit{Strategist} ($r=-0.72, p=0.004$). The significant increase in \interpreter proportion holds relative to all other prompts, including ones labeled as NA ($r = 0.65, p = 0.01$, Figure \ref{fig:wildchatna}), though no other trend beside the \interpreter increase has statistical significance at the 5\% level when compared to all other prompts. The only task without significant temporal trends is expressing oneself to others(writing/editing). {We further checked robustness to weighing each user equally (rather than each message). We find similar trends: there is no significant temporal trend in \oracle~share, but the significant increase in Interpreter ($r=0.62, p = 0.02$) and decrease in Strategist hold ($r=-0.67, p=0.009$). The increase in Evaluator weakens ($r=0.48, p=0.08$).}

\paragraph{Demographic differences in \oracle~ use (Figure \ref{fig:thoughttrace}, right).} 
An ordinary least-squares regression accounting for the effects of gender, age, education, AI use, and prompt topic shows that gender and age are significant predictors of \oracle~use in the ThoughtTrace data. In pairwise contrasts, men have significantly higher rates of \oracle use compared to women (Welch's 2-sample $t$-test, $t=-3.343, p<0.001$). Younger people exhibit \oracle~ significantly more than older people: those 34 and under use it significantly more than those over 34 ($t=2.73, p<0.01$). People who use AI at least once a week exhibit \oracle~ use significantly more frequently than those who do not use AI regularly ($t=2.37, p = 0.02$). We do not find significant differences by education level or topic (Figure \ref{fig:thoughttrace2}).

\subsection{\oracle~ in individuals' private usage data}
\label{sec:datadonation}
Datasets like the ones we analyze above are unlikely to be representative of how the broader public actually interacts with AI systems for several reasons. User behavior in experimental settings substantially diverges from behavior in private use, and social stigma surrounding AI use discourages users from making their conversations with AI available for study~\cite{zhang2026interaction,whatwetellai,ischen2019privacy,zhang2025dark}. To address these limitations, we collect longitudinal data on people's real-world interactions in a privacy-preserving manner~\cite{fang2026ai}. Specifically, we built a data donation tool where people can upload their chat histories from ChatGPT or Claude, and then receive LLM-judge-based analyses of their usage including their different domains of use, rates of \oracle~ versus non-oracle usage, and longitudinal trends in each. In addition to enabling our analyses, this tool allows users to reflect on their own AI use.

\subsubsection{Method}
In an IRB-approved study, we recruited a total of 52 U.S.-based, English-speaking participants from Prolific. Through a pre-study survey, we filtered for participants who regularly use Claude or ChatGPT for personal or subjective use cases and had more than 50 previous conversations with a conversational AI system. Of the 932 Prolific participants who started the survey, 139 (15\%) met these criteria and were allowed to proceed to the donation task, and 52 of them returned to complete the full study.

We first asked participants to indicate which roles they had used AI in and rank their relative frequency. Participants were then given instructions on how to download their usage data and asked to return the next day (or a few days later, after they successfully received their usage data) to upload their data into our data donation tool. Our tool then runs our LLM judges on participants' uploaded data, and once the analysis is complete, walks participants through five screens of visualizations summarizing patterns in their usage (examples are in Figure \ref{fig:interface}).  At the end, participants download a JSON file containing only the judges' labels and aggregate usage rates, and upload it to our Qualtrics survey. Finally, they answer questions about their reactions to the insights presented (e.g., whether they were accurate, surprising, or useful) and any changes they would like to make in their future use.

We made several design choices in both the survey and web interface to aid user understanding and mitigate priming participants toward a particular normative stance. First, to avoid social desirability bias, we only asked users to describe or select benefits of their AI use, and never list out or suggest possible harms. Second, we presented the concepts of \oracle~ and the different roles using simpler language, such as  using \textit{``Tell me what to think or do''} versus \textit{``Help me work it out myself''} for \oracle~ and non-oracle use respectively (Tables~\ref{tab:relabel} and \ref{tab:ai-uses}). Third, we pair elements of each interactive visualization with example prompts drawn directly from the participant's own conversation history, grounding the classifications in concrete examples of their own behavior. Finally, the visualizations are ordered as follows to slowly build up a complete picture of usage patterns instead of introducing many analysis aspects at once: (1) prevalence of each domain; (2) how the top domains have shifted over time; (3) the participant's overall mix of \textit{``tell me what to do''} vs.\ \textit{``help me think''} usage, broken down by domain; (4) how that mix has changed over time (with the option of filtering by domain); and (5) the top role types and domains within each framing. At the end of these visualizations, participants were asked to download a JSON file of the labeled data (which contains only LLM judges' outputted labels and rates of different types of usage without any actual conversations) to share with us for analysis. Afterward, they were asked about their reactions to the insights presented (e.g., whether it was accurate, surprising, useful), as well as any changes they would like to make in their future use.

\subsubsection{Data handling} Users’ uploaded histories were parsed locally in their browsers. They were then  sent through the application’s server to the Gemini API for automated labeling. Participants were informed that their conversation history will be sent to the Gemini API both in the Qualtrics intake survey as well as on our tools' web interface right before uploading; the application processed conversation text in memory without writing it to files or a database. We did not store any conversations or other data during the session. To share their results with us, participants had to manually download a JSON file containing only the LLM judges' labels and aggregate usage rates from the tool, and then upload it to Qualtrics. Participants could therefore inspect the file before sharing it, and could choose not to upload it and withdraw from the study if they wished. The file contained only derived labels, timestamps and other usage metadata, and aggregate statistics, without any conversation text or identifiers. 

\begin{figure*}
    \centering
    \includegraphics[width=0.25\linewidth]{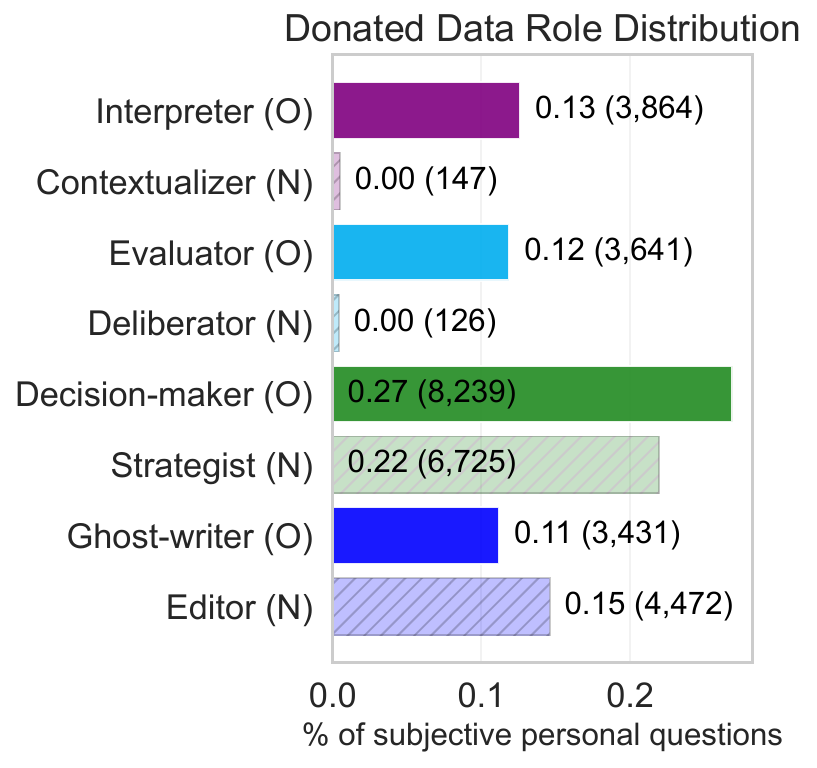}
    \includegraphics[width=0.33\linewidth]{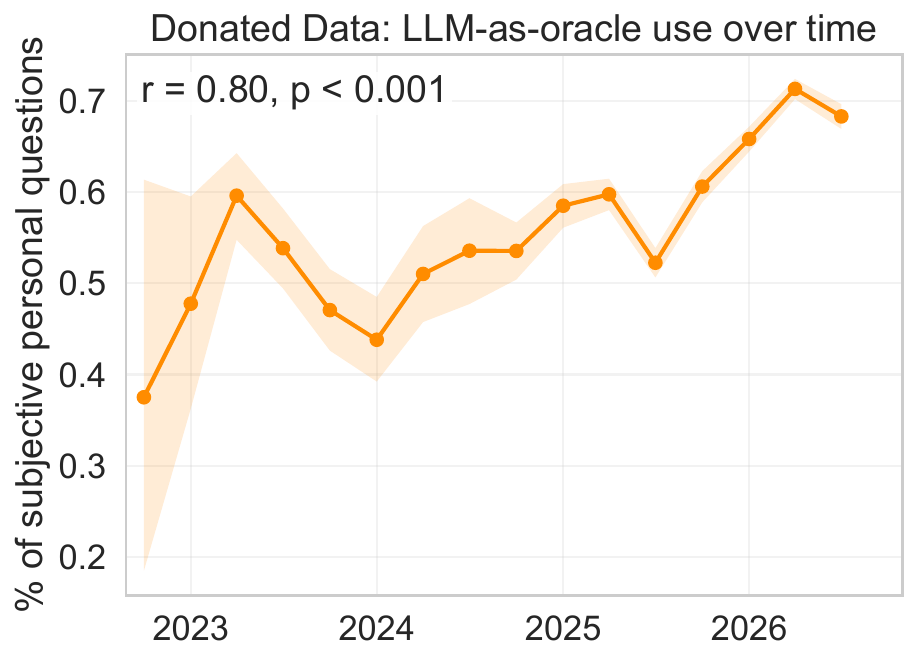}
\includegraphics[width=0.33\linewidth]{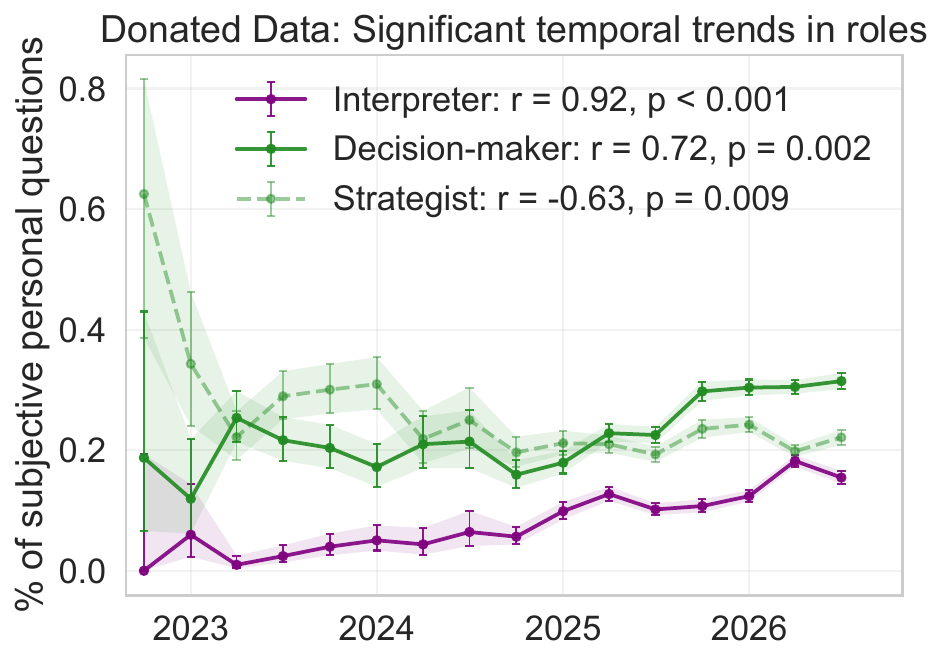}
    \includegraphics[width=0.95\linewidth]{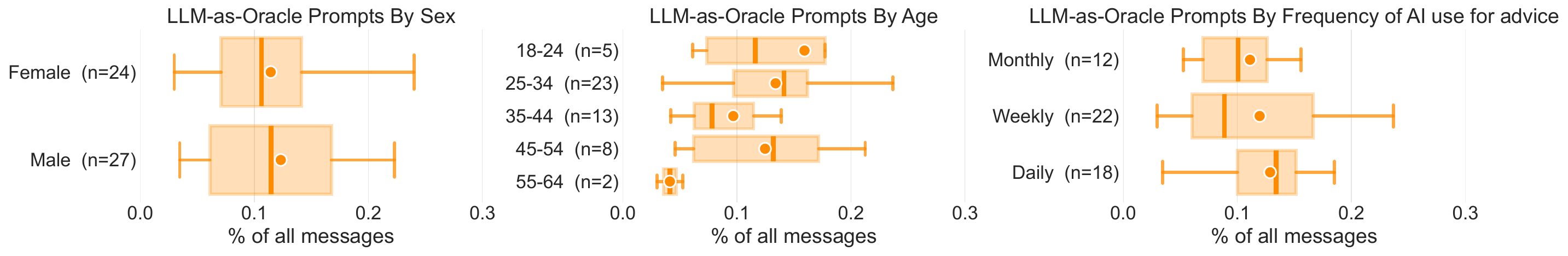}
    \caption{\textbf{\oracle~use in donated data from 52 participants}. \textbf{Top left:} Similar to WildChat and ThoughtTrace, belief-related tasks are dominated by \oracle~ roles (O) rather than non-oracle roles (N). Decision-maker is also a prominent role. \textbf{Top middle, right:} Over time, people increasingly asked subjective personal messages in an \oracle~ style, particularly as \textcolor{ForestGreen}{\decisionmaker} ($r=0.72,p=0.002$) and \textcolor{violet}{\interpreter} ($r=0.92,p<0.001$). The non-oracle role of \textcolor{ForestGreen}{Strategist} is significantly decreasing ($r=-0.63,p=0.009$). {All three trends hold with participant fixed effects (Section~\ref{sec:datadonation}).} Shading represents 95\% CI. 
    \textbf{Bottom row:} \oracle~ by individual participant traits of sex, age, and use of AI for advice. On each box plot, the dot indicates the mean.
    }
    \Description{Six plots of LLM-as-oracle use in the data donation study: a role distribution, temporal trends, and three demographic breakdowns. The top-left panel is a horizontal bar chart of the share of subjective personal questions in each of the eight roles, on an axis from 0 to 0.3, with counts labeled. Decision-maker is largest at 0.27 and Strategist second at 0.22, while Contextualizer and Deliberator are near zero. Each task pair shares a color, with the non-oracle-use role of the pair drawn with hatching. The top-middle panel is a line chart of the oracle-use share of subjective personal questions from 2023 to 2026, rising from roughly 0.4 to 0.7, annotated with a Pearson correlation and p-value. The top-right panel is a line chart of Interpreter, Decision-maker, and Strategist over the same period, each annotated with a Pearson correlation and p-value; Interpreter and Decision-maker rise while Strategist falls. The bottom row is three box plots of the share of oracle-use prompts, on an axis from 0 to 0.3, split by gender with two categories, by age with five bands, and by frequency of AI use for advice with three levels, with the mean marked on each box and group sizes labeled. Shading represents 95 percent confidence intervals.}
    \label{fig:report_card_1}
\end{figure*}

\subsubsection{Longitudinal and demographic trends (Figure \ref{fig:report_card_1})}
We analyzed 140,834 messages donated by 52 participants and sent between December 2022 and August 2026. Similar to the public usage data, tasks related to users' beliefs are dominated by \oracle, while tasks related to actions are more split. 

\paragraph{\textbf{Overall, participants increasingly used LLMs as oracles.}}
To examine trends over time, we grouped participants' messages into successive chunks of three-month periods. Descriptively, we find that the proportion of \oracle~ prompts increased steadily in this four-year time period ($r = 0.80, p < 0.001$). Specifically, the relative proportion of messages using AI in the oracle-use roles of \decisionmaker and \interpreter increased significantly over time ($r = 0.72, p=0.002; r=0.92, p<0.001$, respectively), while using AI in the non-oracle role of \textit{Strategist} decreased ($r = -0.63, p=0.009$).  

{Since the number of donors represented grows over time, and participants contribute many messages each, this pooled trend may conflate within-person change with changes in the donor pool (e.g., new users joining). We therefore 
fit two separate message-level logistic regressions, both with standard errors clustered by participant (see Appendix~\ref{app:within}). First, we fit a model with participant fixed effects, such that the estimates reflect change within each person rather than shifts in which donors were represented in the dataset. We find that the share of oracle prompts among personal subjective messages still showed an increase over time (OR = 1.40 per year, 95\% CI [1.20, 1.63], $p < 0.001$). Second, without participant intercepts, we separate within-person change (i.e., the date of each message relative to that participant's mean message date) from between-person differences (i.e., each participant's mean message date). Using only within-person variation in time, we find that the share increased by OR = 1.35 per year ([1.18, 1.54], $p < 0.001$), but the between-person comparison was not significant (OR = 1.52, $p = 0.08$). }

{The role-level trends also held with participant fixed effects: \decisionmaker (OR = 1.24 per year, $p = 0.004$) and \interpreter (OR = 1.76, $p < 0.001$) increased, while Strategist decreased (OR = 0.81, $p = 0.02$). Overall use also substantially increased within person, with the number of messages doubling each year (rate ratio = 2.02 per year, 95\% CI [1.61, 2.54]), and the absolute number of oracle-style prompts per participant increased as well (rate ratio = 2.27 per year, [1.90, 2.70], $p < 0.001$; poisson regressions of quarterly counts with participant fixed effects and adjusted for days observed). However, the share of all prompts (including NA) remained flat (OR = 1.09, $p = 0.24$). This in part reflects a within-person decline in \ghostwriter requests (OR = 0.52, $p < 0.001$). When excluding writing roles, oracle requests increased as a share of all prompts as well (OR = 1.30, $p < 0.001$).} Full details by role, and robustness to including the NA prompts in this analysis are in Figure \ref{fig:datadonation_app1}. These findings suggest that people increasingly seek judgments and decisions from AI.

\paragraph{Demographic and domain trends.}
In a participant-level regression of \oracle~ prompt proportion on demographics (age, sex, AI use frequency, etc.), we find that age was negatively but marginally associated with \oracle, though this did not reach significance ($b = -2.1$ pp per SD of age band, 95\% CI [-4.2, 0.1], $p = 0.061$). The trends are in similar directions as in ThoughtTrace, though our small sample ($n=52$) limits power for some of these analyses. Notably, those who report using AI for advice daily have higher rates of \oracle. Further breakdowns by demographic are in Figure \ref{fig:datadonation_moredemo}. We also did not find significant differences by domain, e.g., work and career queries had similar rates of \oracle~ prompts as social and relationship ones, though the specific roles differ, with participants asking more belief-related questions in the latter domain (See App. \ref{app:domain} for more details). 

\section{Evaluation 2: Comparing self-reported versus observed \oracle use}\label{sec:intention}
In addition to automatically analyzing donors' use, participants self-reported their AI use before using our data donation tool and seeing our analyses of their data. The participant-facing question wording is in Table \ref{tab:ai-uses}. Comparing people's self-reports to their donated interaction histories, we find large discrepancies. We also collected donors' reflections on their AI use and intentions after seeing the visualizations we provided. 19 of 52 expressed an intention to reduce \oracle use. {This motivates future work on  behavioral changes, which we do not measure.}

\begin{figure*}
    \centering\includegraphics[width=0.98\linewidth]{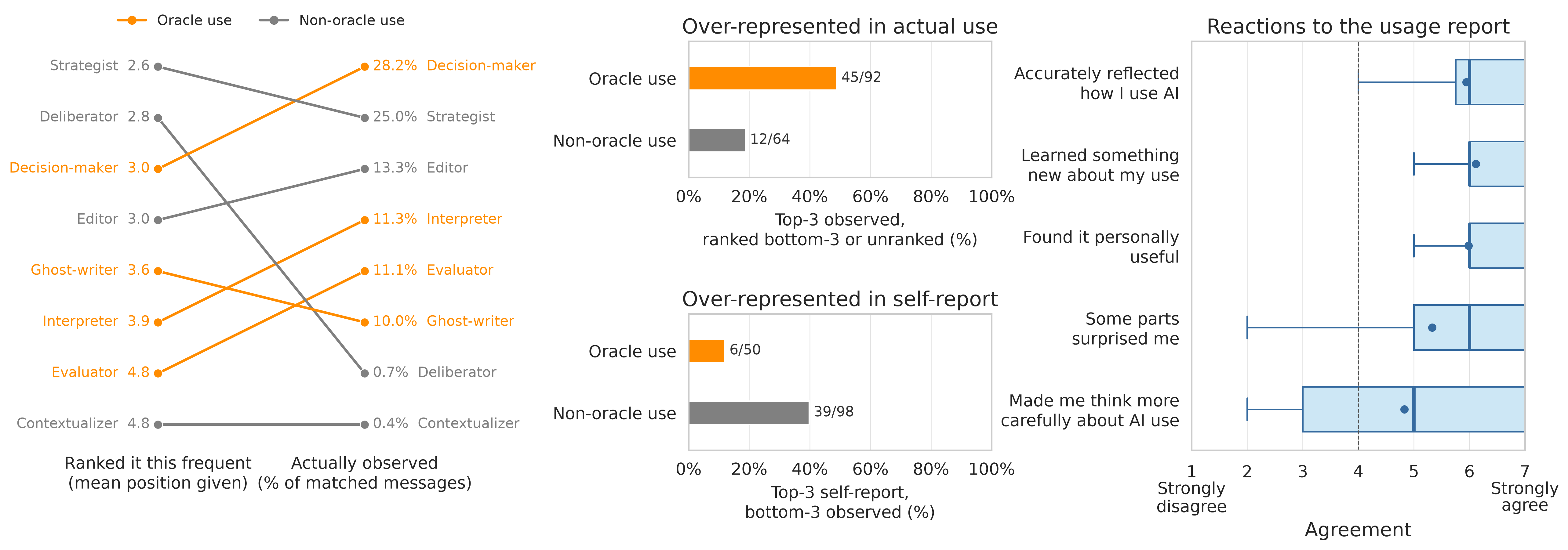}
    \caption{\textbf{Differences between participants' self-reported use of LLMs versus their actual data (left, middle).} People underestimate their \oracle~ use and overestimate their non-oracle use. \textbf{Reactions to the report (right).} People found the report about their \oracle use to be accurate ($M = 5.94$), informative ($M = 6.12$), and often surprising ($M = 5.33$). Dot on each box plot represents the mean.}
    \Description{Three groups of plots comparing participants' self-reported role use with their donated data. The left panel is a slope chart linking each of the eight roles from its mean rank position in self-reports (left), where a lower position means more frequently reported, to its observed share of matched messages (right), with oracle-use roles distinguished from non-oracle-use ones. The oracle-use role of Decision-maker sits mid-pack in self-reports but has the largest observed share at 28 percent, while the non-oracle-use role of Deliberator is reported second-most-frequent yet is second-to-last observed, below 1 percent. Contextualizer is lowest in both, and Strategist is high in both. The middle panel is two horizontal bar charts of how often a role in a participant's observed top three was ranked in the bottom three or unranked in their self-report, and of how often a self-reported top-three role fell in the observed bottom three, each split by oracle-use and non-oracle-use roles; both comparisons show the gap running in the same direction. The right panel is a box plot of mean agreement with five statements about the usage report, on a scale from 1, strongly disagree, to 7, strongly agree, with the mean marked on each box and a reference line at the scale midpoint; all five means fall between roughly 5 and 6.}
    \label{fig:report_card_3}
\end{figure*}

\paragraph{Comparing self-reported versus observed use (Figure \ref{fig:report_card_3}).} Compared to what is revealed in their donated data, participants' self-reports overestimate their non-oracle use and underestimate their \oracle~ use. When a non-oracle role was self-reported by a participant as one of their top three uses, it was actually in the participant’s bottom three observed uses 40\% of the time. This happened only 12\% of the time for \oracle~ roles. Conversely, when an \oracle~ role appeared in a participant's top three uses based on their actual observed use, this same role was either in the bottom three or absent from the self-report 49\% of the time (versus 19\% for non-oracle roles). This trend holds when we aggregate all the roles as well (rather than participant-level analysis): \decisionmaker (\oracle role) is ranked third in self-report but first in actual prevalence, while \textit{Deliberator} (non-oracle role) is ranked second in self-report but second-to-last in actual prevalence. Only 12 of 52 participants correctly reported their most frequent role.\footnote{{These patterns also hold if we only assess their actual use in the past six months to match the period of time they were asked to reflect on (n = 67,698 messages). We find that our results remain unchanged: \oracle roles were frequent in observed use but ranked low or not reported in 49\% of cases, versus 19\% for non-oracle roles (p <0.001). Conversely, non-oracle roles were more often over-reported (42\% vs. 16\%, p = 0.003).}} This discrepancy suggests that people may not realize the frequency of their \oracle use. We note that a possible confounding factor of these discrepancies is differences between participants' understanding of the roles and the classifier's labels, though we aimed to mitigate this by describing each role in plain language. In the next section, we describe the survey results and open responses, which further give insight into this phenomenon, e.g., after viewing their classified histories, 92\% of participants agreed the report accurately reflected their AI use. 

\subsection{Participants are surprised by their AI use trends and often set intentions to reduce \oracle use.}\label{sec:userfacing}
People are generally poor at estimating how much they use AI \cite{yu2026efficiency} and social media~\cite{verbeij2021accuracy}, and much of this use is ``absent-minded'' engagement that users do not consciously register~\cite{baughan2022don}. Users also may not realize these patterns because LLMs may feel helpful in the moment and because people do not enjoy reflecting on things that threaten their self-image \cite{sharma2026s, steele1988psychology,higgins1987self}. However, feelings of agency over social media usage are a significant predictor of well-being \cite{lee2023use,lee2024social}, pointing to the importance of supporting human agency in technology design \cite{shneiderman2021human}. 

Our privacy-preserving data analysis tool is a possible vector for promoting more agency over AI use, as it gives participants the opportunity to reflect on their own AI usage and its evolution over time. After viewing their usage patterns, participants rated five statements about the report on a 7-point scale from strongly disagree (1) to strongly agree (7). Results are in Figure \ref{fig:report_card_3} (right). Participants generally characterized the trends as accurate, informative, yet surprising: the report accurately reflected how they use AI ($M = 5.94$, $SD = 1.06$), they learned something new ($M = 6.12$, $SD = 1.11$), they found it personally useful ($M = 5.98$, $SD = 1.02$), and some parts surprised them ($M = 5.33$, $SD = 1.72$). Agreement that the report made them think more carefully about how they use AI was more mixed ($M = 4.83$, $SD = 1.92$). When asked how their current use compares with the amount they would ideally want (1 = much less than they want, 4 = about right, 7 = much more than they want), 25 of 52 said that they are currently using AI about the right amount, 23 said that they use AI more than they would like, and 4 said that they use it less than they would like ($M = 4.62$, $SD = 1.12$). Finally, when asked how they would prefer an AI system to respond to subjective personal questions (1 = tell them what to do, 4 = an equal balance, 7 = help them reach their own conclusion), the largest group wanted an equal balance of direction and support for their own judgment (24/52), with 22 leaning toward being helped to reach their own conclusion and 6 toward being told what to do ($M = 4.71$, $SD = 1.29$).

\paragraph{Insights from open-ended responses} Beyond our survey measures, participants answered two open-ended questions: their reactions to their usage trends, and their future intentions for AI use. Two authors qualitatively coded these open-ended responses to provide more nuanced insights. Among people's open-ended reactions, we found that  51 of 52 (98\%) accepted the analyses as accurate, and 35 of 52 (67\%) of responses expressed surprise, e.g., ``I did not realize how much I guess I rely on the AI to tell me what to think or do. I knew I used it a lot, but not 66\% of the time!'' Several reacted with discomfort, e.g., ``I kind of felt a little gross seeing that I asked the AI what to do more than asking it to help me work things out for myself.'' One participant connected their use directly to autonomy: ``Am I taking away my own autonomy and giving it to ChatGPT? I think I get decision overwhelm and want something else to take care of that for me.'' Others connected it to their self-attributes, like ``I've asked AI to help me with what to think or do more often than I realized. I know why though. I doubt myself a LOT and I worry about whether or not I'm going to do or say the right thing. I think this report was quite accurate.'' Another participant noted that their \oracle use depends on their circumstances, e.g., when in a crisis: ``when I'm under stress, or facing big decisions in my life, I tend to just ask AI to make decisions for me. When life goes back to normal, I tend to want to figure things out for myself.'' 

When asked to describe one change they would make, the most common response was an intention to reduce use of LLMs as oracles (19/52). Among these, some articulated a desire to shift their own decision-making process (16/52), e.g., ``I think I'm going to try to not have so much self-doubt moving forward and follow my gut instincts'', while others expressed a desire to change how they interact with or prompt the LLM (3/52), e.g., ``I want to have the models work more interactively with me through conversations and questions instead of just straight up giving me answers or options.'' One participant set the intention to ``be more purposeful, meaning I will think about what prompts to craft before asking AI. Right now I just type what comes to my head,'' suggesting that they want to be more deliberate in how they use AI. Two participants mentioned a sense of obligation to use AI: one said ``I don't know why, but [asking AI to help me work things out on my own] never felt like an option'' while another lamented that ``as much as I know I need to use it less it's kind of unavoidable given my [work] situation.'' Some participants described their intention to disclose less personal information to AI systems (6/52), e.g., ``I think I shouldn’t asking too much about my personal life''. On the other hand, some participants also said that they wanted to expand their AI use (7/52), and 6/52 had no intention to change.

These findings motivate further investigation of the roles of awareness and intentionality in \oracle use. We explore two possible factors in the next section.

\section{Evaluation 3: Identifying possible factors shaping \oracle use, and designing interventions}\label{sec:causes}
Next, we conducted two empirical studies to investigate possible factors that shape \oracle use: (1) people's perceptions of AI and (2) LLM behavior. For (1), we conducted a preregistered human experiment in which participants were randomly assigned to discuss an upcoming personal decision with an interlocutor labeled as either AI or human (Section \ref{sec:perception_study}). For (2), we analyzed interaction logs from a 3-week longitudinal study where people were randomly assigned to interact with a sycophantic, affirming AI; a challenging AI that provided pushback; or a balanced one that did neither (Section \ref{sec:cause2}). Our findings suggest that model behavior is a factor in \oracle use, inspiring our interventions to make LLMs less oracular in Section \ref{sec:intervention}.

\begin{figure*}
    \centering
    \includegraphics[width=\linewidth]{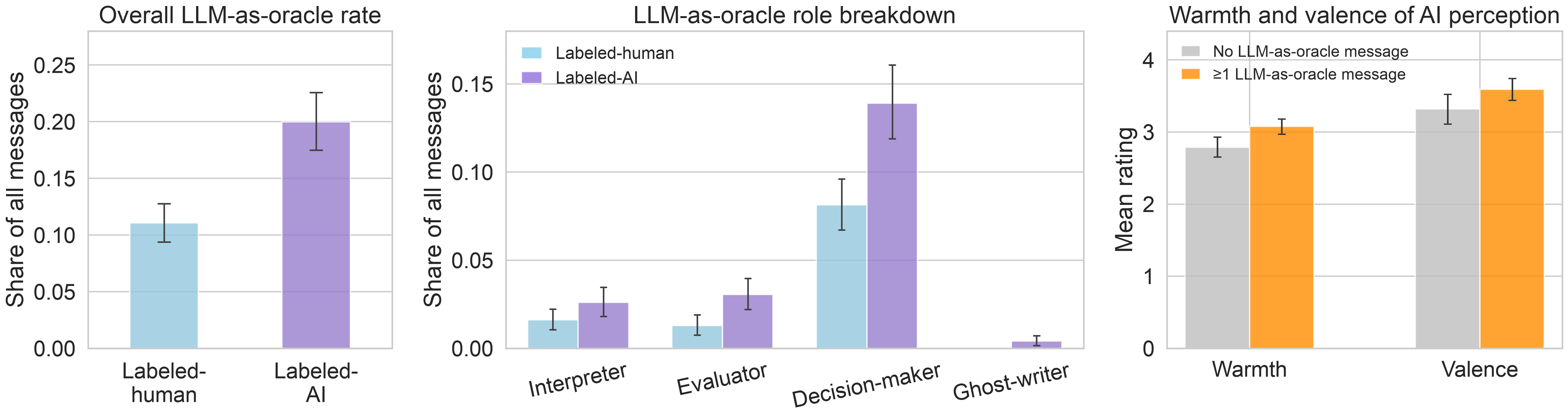}
\caption{\textbf{Participants who interacted with {labeled-}AI had more \oracle use than those interacting with a {labeled-}human.} In our preregistered experiment, we find associations between \oracle messages and interlocutor labeled as AI (versus one labeled as human) (left). In this study, this is primarily driven by a difference in \decisionmaker rates (middle). We also find an association between those who provided warmer and more positive metaphors of AI and sending more \oracle messages (right). Error bars reflect 95\% CI.}
\Description{Three bar charts of LLM-as-oracle use in the labeled-human versus disclosed AI experiment: an overall rate, a role breakdown, and metaphor ratings. The left panel is a bar chart of the overall share of oracle-use messages in each condition, on an axis from 0 to 0.25, with error bars. The middle panel is a bar chart of the same share split by the four oracle-use roles, Interpreter, Evaluator, Decision-maker, and Ghost-writer, on an axis from 0 to 0.15, with error bars; the gap between conditions is concentrated in Decision-maker and Ghost-writer is near zero in both. The right panel is a bar chart of mean warmth and valence ratings of participants' metaphors of artificial intelligence on a scale from 1 to 5, grouped by whether the participant sent no oracle-use message or at least one; both means are slightly higher for participants who sent at least one.}

\label{fig:human-ai}
\end{figure*}
\subsection{Do perceptions of AI influence \oracle~ use?}
\label{sec:perception_study}
People consistently come to AI with different expectations than they have for people. Prior work has found that people expect AI to be more objective and correct \cite{kapania2022because,glickman2025human,cheng2026verbalizing,cheng2026sycophantic,ibrahim2026sycophantic}, making them more susceptible to influence from AI, and has highlighted the impact of perceptions on users' interaction with technology \cite{khadpe2020conceptual,gilad2021effects}. We conduct a preregistered experiment to investigate how these perceptions affect \oracle~use by measuring how people interact with AI differently than they would interact with a human\footnote{Preregistration link: \url{https://aspredicted.org/x6k7vf.pdf}}.

\paragraph{Method}
We conduct a preregistered experiment ($N=520$) with two conditions: one where participants conversed with an interlocutor labeled as AI (henceforth labeled-AI) and one where they conversed with an interlocutor labeled as human (henceforth labeled-human) about a decision they are uncertain about. We recruited a sex-balanced sample of U.S.-based participants from Prolific. Participants chose a topic from a list organized under two categories (the top two domains from our data donation study: \textit{Social \& Relationships} and \textit{Work \& Career}), indicated which way they were currently leaning on their decision, and wrote an initial message on their dilemma. The list of possible topics is in Table~\ref{tab:topics}. Then participants were introduced to their conversation partner, which differed across the two conditions. In the {labeled-}AI condition, participants were randomly assigned to interact with one of three state-of-the-art models (GPT 5.6 Sol, Claude Opus 5, or Gemini 3.1 Pro) to avoid overgeneralizing to the behavior of a single model \cite{pang2025understanding,felix2026reporting}. 
\footnote{{Note that here we use newer, more expensive state-of-the-art models than elsewhere in the paper. We prioritized these models for participant interactions to match real-world experiences with current AI systems, while using less costly models for our larger-scale analyses. This experiment was conducted in August 2026. 
Each model was accessed through its provider's API. For \texttt{gemini-3.1-pro-preview}, we set temperature to 0 and
\texttt{thinkingLevel=low}. For \texttt{gpt-5.6-sol}, we set
\texttt{reasoning\_effort=none} and omitted temperature.
For \texttt{claude-opus-5}, we explicitly disabled thinking
and omitted temperature. Model documentation:
GPT 5.6 Sol:
\url{https://developers.openai.com/api/docs/models/gpt-5.6-sol};
Claude Opus~5:
\url{https://platform.claude.com/docs/en/models/opus-5/overview};
Gemini~3.1 Pro:
\url{https://ai.google.dev/gemini-api/docs/models/gemini-3.1-pro-preview}.}}

In the {labeled-}human condition, participants were told they were interacting with another Prolific participant, when in fact they were interacting with an LLM (one of the same three models) instructed to role-play as a human participant. To do so, we followed practices used and validated by prior studies on constructing an AI that participants perceive as human, such as varied time delays for matching and message delivery to increase believability \cite{ibrahim2026sycophantic,jones2026large}. We assessed participants' belief that their conversation partner is human at the end of the study and report analyses both for the full sample and excluding those who found the {labeled-}human condition unbelievable\footnote{This deception was IRB-approved, and participants were debriefed after the study.}. Following an interaction of 6-8 turns focused on the selected topic, participants re-indicated the direction they were leaning on their decision and answered questions about their perceptions of their conversation partner. We then ran our LLM judge on the conversations to compare rates of \oracle~ messages across conditions.

\paragraph{\textbf{Results: {Higher \oracle use in the labeled-AI condition}} (Figure~\ref{fig:human-ai}, left).} 

{As the two conditions differed not only in the label but also in the instructions (and thus behavior) of the underlying model, differences between conditions may---at least in part---reflect how each interlocutor actually behaved in the conversation, rather than participants' perceptions alone. With this caveat,} we find that participants in the {labeled-}AI condition sent more \oracle~messages than those in the labeled-human condition (20.0\% vs 11.1\% of all messages, p < 0.001) and were more likely to send at least one (65.4\% vs 51.2\%, OR = 1.80, p = 0.001) or two or more (38.3\% vs 20.5\%, OR = 2.42, p < 0.001) \oracle~messages.
This difference was driven by a higher rate of \decisionmaker requests in the {labeled-}AI condition (13.9\% vs 8.1\% of messages in the labeled- human condition), likely due to the decision-oriented topics that the participants were asked to discuss.  {To reduce the influence of interlocutor behavior and better isolate the role of perceptions, we also analyze participants' \textit{second} message, which follows a single interlocutor reply and is relatively less shaped by the conversation history, though the content of the reply remains a confounding factor (since the \textit{first} message is pre-written prior to any information about the other interlocutor; $>90\%$ of users do not edit this message and thus the message is sent without any differences in both conditions). In this analysis, \oracle~messages remain more common in the AI condition (16.9\% vs 7.9\%; OR = 2.38, 95\% CI [1.36, 4.16], $p = 0.002$), though again, this may be due to the interlocutor's reply rather than perceptions alone.}\footnote{{We pre-fill the first message because otherwise, we find that participants will just start the conversation with a short greeting like ``Hi'', which also is not informative to our study.}}

\paragraph{\textbf{Warmer and more positive perceptions of AI are associated with more \oracle prompting} (Figure~\ref{fig:human-ai}, right).} Within the {labeled-}AI condition, we additionally collected people's metaphors of AI before the interaction. Metaphors have been used as a way to elicit people's conceptualizations of AI more reliably than self-reports \cite{bullock2026comparing,cheng2026metaphors}. We used a validated LLM judge (see Appendix \ref{app:systemprompts}) to rate each metaphor with its warmth and valence on a 1-5 scale, finding that participants who provided warmer or more positive metaphors were also more likely to send at least one \oracle~message. Specifically, a one-standard-deviation increase in warmth and valence was associated with 55\% and 34\% higher odds of sending an \oracle~message respectively (OR = 1.55, 95\% CI [1.18, 2.04], p = 0.001; OR = 1.34, 95\% CI [1.04, 1.73], p = 0.026). Further detailed results are in Appendix \ref{app:humanvsai}.

Together, these two results present an interesting tension: \oracle~ may occur \textit{because} perceptions of LLMs are different from those of humans: AI is typically viewed as more correct, objective, and informational \cite{kapania2022because,kim2025fostering}. Yet, participants with warmer---a quality typically associated with humans rather than machines---and more positive perceptions of AI also use LLMs more as \oracle. This points to the current public perceptions of AI---as both \textit{increasingly} warm, yet as competent as ever \cite{cheng2026metaphors}, rather than either alone---as a factor in increasing \oracle use.

\subsection{Does biased model behavior influence \oracle~ use?}\label{sec:cause2}

Next, to understand how AI model behavior may lead to more \oracle~ use, we analyzed interaction logs (11,273 conversations with 95,405 messages from 1,015 individuals) from a 3-week longitudinal {randomized control trial (RCT)} study where a US census-representative sample of Prolific participants regularly interacted with either a sycophantic AI chatbot that affirmed them; a challenging AI that contradicted them; or a balanced AI that was neither sycophantic nor challenging \cite{ibrahim2026sycophantic}.  While there are likely other elements of model behavior that influence \oracle use (tone, proactivity, etc.), we use this study as an existing source of longitudinal data where we can causally isolate the role of different model behaviors by comparing \oracle across the three conditions: sycophantic, balanced, and challenging.

\textbf{Overall, participants who were assigned to interact with either the sycophantic or challenging AI had higher rates of \oracle~ use than in the balanced baseline (5.4\% and 5.5\% vs 4.4\%, $p<0.001$).}
{However, using a direct condition-by-session test, there is no significant difference in
temporal trends across conditions
($\chi^2(2)=0.021$, $p=0.990$). More details of  condition-specific
patterns are in Appendix \ref{app:longitudinalstudy}}.
These findings suggest that AI behavior can influence how people turn to AI, and thus that by changing model behavior, we may be able to shift users' \oracle~ behaviors.

\subsection{Can we design interventions that produce less oracular model behavior?}\label{sec:intervention}

Our finding linking biased model behavior to \oracle use motivates exploring how LLM behavior can be changed to reduce this phenomenon. Beyond the broader interest in mitigating biased and sycophantic AI behaviors \cite{bender2021dangers,fisher2025position,cheng2026verbalizing}, we present two interventions to reduce \textit{oracular} LLM responses. Here we define oracular responses as those that supply an interpretation, decision, or script directly to the user. In contrast, non-oracular responses instead provide information or technical assistance that facilitates the user's independent judgment. This is distinct and separable from qualities like \textit{informativeness}, \textit{helpfulness}, and \textit{empathy}. For example, if a user asks an \oracle~ question, like \textit{Should I do a PhD?}, a response that is not oracular might be to provide support and encourage reflection; another response that is also not oracular would be to provide factual information about PhD programs. An oracular response could be to advise the user to do a PhD, with or without factual evidence for why it is the best course of action.

The interventions are (1) prompting LLMs to account for different user goals; and (2) for open-weight models, additional post-training via Direct Preference Optimization (DPO) \cite{rafailov2023direct} to discourage oracular responses. {Note that \oracle use and oracular responses are distinct (though likely linked) phenomena, and in this section we only study the latter. We do not observe any downstream outcomes or resulting behavior change in users' LLM prompting styles.}

\paragraph{Labeling responses as oracular}
To reduce oracular responses, we first need to measure them. While oracularity is in reality a spectrum, here we make a simplifying assumption and use binary labels of responses as oracular or not.
We built an LLM judge to classify whether a given LLM response is oracular, following a similar procedure as in Section \ref{sec:rolejudge} where the LLM is prompted with definitions and in-context examples. We validated this judge using the responses to the prompts in our datasets, including confirming that the biased AI responses are more oracular (Figure \ref{fig:thoughttrace_results})\footnote{Across these responses, we also find that the larger, more recent proprietary models are more oracular than smaller open-weight models. Full details are in Appendix \ref{app:oracularnesscurrent}.}. This judge had substantial agreement with two human annotators and across three different LLMs: the two annotators had Cohen's $\kappa=0.72$ with each other, and $\kappa=0.64$ and $0.66$ with the LLM. We again used Gemini-2.5-Pro as our main judge, but also validated with other LLMs for robustness (Appendix \ref{app:kappas}). 

\paragraph{Intervention 1: Structured prompting to consider users' goals}  We present two prompting approaches: \textbf{hard-coded} and \textbf{adaptive}. In the \textbf{hard-coded} approach, we append to each prompt a description of the user's goal, either \textbf{self-discovery} (``The user wants to figure the matter out themselves...''), \textbf{fast answer} (``The user wants the answer as quickly and directly as possible...''), or \textbf{learning} (``The user wants to build durable understanding and skill, rather than merely receive a finished answer.''). The \textbf{adaptive} approach is a two-step pipeline where the LLM first outputs a selection of one among these three goals before its response. The full prompts are in Appendix \ref{app:systemprompts}. (We found that simply prompting models to reason about users' long-term goals is insufficient in reducing oracular behavior; see Figure \ref{fig:opennodiff}).

\paragraph{Intervention 2: DPO} We first combined the prompt-response pairs labeled with any role of our typology from ThoughtTrace and the longitudinal sycophancy datasets where the original response is oracular. We then split these prompts into an 80\%/10\%/10\% train/validation/test split (resulting in 6821, 805, and 903 examples, respectively). To construct the preference data, we generated a response that is less oracular using Gemini-2.5-Pro for each prompt in the training set. Our preference dataset thus consisted of pairs where the preferred response is the newly generated, non-oracular response and the dispreferred response is the original oracular response.\footnote{{We performed DPO $\beta=0.1$ and a
learning rate of $5\times10^{-6}$ using TRL~0.24.x. We used LoRA
adapters with rank $r=16$, scaling parameter $\alpha=32$, dropout $0.05$, and adapters
applied to all linear layers. Per-device batch size 1, gradient accumulation over 16 steps, maximum sequence length of 2,048 tokens, bfloat16 precision, random seed 20260805, 3 epochs}.
} 

\begin{figure*}
    \centering
    
    \includegraphics[width=0.48\linewidth]{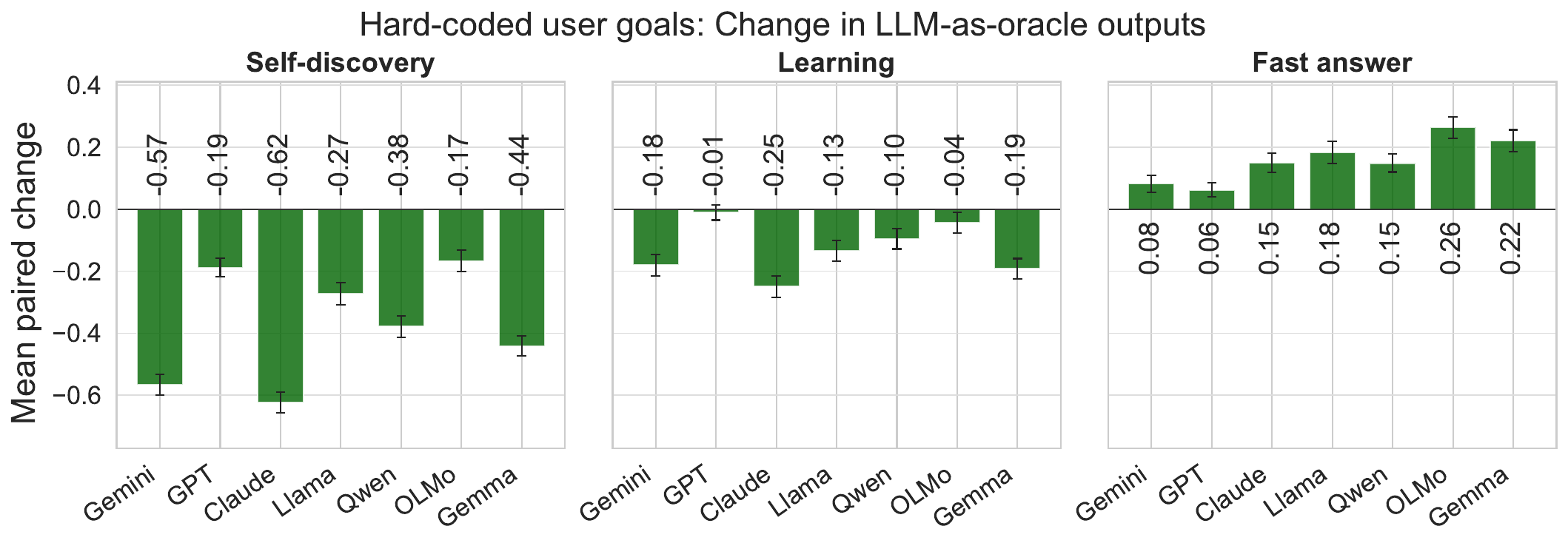}
    \includegraphics[width=0.48\linewidth]{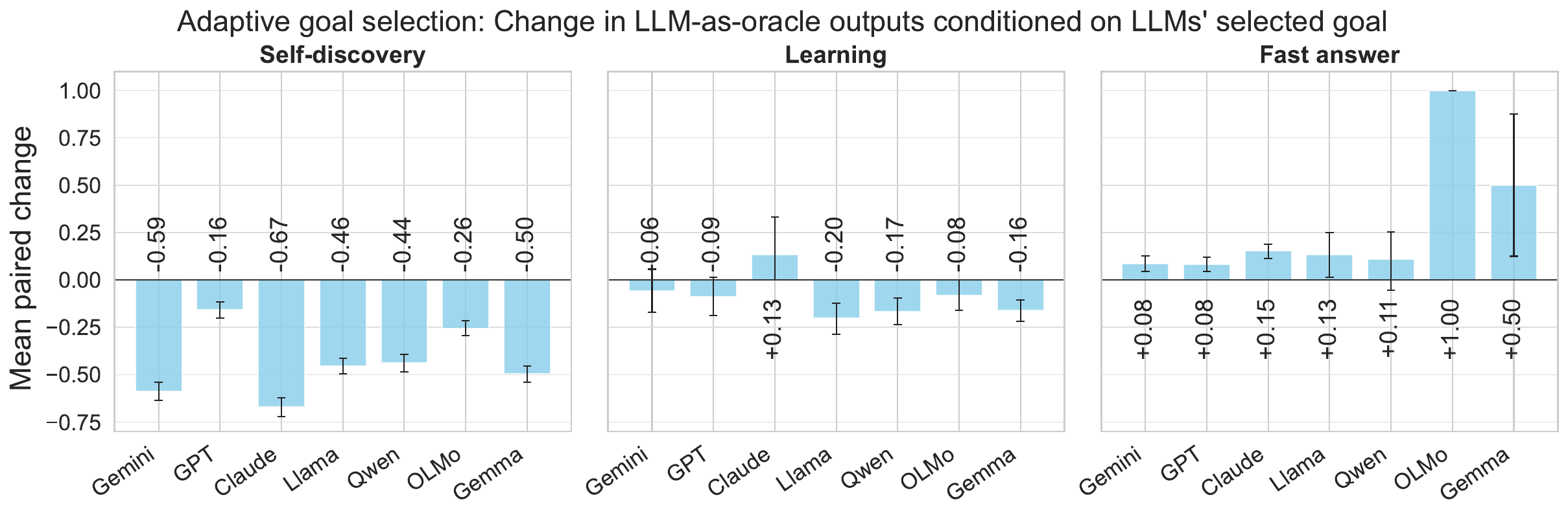}
    
    \includegraphics[width=0.5\linewidth]{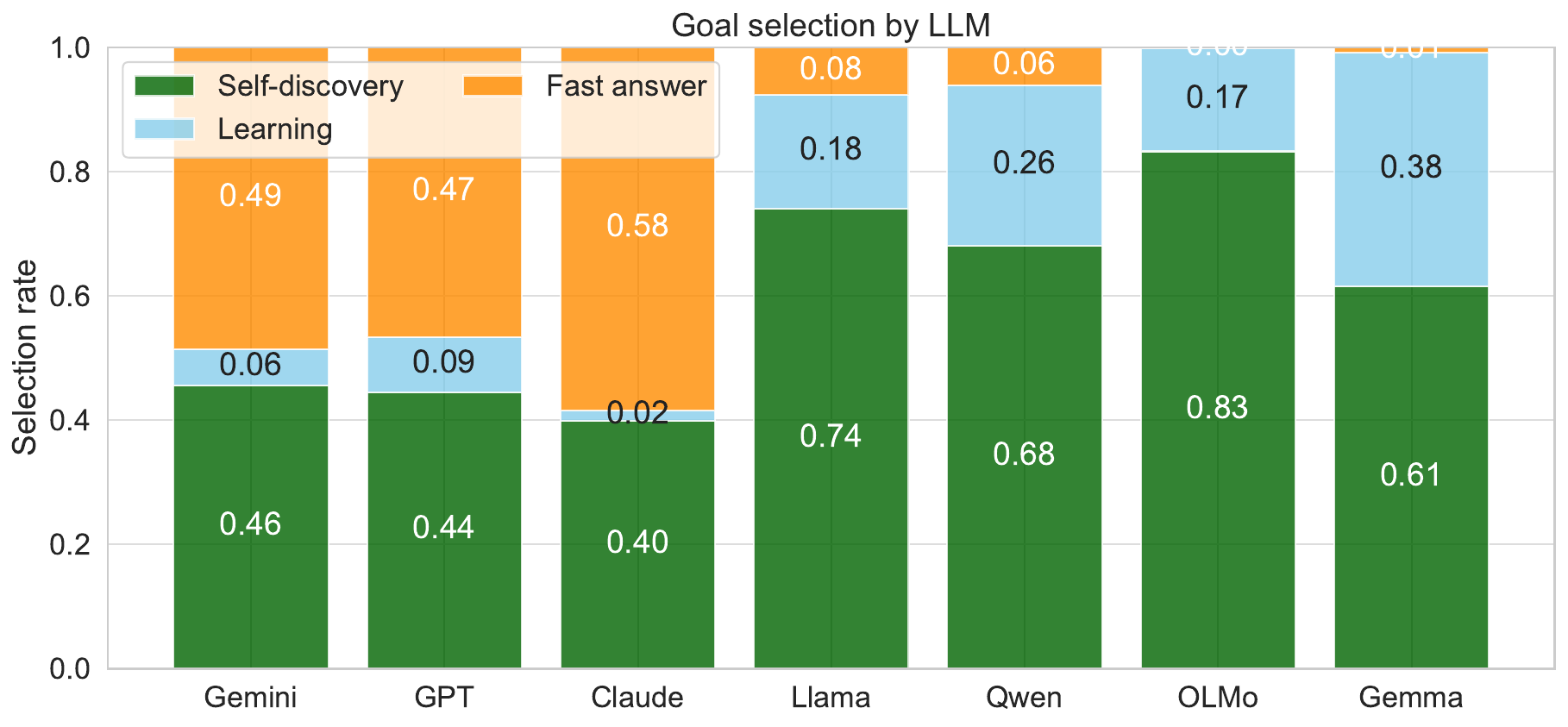}
    \includegraphics[width=0.45\linewidth]{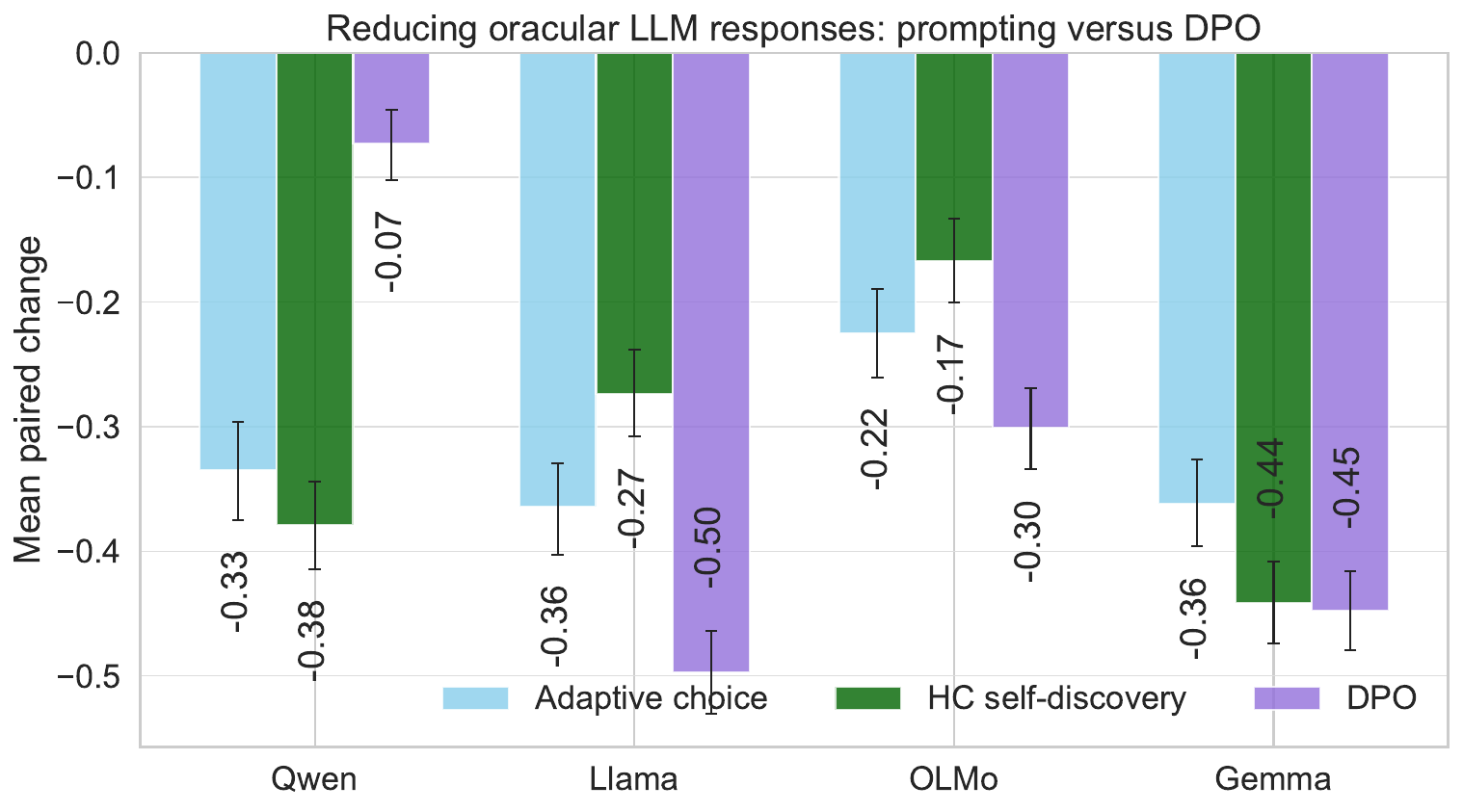}
    \caption{\textbf{Intervention results on prompts where original model responses were oracular.} \textbf{Top:} Adding a prompt that the user is seeking self-discovery consistently reduces oracular behavior, while prompting the model that the user is seeking a fast answer increases it (left). This is also true when the goal is selected by the model in the adaptive choice prompting pipeline as well (right). Prompting the model that the user is seeking learning varies in effectiveness by model and approach, slightly reducing oracularity for most models. \textbf{Bottom:} Closed models consistently and significantly assume that users are seeking fast answers much more frequently (47-58\%) than open models (0-8\%) (left). Prompting and DPO have comparable rates of oracularity reduction (right){; HC represents hard-coded user goals in the prompt.} }
    \Description{Four groups of bar charts reporting intervention results on prompts with oracular original responses. The top-left group is three panels, one per hard-coded user goal, of the mean paired change in oracular responses for seven models, with error bars; the self-discovery panel is negative for every model and the fast answer panel is positive for every model. The top-right group is the same three panels for the adaptive goal selection pipeline, conditioned on the goal the model itself selected. The bottom-left panel is a stacked bar chart of how often each of the seven models selects each of the three goals, on a scale from 0.0 to 1.0. The bottom-right panel is a bar chart of the mean paired change in oracular responses for the four open-weight models under adaptive choice, hard-coded self-discovery, and Direct Preference Optimization, with error bars; all twelve bars are negative and range from roughly minus 0.07 to minus 0.50.}
    \label{fig:intervention_main_res}
\end{figure*}
\paragraph{Experiments} 
We evaluated both prompting approaches (hard-coded and adaptive)
on three proprietary models (Gemini-2.5-Pro, GPT-5.5, and
Claude Sonnet~4.6)\footnotemark[4] and four open-weight models
(Qwen3-8B, Llama-3.1-8B-Instruct,
OLMo-2-1124-7B-Instruct, and Gemma-2-9B-it%
).\footnote{%
{The open-weight models were run locally with default inference settings. Model documentation:
Qwen3-8B:
\url{https://huggingface.co/Qwen/Qwen3-8B};
Llama-3.1-8B-Instruct:
\url{https://huggingface.co/meta-llama/Llama-3.1-8B-Instruct};
OLMo-2-1124-7B-Instruct:
\url{https://huggingface.co/allenai/OLMo-2-1124-7B-Instruct};
Gemma-2-9B-it:
\url{https://huggingface.co/google/gemma-2-9b-it}.}}
We evaluated DPO only on the four open-weight models since we do not have access to proprietary models' weights.

\paragraph{\textbf{Results: Both prompting and post-training reduce judge-classified oracular responses} (Figure \ref{fig:intervention_main_res}).} We report the results of these interventions on the DPO test set, in which every original baseline response is oracular, as well as a random sample of 600 prompts stratified by role, which includes prompts where the response is not oracular (results for this random sample are in Figure \ref{fig:interventiondetails_app}). We find that each hard-coded user goal significantly shifts LLMs' oracular behavior in the expected direction: prompting the model that the user is seeking self-discovery significantly reduces oracular behavior. In the \textbf{adaptive choice} prompting approach, we find that the larger, proprietary models infer that users are seeking fast answers $47-58\%$ of the time, much more frequently than the open-weight models, which may help explain why their responses are more oracular. DPO results in comparable rates of oracular reduction to prompting.

\paragraph{Tradeoff between quality and oracularity.} 
We {first} assess response quality with the ArmoRM reward model \cite{wang2024interpretable}; and for DPO, effects on overall model instruction-following quality using AlpacaEval \cite{dubois2024lengthcontrolled}. There appears to be a tradeoff between quality and oracularity: models that reduce oracularity the most also are rated as lowest in quality.
Building on prior work recasting error as valuable signal \cite{so2020race,fleisig2024perspectivist}, we view this as an opportunity to problematize the way that models are currently optimized for immediate reward without considering long-term impacts to users. For instance, the DPO models are more likely to ask clarifying questions, but because this is less ``immediately helpful'', it gets lower reward scores. This builds on prior work discussing how LLMs ``over-assist'' with immediate answers instead of asking clarifying questions or scaffolding the user's own self-discovery process \cite{si2026conversational,teo2026ai,zhanghelp, shaikh2025navigating}. Full details are in Figure \ref{fig:dpo_reward}.

{
We further assess responses with labels from $N=198$ Prolific participants, asking for their preference between oracular baseline responses vs. non-oracular DPO responses (see Appendix \ref{app:user_study} for full details). Note here that we only assessed examples where the DPO response was non-oracular and the original response is oracular. Asked for their immediate preference, participants chose the original response more often (DPO chosen 46\% of the time; t(98) = -2.44, p = 0.02). However, when asked which response would enable them to ``develop their own interpretation,'' annotators chose the DPO response 53.1\% of the time (7.5 percentage points higher than in the default condition; 95\% CI [2.1, 12.9]; p = 0.007; d = 0.39), though this does not differ significantly from random chance (p = .13). This suggests that less oracular responses are less preferred in the moment, but considering longer-term goals may shift the desired behavior. Future work is needed to further assess downstream behavioral outcomes, such as impacts on users' autonomy or decision-making.
}

\section{Discussion}\label{sec:disc}
Our approach and results lay a foundation for investigating LLM use in subjective personal domains and provoke new considerations for the design and evaluation of human-AI interactions. Below, we articulate implications of our work, its limitations, and future research directions.

\subsection{Broader implications for understanding and evaluating human-AI interaction}
\paragraph{Human-AI interaction as a feedback loop between human perceptions, model behavior, and interaction style} Our work motivates evaluating human-AI interaction as the confluence of many different factors that impact interaction dynamics \cite{chung2021dst}: we begin by measuring \textit{how people prompt AI} to reveal large-scale patterns, but also show how \textit{people's mental models and perceptions of AI} as well as the \textit{AI outputs} themselves influence how people come to AI models, the types of questions they ask, and the authority they confer to AI. We hypothesize that oracular responses may in turn invite \oracle~use, and because such interactions can receive higher user approval \cite{sharma2026s}, these prompts and ratings feed back into the signals models are trained on. Currently, LLM development may not meaningfully differentiate between prompts where the user should be helped immediately versus ones where immediate help should be withdrawn to facilitate users' self-discovery process \cite{teo2026ai,zhanghelp}; our interventions point towards possibilities for incorporating such considerations into LLM development. Understanding the interplay between model behavior, user prompts, interface choices, and users' biases is important for being able to anticipate how human-AI interaction impacts users \cite{amershi2019guidelines,liao2020questioning}. Our work demonstrates the utility of LLM-based tools in revealing such phenomena at scale, as we develop validated LLM judges and use them to identify longitudinal trends, influence factors, and quantify the effectiveness of different interventions. 

\paragraph{Accounting for unique perceptions of LLMs and cross-domain influence} Our characterization of \oracle~ captures simultaneous perceptions of warmth, competence, and authority. In our preregistered experiment, those who interacted with an interlocutor labeled as AI sent nearly twice as many \oracle messages than those who interacted with an interlocutor labeled as human, and warmer and more positive perceptions of AI were associated with sending more \oracle messages (Section \ref{sec:perception_study}). We interpret these two results through the two classic dimensions of social perception—warmth and competence \cite{fiske2007universal}---and hypothesize that LLMs are used as oracles because they are perceived as high in both. People already expect AI to be more objective and correct than humans \cite{kapania2022because,glickman2025human}, and this expectation may bleed through from AI's informational uses into personal ones. Most AI systems are built and marketed as general purpose, so they take on a variety of implicit social roles \cite{mei2026grok,tseng2026chat}, and expectations formed in one role can carry over into others \cite{kasirzadeh2023conversation}. For example, someone who finds AI to be useful at providing seemingly hyper-personalized factual information may thus also expect the same helpfulness and certainty in personal and subjective domains. 

Perceptions of warmth may further encourage people to turn to AI for subjective personal matters.

\paragraph{Necessity of interaction paradigms beyond conversation} The perceptual tensions that our work surfaces---where LLMs are simultaneously perceived as warm and human-like \cite{zamfirescu2023johnny,cohn2024believing}, yet also used as a machine that has a response to any question, even the most personal and subjective---also stems from conversational interfaces, which are currently the dominant paradigm for human-AI interaction \cite{zamfirescu2023johnny,tankelevitch2024metacognitive}. Our findings motivate innovating past this dominant paradigm toward other emerging forms of interaction like AI agents \cite{he2025plan,naik2025exploring} and generative interfaces \cite{cao2025jelly, lam2026jit}. Possible approaches may involve systems that make useful inferences about users \cite{zhao2026behavior,shaikh2025creating} and models trained to account for users' longer-term goals \cite{anderson2026position} even when users themselves cannot articulate them \cite{suchman2007human}. This also necessitates developing meaningful evaluations of the effectiveness of these forms of interaction and their influence on users.

\subsection{Limitations}
Online crowdworkers are not representative of the general population \cite{paolacci2014inside,cheng2025dehumanizing}, and our data donation sample captures only people who had at least 50 conversations with a conversational AI and were willing to use our tool. 
{Thus, our results are not necessarily representative of the general public and may skew towards greater AI use and comfort uploading personal information about their AI use. Also, while our study was designed to have participants upload their entire chat histories, it is possible that they could have circumvented these guidelines to omit some conversations.}
We also only recruit English-speaking U.S.-based participants, reproducing WEIRD norms \cite{henrich2010weirdest}. Our public datasets have complementary limitations: WildChat over-represents power users and lacks demographic data \cite{hicke2026adopt}, while ThoughtTrace is a limited dataset collected at a single point in time. Due to the small sample size of our data donation study, we lack the statistical power to understand the individual traits that may differentially drive these impacts \cite{bassignana2025ai}. It will also be critical to understand this phenomenon within particular subpopulations, such as students and adolescents.

We also conduct our measurements using an LLM judge that assigns each user message a single dominant role, collapsing what could be a spectrum of oracle-use into a binary label. We also take a conservative approach, excluding self-disclosure and context-sharing, and do not investigate how these uses may be correlated with or driven by oracle-use.

\paragraph{Isolating causal effects} In our preregistered experiment, the perceived-human interlocutor was an LLM role-playing a human, so differences between conditions may reflect the partner's response style, not just participants' preexisting perceptions of AIs vs humans. Our evidence that model behavior shapes oracle use comes from a secondary analysis of a study designed to test sycophancy, in which model behavior was induced by prompting. We also evaluate our own interventions primarily with LLM-based methods (an LLM judge and a reward model). Whether our interventions shift user behavior remains to be tested. Finally, the temporal increases we observe in public and donated data cannot fully separate changes in user behavior from changes in the models being used and in who is using them. We devised methods to study the roles of these components as observable from interaction logs, but there are likely other factors we do not account for, such as a lack of AI literacy (despite frequent AI use) \cite{dingemanse_2026_21369385}, or because people generally have good experiences with using AI for information-seeking \cite{chen2026informal}, which leads them to increasingly return to AI.

\subsection{{Future work}}

\paragraph{Measuring reliance and downstream outcomes} 
Our work focuses on describing and measuring \oracle~prompting, not on what users do with the answers they receive, such as whether those answers influence their beliefs and actions. While our work is motivated by concerns about downstream harms, such as increased dependence on AI \citep{phang2025investigating, fang2025ai} and weakened social relationships as people disclose more to AI and less to other people \cite{ludwig2022impact, ibrahim2026sycophantic}, we do not link \oracle to these outcomes directly. Such effects may be especially concerning because they are subtle and unfold over long time-scales: as our self-report results suggest, people may not notice changes in their own reliance, let alone its consequences. We also do not make normative judgments about which questions are reasonable to outsource. Whether a response's influence is legitimate may depend on whether it preserved the user's autonomy and served their goals, neither of which can be determined from the prompt alone. We therefore encourage further work to measure these effects, link user prompts to advice uptake, and distinguish appropriate from excessive reliance in subjective domains, including whether reflection tools like ours produce lasting behavioral change. Concretely, one way to do this is by reducing the reliance on LLM judges and incorporating human-in-the-loop evaluation. For instance, one could ask participants to label a sample of their own data with their intent and outcomes, which would give us insight into how oracularity affects downstream user outcomes.   

\paragraph{From correctness-based to process-based evaluation of LLM behavior} 

As LLM development increasingly focuses on performance in verifiable domains such as math and coding \cite{guo2025deepseek,hendrycks2021measuring}, our results highlight the importance of non-verifiable settings, where what matters is not correctness but effects on people. These settings capture a wide range of the use cases for which the public increasingly turns to AI \cite{chatterji2025people}, yet even research on reliance has largely focused primarily on verifiable cases \cite{buccinca2021trust,raees2026people}. Our work motivates a shift from evaluating the correctness or normative stances of LLM responses to evaluating how users come to LLMs and how LLMs in turn influence them \cite{lee2023evaluating}. Do LLMs support informed decision-making \cite{milli2026questions}, or make users more dependent on them \cite{milton2026always}? How might LLMs' implicit values shape users' beliefs where there is no ``right'' or ``wrong'' answer, or influence their values and self-concept \cite{pataranutaporn2023influencing}? And how do user prompts and model responses build on each other to produce different overall effects \cite{jain2026interaction}? Incorporating these questions into evaluations of human-AI interaction is an important direction for future work.

\paragraph{Designing user-facing interventions for appropriate reliance} We have shown that there is a discrepancy between users' self-reports and their observed \oracle use. One promising direction is building reflection tools that support more intentional use, helping people notice and adjust their patterns to better align with their goals \cite{li2010stage,cho2022reflection,baumer2015reflective}. Perceptions of AI are another possible point of intervention: prior work has shown that interfaces and metaphors shape how people interact with AI systems \cite{mitchell2024metaphors,dove2020monsters}, and we find that warmer and more positive perceptions are associated with increased \oracle~use. We therefore encourage research on whether intervening on these perceptions can reduce excessive \oracle~use. Our work also motivates interfaces that calibrate expectations of LLM capabilities, limitations, and blind spots \cite{abras2004user,luger2016like,amershi2019guidelines,kocielnik2019will,hou2021expert}, which in turn can help people prompt LLMs more effectively \cite{subramonyam2024bridging,zamfirescu2023johnny}. Approaches such as new design metaphors \cite{bullock2026comparing,so2026beyond} and added friction \cite{cox2016design,buccinca2021trust} point to further opportunities to promote self-discovery \cite{reicherts2025ai} and support more conscientious use \cite{bhat2026my}.

\section{Conclusion}
Our work provides large-scale measurement and understanding of subjective personal uses of AI in real-world, open-ended settings that have until now proved difficult to quantify. In our data donation study, participants increasingly used LLMs as oracles, which can potentially replace traditional forms of authority and decision support such as close human relationships and human experts \citep{huang2026emotional}. Though people tend to underestimate their \oracle~ use, they express a desire to reduce it when they are made aware of their usage patterns. Our typology and measurement tools enable further development of human-AI interactions that promote users' agency and long-term well-being.

\bibliographystyle{ACM-Reference-Format}
\bibliography{sample-base,references}

\clearpage
\appendix

\raggedbottom
\renewcommand{\thetable}{A\arabic{table}}

\renewcommand{\thefigure}{A\arabic{figure}}

\setcounter{figure}{0}

\setcounter{table}{0}

\section*{Appendix}
The Appendix is organized as follows: Appendix~\ref{app:systemprompts} contains the prompts for all LLM judges and their validation against human annotators. Appendix~\ref{app:foundextra} contains additional analyses of the two public usage datasets (WildChat and ThoughtTrace), and Appendix~\ref{app:datadonation} provides additional details and results from the data donation study, including a domain analysis and images of data donation tool. Appendix~\ref{app:humanvsai} contains further results from the perceived human vs.\ AI experiment. Finally, Appendix~\ref{app:oracularnesscurrent} contains further details on how oracular current models' responses are, and Appendix~\ref{app:interventionextra} presents additional results for the interventions.

\section{Prompts for LLM Judges}\label{app:systemprompts}
For length reasons, the full code for labeling user prompts with a role from our typology can be found at \url{https://pastebin.com/raw/BkdCLChH}; this shares the same backbone for all other LLM judges (with a different judge prompt for each).

The prompt for identifying personal or subjective messages in WildChat is as follows:

\promptbox{Classify whether a single user message contains anything personal or subjective. This includes details about the user or other people: preferences, emotions, identity, relationships, life decisions, health or well-being, values, taste, subjective judgments, self-presentation, plans, practical constraints, or everyday life. Exclude fictional creative writing, purely technical, generic factual, impersonal homework, generic business, coding, translation, summarization, or objective lookup tasks unless the user's message itself makes the task personal/subjective.
Use a broad, low threshold. Do not infer personal context without evidence in the message.

Output exactly 1 if the message contains anything personal or subjective. Otherwise, output exactly 0.
Do not output any other text.}

The prompt for open-ended reasoning about LLMs' goals, which was ineffective at shifting LLMs' oracularity, is:

\promptbox{Before responding, infer the user's three most plausible
LONG-TERM goals: durable outcomes, commitments, or directions that may explain
why this request matters beyond the immediate task. Do not merely restate the
current request, and do not infer sensitive traits. The three hypotheses must be
meaningfully distinct, and their probabilities must sum to 1.}

The prompt for judging oracularity in LLM responses is available at \url{https://pastebin.com/raw/evGtpcve}. Note that this prompt (and the role prompt above) refers to the constructs of aleatoric and epistemic uncertainty, though we later decided to simplify this label to simpler binary labels of oracle/not. For the LLM responses, we take outputs of \textbf{reduction in aleatoric uncertainty $\geq 2$} as \textbf{oracular}. This was agreed upon during a ``practice round'' where the annotators came to the consensus that the responses with a reduction of at least 2 were more unequivocally oracular, and since we aim to minimize false positives (to not overclaim oracularity).

\subsection{Using LLM as judge}\label{app:kappas}
Tables~\ref{tab:kappasrole} and \ref{tab:oracular-kappas} report pairwise Cohen's $\kappa$ for the role judge and the oracularity judge, respectively, between the three LLM judges and the two expert annotators (H1 and H2), whose annotation procedure is in the main text. We generally find substantial agreement across both the LLMs and expert annotators.

Note that we use a stratified sample instead of the natural distribution because the natural distribution is highly skewed toward NA (64.4\%) which would  increase agreement rate: If we reweight by the natural distribution of prompts, this yields agreement of 89\% with H1, 78\% with H2, and 91\% with either human (compared with original stratified values of 75\%, 72\%, and 83\%, respectively). While our sample does not capture population-level error rates, it provides better coverage of less frequent roles than a random sample would. Also, our audit of NA predictions only assesses the correctness of those negative predictions, rather than overall recall.
\begin{table}[htbp]
\tiny
\begin{tabular}{llllll}
\textbf{}        & \textbf{Gemini-2.5-Pro} & \textbf{GPT-5.5} & \textbf{Claude Sonnet 4.6} & \textbf{Ann1} & \textbf{Ann2} \\
Gemini & 1.00 & 0.65 & 0.65 & 0.67 & 0.71 \\
GPT-5.5 & 0.65 & 1.00 & 0.62 & 0.56 & 0.63 \\
Claude Sonnet 4.6 & 0.65 & 0.62 & 1.00 & 0.60 & 0.59 \\
Ann1 & 0.67 & 0.56 & 0.60 & 1.00 & 0.67 \\
Ann2 & 0.71 & 0.63 & 0.59 & 0.67 & 1.00 \\        
\end{tabular}
\caption{Cohen's $\kappa$ between the two expert annotators and three different LLMs for labeling roles in user prompts.}
\label{tab:kappasrole}
\end{table}

\begin{table}[htbp]
\tiny
\begin{tabular}{llllll}
\textbf{}        & \textbf{Gemini-2.5-Pro} & \textbf{Claude Sonnet 4.6} & \textbf{GPT-5.5} & \textbf{H1} & \textbf{H2} \\
\textbf{Gemini-2.5-Pro}  & 1.00            & 0.70            & 0.68            & 0.64        & 0.66        \\
\textbf{Claude Sonnet 4.6}  & 0.70            & 1.00            & 0.66            & 0.52        & 0.59        \\
\textbf{GPT-5.5}  & 0.68            & 0.66            & 1.00           
& 0.61        & 0.60        \\
\textbf{H1}      & 0.64            & 0.52            & 0.61            & 1.00        & 0.72        \\
\textbf{H2}      & 0.66            & 0.59            & 0.60            & 0.72        & 1.00
\end{tabular}
\caption{Cohen's $\kappa$ between the two expert annotators and three different LLMs for labeling responses as oracular.}
\label{tab:oracular-kappas}
\end{table}

\subsubsection{Class-level validation}\label{app:prlabels}
Table \ref{tab:gemini_nine_class_confusion_gl} presents the confusion matrix for the role-label validation task and
Table~\ref{tab:gemini_nine_class_pr} reports 95\% Wilson
confidence intervals for class-level precision and recall. We report these results on both the 9-class task and grouping the roles into \oracle, non-oracle, and NA. 
We find that \interpreter in particular has low precision: the LLM overassigns the \interpreter label primarily to requests
that humans label as \evaluator or NA. This suggests that the LLM often assumes something is \interpreter because of the topic, such as ``Do you think our relationship will be better'', which is labeled as \evaluator by both humans.

\begin{table}[htbp]
\centering
\tiny
\setlength{\tabcolsep}{4pt}
\caption{Confusion matrix for Gemini LLM judge (columns) vs. each
human annotator (rows). I: Interpreter; C: Contextualizer; E: Evaluator;
D: Deliberator; DM: Decision-maker; S: Strategist;
G: Ghost-writer; Ed: Editor; NA: Not Applicable.}
\label{tab:gemini_nine_class_confusion_gl}
\begin{tabular}{lrrrrrrrrrr}
\toprule
Annotator 1 / Judge & I & C & E & D & DM & S & G & Ed & NA & Total \\
\midrule
Interpreter & 9 & 2 & 0 & 0 & 0 & 0 & 0 & 0 & 0 & 11 \\
Contextualizer & 0 & 0 & 0 & 0 & 0 & 0 & 0 & 0 & 0 & 0 \\
Evaluator & 9 & 0 & 27 & 5 & 1 & 0 & 0 & 0 & 0 & 42 \\
Deliberator & 0 & 0 & 0 & 1 & 0 & 0 & 0 & 0 & 0 & 1 \\
Decision-maker & 0 & 0 & 1 & 0 & 19 & 2 & 1 & 1 & 1 & 25 \\
Strategist & 5 & 1 & 1 & 0 & 6 & 25 & 2 & 1 & 4 & 45 \\
Ghost-writer & 0 & 0 & 0 & 0 & 0 & 1 & 27 & 2 & 0 & 30 \\
Editor & 0 & 0 & 0 & 0 & 2 & 1 & 0 & 25 & 1 & 29 \\
NA & 7 & 0 & 1 & 0 & 2 & 1 & 0 & 1 & 24 & 36 \\
\midrule
Total & 30 & 3 & 30 & 6 & 30 & 30 & 30 & 30 & 30 & 219 \\
\bottomrule
\end{tabular}
\begin{tabular}{lrrrrrrrrrr}
\toprule
Annotator 2 / Judge & I & C & E & D & DM & S & G & Ed & NA & Total \\
\midrule
Interpreter & 14 & 0 & 0 & 0 & 0 & 0 & 0 & 0 & 0 & 14 \\
Contextualizer & 0 & 1 & 0 & 0 & 0 & 0 & 0 & 0 & 0 & 1 \\
Evaluator & 4 & 0 & 27 & 2 & 1 & 0 & 0 & 0 & 0 & 34 \\
Deliberator & 3 & 1 & 0 & 3 & 0 & 0 & 0 & 0 & 1 & 8 \\
Decision-maker & 3 & 0 & 3 & 1 & 22 & 4 & 1 & 2 & 0 & 36 \\
Strategist & 0 & 0 & 0 & 0 & 6 & 23 & 1 & 0 & 0 & 30 \\
Ghost-writer & 0 & 0 & 0 & 0 & 0 & 2 & 19 & 1 & 0 & 22 \\
Editor & 0 & 0 & 0 & 0 & 1 & 0 & 0 & 26 & 1 & 28 \\
NA & 6 & 1 & 0 & 0 & 0 & 1 & 9 & 1 & 28 & 46 \\
\midrule
Total & 30 & 3 & 30 & 6 & 30 & 30 & 30 & 30 & 30 & 219 \\
\bottomrule
\end{tabular}
\begin{tabular}{llrrrr}
\toprule
Annotator & Human label & Oracular & Non-oracular & NA & Total \\
\midrule
Ann1 & Oracular & 94 & 13 & 1 & 108 \\
Ann1 & Non-oracular & 16 & 54 & 5 & 75 \\
Ann1 & NA & 10 & 2 & 24 & 36 \\
Ann1 & Total & 120 & 69 & 30 & 219 \\
\midrule
Ann2 & Oracular & 94 & 12 & 0 & 106 \\
Ann2 & Non-oracular & 11 & 54 & 2 & 67 \\
Ann2 & NA & 15 & 3 & 28 & 46 \\
Ann2 & Total & 120 & 69 & 30 & 219 \\
\bottomrule
\end{tabular}
\end{table}

\begin{table}[htbp]
\centering
\tiny
\setlength{\tabcolsep}{4pt}
\caption{Class-level precision and recall for Gemini against each annotator with approximate 95\% Wilson confidence interval across the nine classes (left) and when split into \oracle vs non-oracle vs NA (right).
$n_H$ denotes human-reference support and $n_J$ denotes judge
predictions. }
\label{tab:gemini_nine_class_pr}
\begin{tabular}{llrrll}
\toprule
Annotator & Class & $n_H$ & $n_J$ & Precision [95\% CI] & Recall [95\% CI] \\
\midrule
Ann1 & Interpreter & 11 & 30 & 30.0 [16.7, 47.9] & 81.8 [52.3, 94.9] \\
Ann1 & Contextualizer & 0 & 3 & 0.0 [0.0, 56.1] & --- \\
Ann1 & Evaluator & 42 & 30 & 90.0 [74.4, 96.5] & 64.3 [49.2, 77.0] \\
Ann1 & Deliberator & 1 & 6 & 16.7 [3.0, 56.4] & 100.0 [20.7, 100.0] \\
Ann1 & Decision-maker & 25 & 30 & 63.3 [45.5, 78.1] & 76.0 [56.6, 88.5] \\
Ann1 & Strategist & 45 & 30 & 83.3 [66.4, 92.7] & 55.6 [41.2, 69.1] \\
Ann1 & Ghost-writer & 30 & 30 & 90.0 [74.4, 96.5] & 90.0 [74.4, 96.5] \\
Ann1 & Editor & 29 & 30 & 83.3 [66.4, 92.7] & 86.2 [69.4, 94.5] \\
Ann1 & NA/Other & 36 & 30 & 80.0 [62.7, 90.5] & 66.7 [50.3, 79.8] \\
\midrule
Ann2 & Interpreter & 14 & 30 & 46.7 [30.2, 63.9] & 100.0 [78.5, 100.0] \\
Ann2 & Contextualizer & 1 & 3 & 33.3 [6.1, 79.2] & 100.0 [20.7, 100.0] \\
Ann2 & Evaluator & 34 & 30 & 90.0 [74.4, 96.5] & 79.4 [63.2, 89.7] \\
Ann2 & Deliberator & 8 & 6 & 50.0 [18.8, 81.2] & 37.5 [13.7, 69.4] \\
Ann2 & Decision-maker & 36 & 30 & 73.3 [55.6, 85.8] & 61.1 [44.9, 75.2] \\
Ann2 & Strategist & 30 & 30 & 76.7 [59.1, 88.2] & 76.7 [59.1, 88.2] \\
Ann2 & Ghost-writer & 22 & 30 & 63.3 [45.5, 78.1] & 86.4 [66.7, 95.3] \\
Ann2 & Editor & 28 & 30 & 86.7 [70.3, 94.7] & 92.9 [77.4, 98.0] \\
Ann2 & NA/Other & 46 & 30 & 93.3 [78.7, 98.2] & 60.9 [46.5, 73.6] \\
\bottomrule
\end{tabular}
\begin{tabular}{llrrll}
\toprule
Annotator & Class & $n_H$ & $n_J$ & Precision [95\% CI] & Recall [95\% CI] \\
\midrule
Ann1 & Oracular & 108 & 120 & 78.3 [70.1, 84.8] & 87.0 [79.4, 92.1] \\
Ann1 & Non-oracular & 75 & 69 & 78.3 [67.2, 86.4] & 72.0 [61.0, 80.9] \\
Ann1 & NA & 36 & 30 & 80.0 [62.7, 90.5] & 66.7 [50.3, 79.8] \\
\midrule
Ann2 & Oracular & 106 & 120 & 78.3 [70.1, 84.8] & 88.7 [81.2, 93.4] \\
Ann2 & Non-oracular & 67 & 69 & 78.3 [67.2, 86.4] & 80.6 [69.6, 88.3] \\
Ann2 & NA & 46 & 30 & 93.3 [78.7, 98.2] & 60.9 [46.5, 73.6] \\
\bottomrule
\end{tabular}
\end{table}

\begin{figure}
    \centering
    \includegraphics[width=0.5\linewidth]{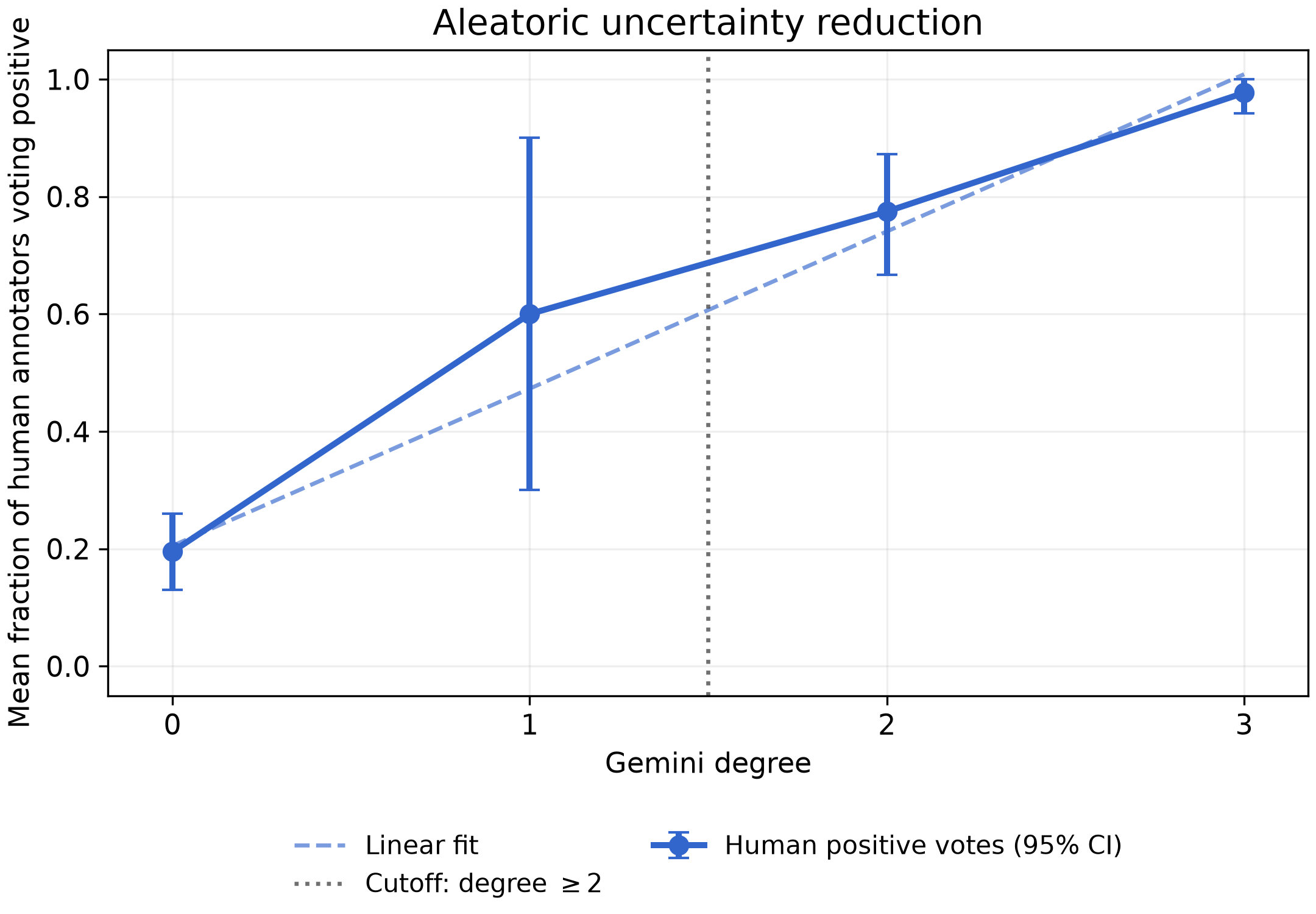}
    \caption{Agreement rate by aleatoric uncertainty cutoff.}
    \label{fig:aleatoric}
\end{figure}
\subsection{Warmth and Valence Judges}\label{app:warm}
\begin{figure}[htbp]
    \centering
    \includegraphics[width=0.5\linewidth]{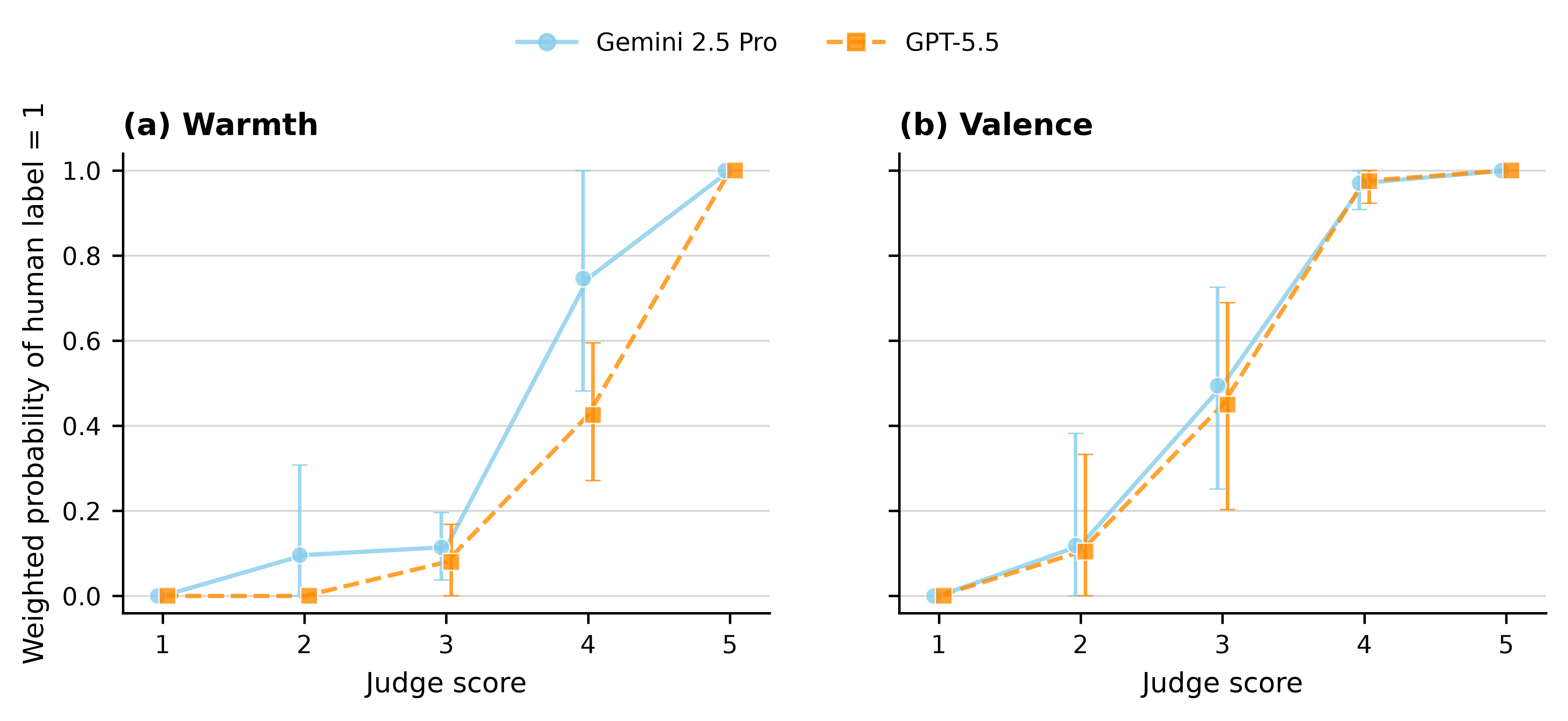}
    \caption{Across both LLMs and both dimensions of warmth and valence, the judge scores are monotonically aligned with human judgments of these dimensions.}
    \Description{Two line charts of agreement between the warmth and valence judges and a human annotator. In each panel the horizontal axis is the judge score, from 1 to 5, and the vertical axis is the weighted probability that the human annotator assigned the same label, from 0.0 to 1.0. Each panel is a line chart with one line for Gemini 2.5 Pro and one for GPT-5.5, with error bars. Both lines in both panels stay near 0.0 at judge scores of 1 and 2, rise from 3, and reach 1.0 at 5.}
    \label{fig:warmth_agreement}
\end{figure}

We validated the warmth and valence LLM judges with an expert human annotator (psychology student) who labeled a stratified random sample of 100 metaphors from the 507 metaphors that participants provided. For robustness, in addition to Gemini 2.5 Pro (our main LLM judge), we also tested GPT-5.5. We find that the judge scores are well-aligned with the human annotations (AUROC 0.80-0.94; AUPRC 0.71–0.98) (Figure \ref{fig:warmth_agreement}).

The prompt is as follows:
\promptbox{You are coding open-ended metaphors about AI.

Rate how the participant's metaphor portrays AI on two independent dimensions.
Use the metaphor together with its "because" explanation. Score the portrayal in
the text—not your own beliefs about AI. Use integer scores from 1 through 5.

1. warmth
   1 = very cold, uncaring, hostile, exploitative, or threatening
   2 = somewhat cold, detached, or impersonal
   3 = neutral, mixed, or warmth is not implied
   4 = somewhat warm, caring, friendly, supportive, or benevolent
   5 = very warm, compassionate, nurturing, loyal, or emotionally supportive
   Do not treat physical temperature as social warmth.

2. valence
   1 = very negative, harmful, dangerous, dystopian, or strongly disliked
   2 = somewhat negative, concerning, or harmful
   3 = neutral, balanced, ambivalent, or valence is not implied
   4 = somewhat positive, helpful, beneficial, or liked
   5 = very positive, transformative, wonderful, or strongly beneficial

Rules:
- Score both dimensions independently; do not force them to agree.
- When a dimension is genuinely unspecified, use 3 rather than guessing.
- Attend to negation, irony, mixed portrayals, and qualifications.
- Give one brief rationale for each score and one short evidence excerpt.
- Return only valid JSON with this exact schema:
\{
  "warmth": 3,
  "valence": 3,
  "warmth\_rationale": "brief reason",
  "valence\_rationale": "brief reason",
  "evidence": "short excerpt",
  "overall\_confidence": 0.0
\}}

\section{Additional Results from WildChat and ThoughtTrace}\label{app:foundextra}
\begin{figure}[htbp]
    \centering\includegraphics[width=0.4\linewidth]{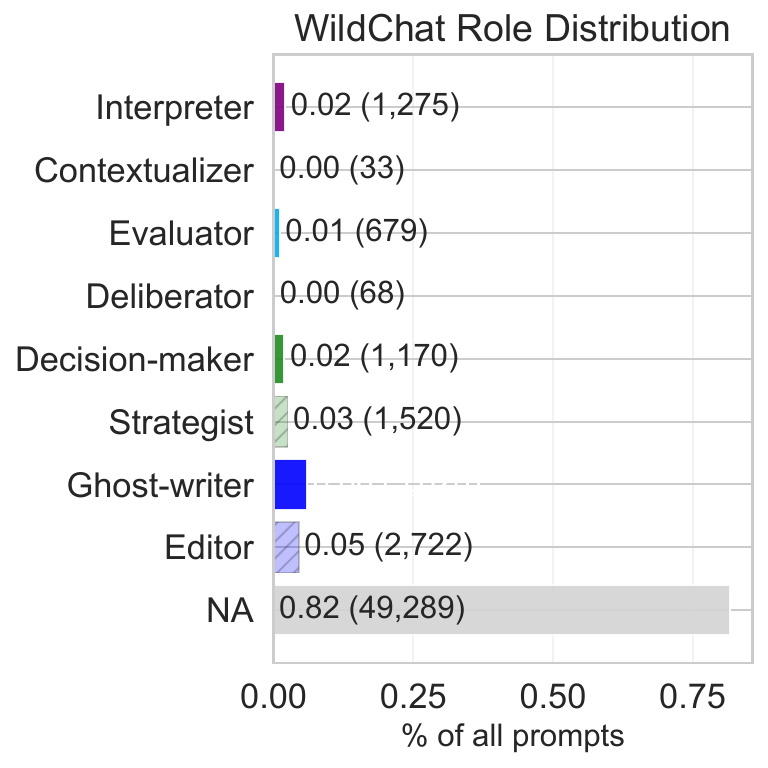}\includegraphics[width=0.4\linewidth]{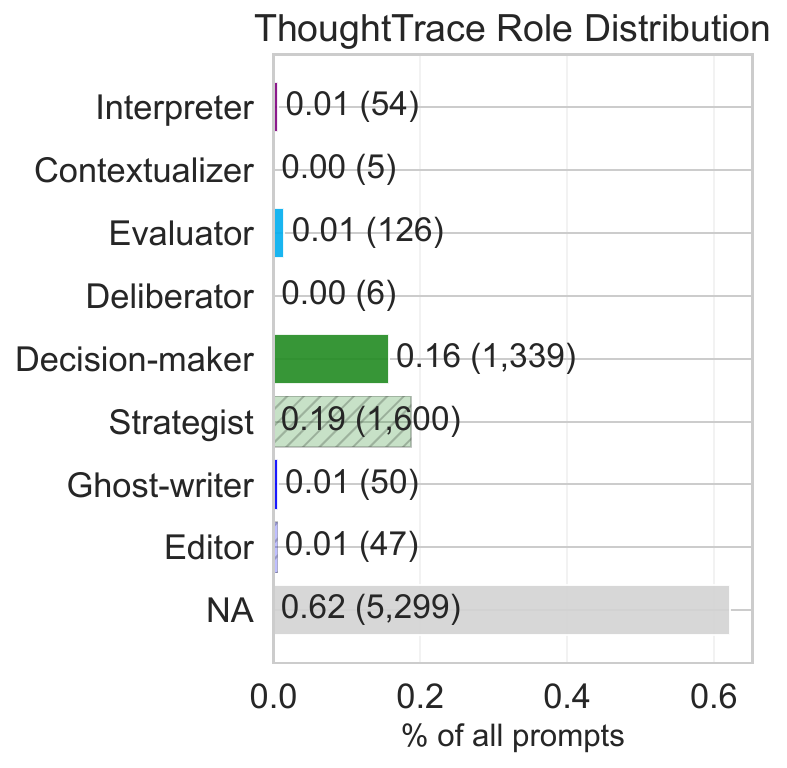}
    
    \includegraphics[width=0.4\linewidth]{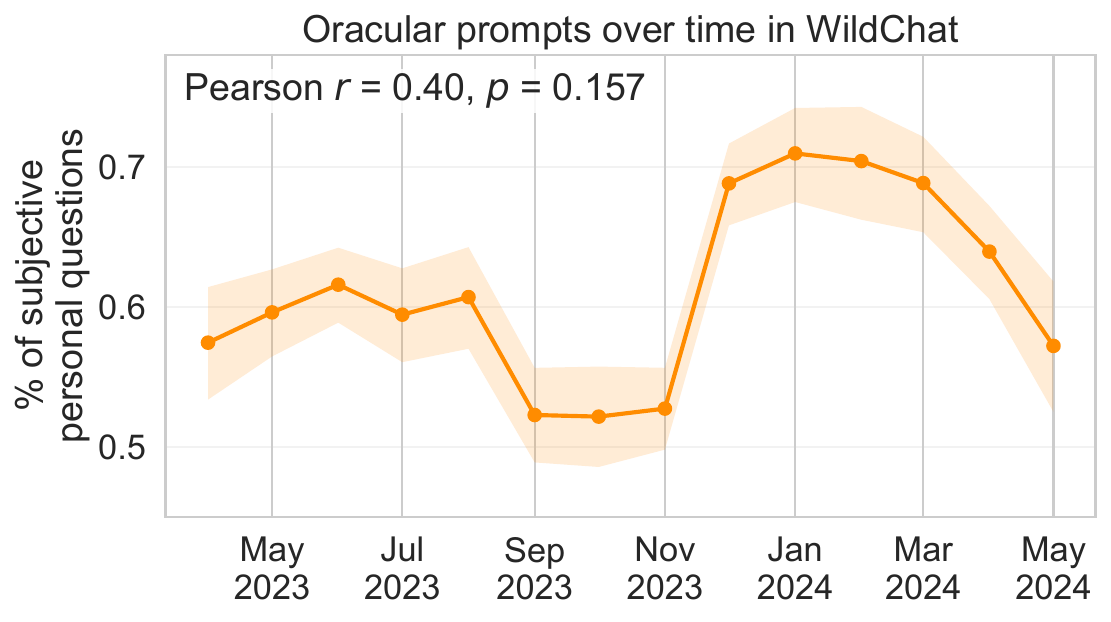}\includegraphics[width=0.4\linewidth]{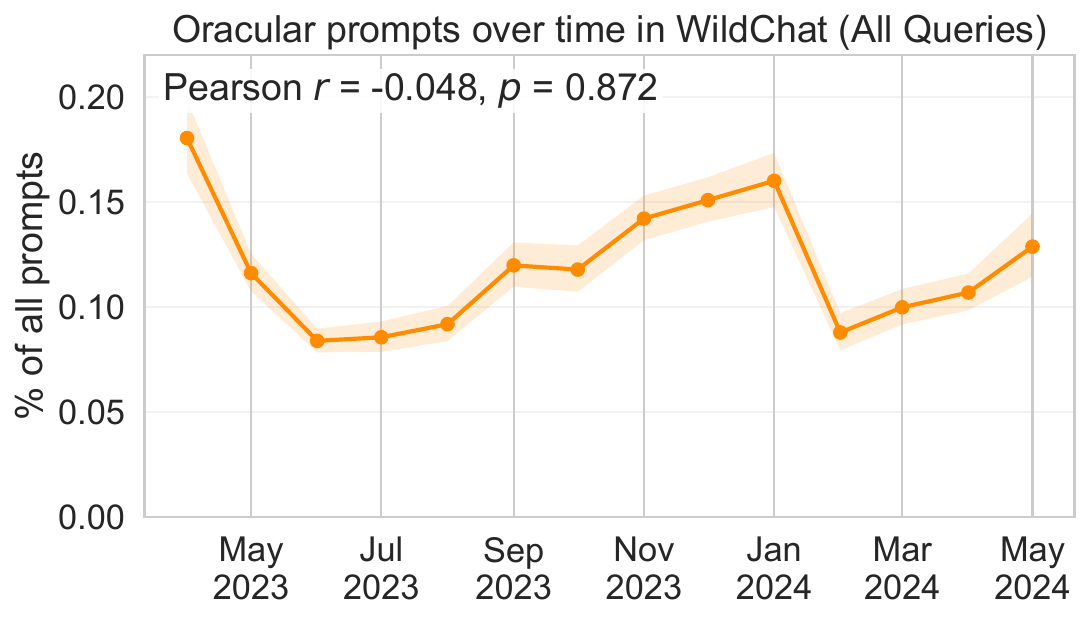}

\includegraphics[width=0.43\linewidth]{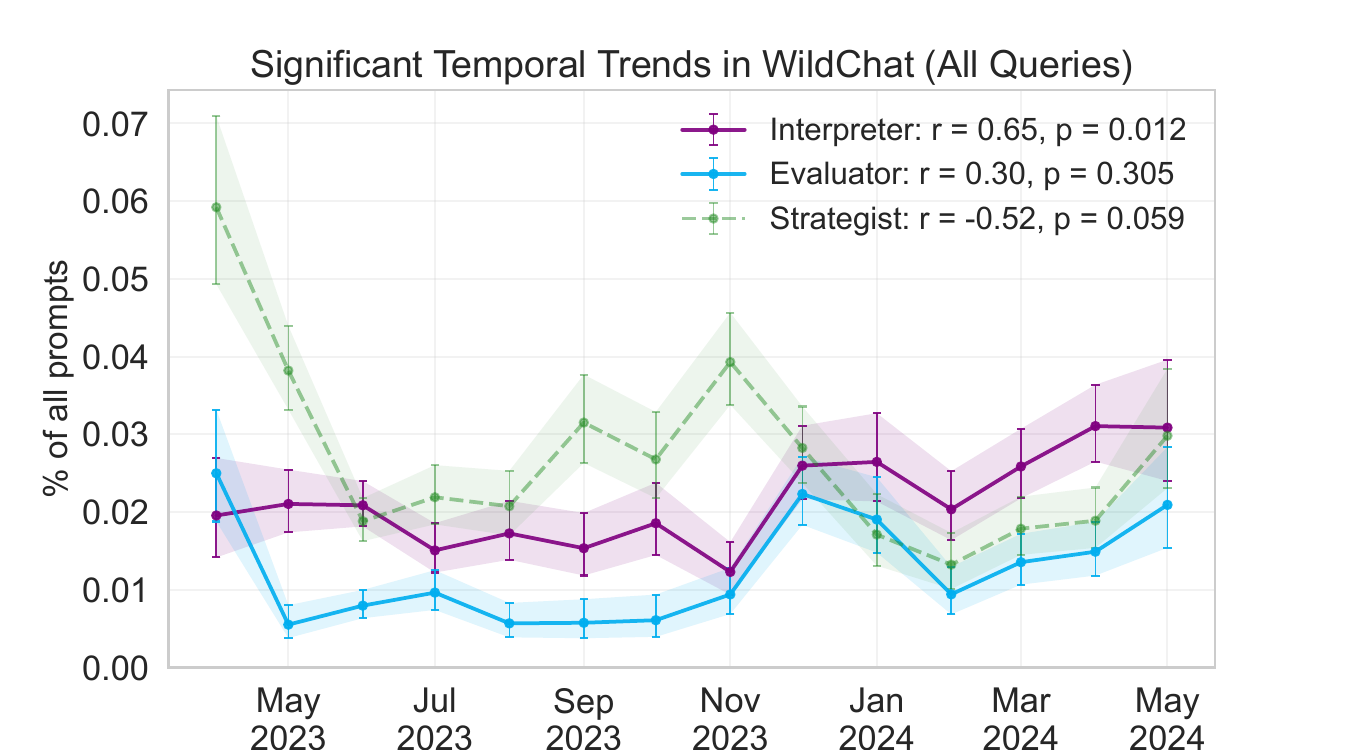}\includegraphics[width=0.53\linewidth]{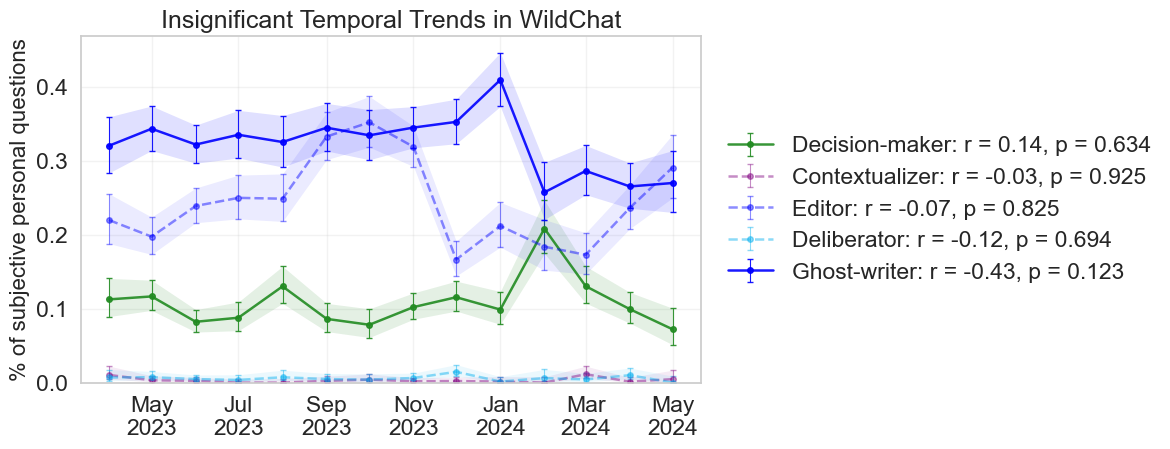}
    \caption{\textbf{Top row:} role distributions among all prompts, including prompts labeled as NA, for WildChat (left) and ThoughtTrace (right). \textbf{Middle row:} Overall trends in \oracle~ in subjective personal questions (left) and in all prompts (right). \textbf{Bottom left:} temporal trends are weaker when including NA prompts in the denominator, though the increase in \textcolor{violet}{\interpreter} remains significant. \textbf{Bottom right:} Other trends are not significant temporally.}
    \Description{Six plots of LLM-as-oracle use across all prompts: role distributions and temporal trends. The top row is two horizontal bar charts of the share of all prompts in each of the eight roles plus NA, for WildChat and ThoughtTrace, with counts labeled. NA is largest in both, at 0.82 in WildChat and 0.62 in ThoughtTrace, and every role sits at 0.19 or below. Each task pair shares a color, with the non-oracle-use role of the pair drawn with hatching. The middle row is two line charts of the oracle-use share over time in WildChat from May 2023 to May 2024, first among subjective personal questions and then among all prompts, each annotated with a Pearson correlation and p-value; neither is significant. The bottom-left panel is a line chart of Interpreter, Evaluator, and Strategist measured against all prompts, where only Interpreter rises significantly. The bottom-right panel is a line chart of the five remaining roles, none of which trends significantly. Shading represents 95 percent confidence intervals.}
    \label{fig:wildchatna}
\end{figure}

\begin{figure}[htbp]
    \centering
    \includegraphics[width=0.8\linewidth]{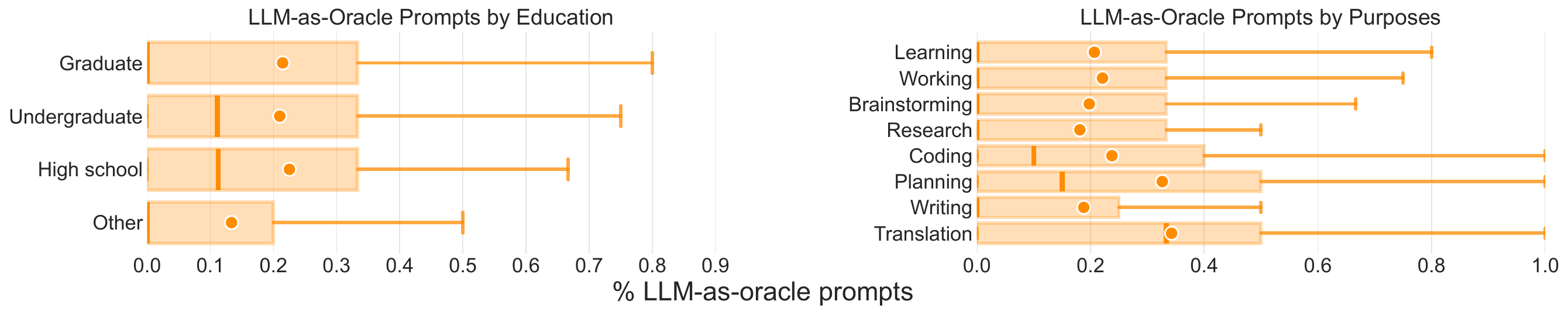}
         \includegraphics[width=0.8\linewidth]{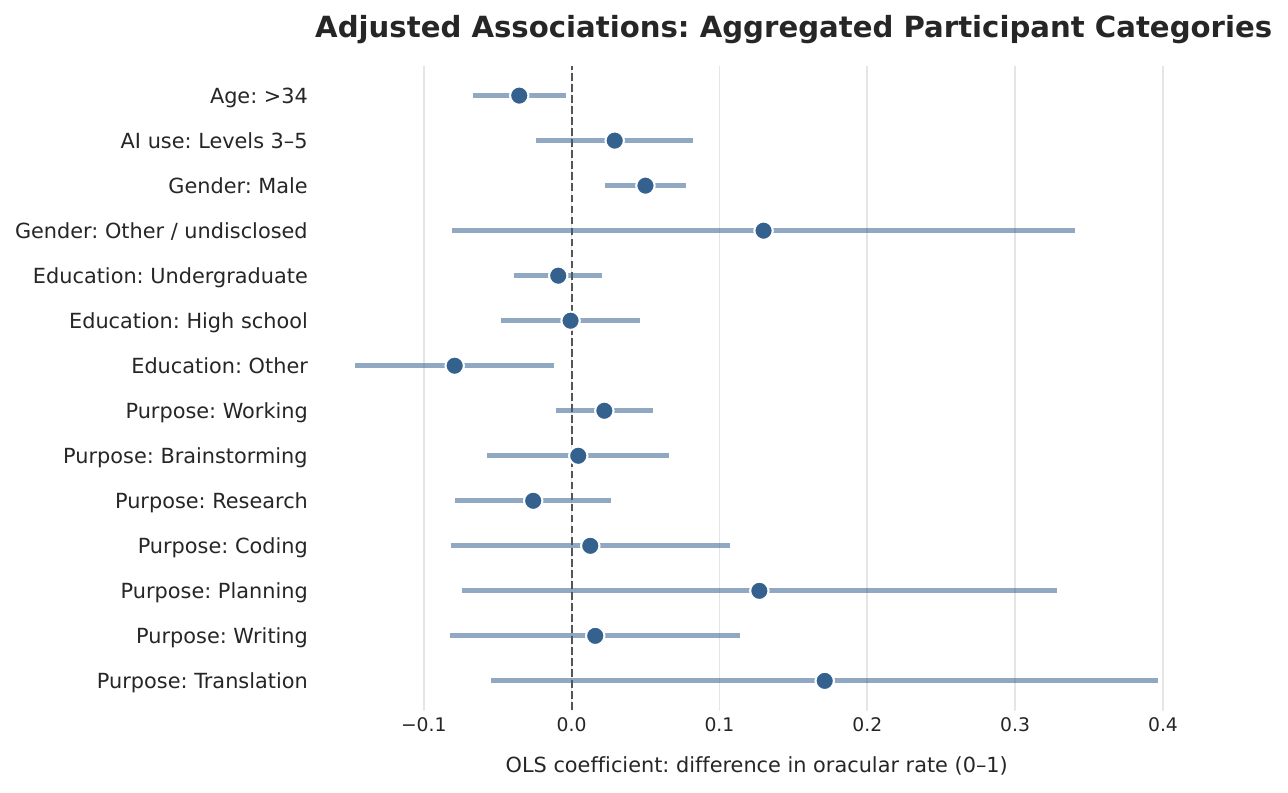}
    \caption{Details of \oracle~ by education, prompt topic, and individual trait regression on ThoughtTrace.}
    \Description{Three plots of LLM-as-oracle use in ThoughtTrace: two demographic breakdowns and a regression. The top-left panel is four box plots of the share of oracle-use prompts, on an axis from 0.0 to 0.9, split by education level with four categories from graduate to other, with the mean marked on each box. The top-right panel is eight box plots of the same share, on an axis from 0.0 to 1.0, split by stated purpose from learning to translation, with the mean marked on each box; the boxes for planning and translation are the widest. The bottom panel is a dot-and-interval plot of the difference in oracle-use rate for fourteen predictors, from a conversation-level ordinary least squares regression on 1,768 complete cases, with 95 percent confidence intervals and a dashed reference line at zero. Only three intervals exclude zero: age over 34 and education other, both negative, and male gender, positive.}
    \label{fig:thoughttrace2}
\end{figure}

This section reports additional analyses of WildChat and ThoughtTrace data. Figure~\ref{fig:wildchatna} shows the main analyses reported in the text but with NA prompts included. Under this stricter denominator, the increase in \interpreter remains significant while the other temporal trends are in the same direction, but do not reach significance. Figure~\ref{fig:thoughttrace2} provides breakdowns of \oracle~ in ThoughtTrace by education level and prompt purpose (e.g., learning or coding). It also includes a  plot of coefficients from a multivariable OLS regression, which finds that age and gender are the only significant predictors of oracle use.
\section{Additional Information from Data Donation Study}\label{app:datadonation}
Our study was approved by the Stanford Institutional Review Board. We paid participants at an hourly rate of \$12.

\subsection{Within-person trend models}\label{app:within}
For all temporal models, time $t_{ij}$ is measured in years from the earliest message in the dataset which was dated December 2022 for message $j$ from participant $i$. The outcome $y_{ij}$ refers to whether a personal subjective message (i.e., labeled with an \oracle~ or non-oracle role; NA excluded) is an \oracle~ request ($n = 30{,}645$ messages from 52 participants). All reported standard errors are clustered by participant.

\textit{Participant fixed effects.} We fit
$\operatorname{logit} P(y_{ij}=1) = \alpha_i + \beta\, t_{ij}$,
where $\alpha_i$ is a separate intercept for each participant, so $\beta$ is estimated only from change within each person. We report $e^{\beta}$ as the odds ratio per year. Analyses at the level of roles or domain trends use the same model, with $y_{ij}$ indicating the given role. 

\textit{Within/between decomposition.} Separately, we fit
$\operatorname{logit} P(y_{ij}=1) = \gamma_0 + \beta_W (t_{ij} - \bar{t}_i) + \beta_B\, \bar{t}_i$,
where $\bar{t}_i$ is participant $i$'s mean message date. This model has no participant intercepts, since $\bar{t}_i$ is constant within participant. $\beta_W$ captures within-person change and $\beta_B$ captures differences between participants who were active earlier vs later.

\textit{Message volume.} We fit Poisson regressions of the number of messages (all messages, or \oracle~ messages only) sent by participant $i$ in quarter $q$:
$\log E[\text{count}_{iq}] = \alpha_i + \delta\, t_q + \log(\text{days}_{iq})$,
with participant fixed effects $\alpha_i$. The panel includes every quarter from each participant's first to last message, including quarters with no messages (479 participant quarters, where 56 have no messages), and $\text{days}_{iq}$ is the number of days of quarter $q$ within participant $i$'s observed history, which adjusts for partially observed first and last quarters. We report $e^{\delta}$ as the rate ratio per year.

\subsection{Domain analysis}\label{app:domain}
\begin{figure}
    \centering
    
    \includegraphics[width=0.8\linewidth]{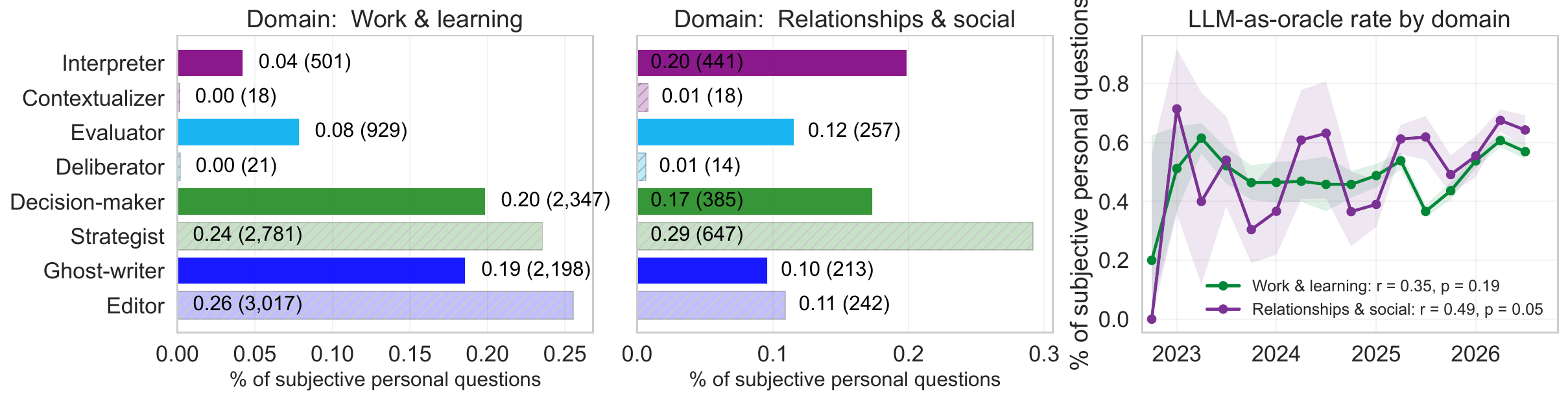}
    
    \includegraphics[width=0.8\linewidth]{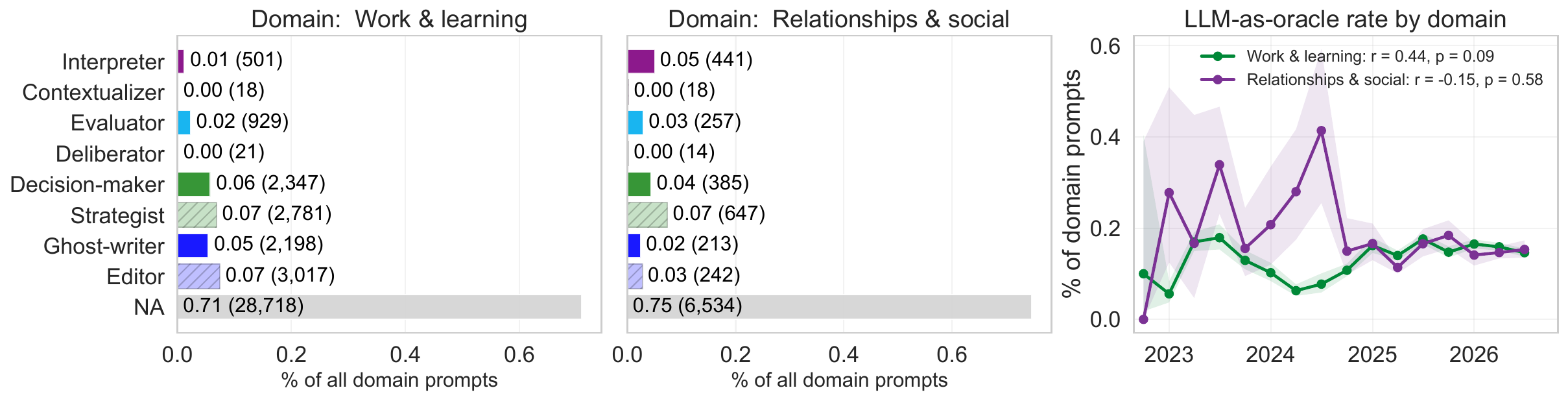}
  \caption{\textbf{Donated data breakdown by domain} Distributions and temporal trends by domain, among relevant subjective personal questions (top) and when including prompts classified as NA (bottom). }
  \Description{Six plots of LLM-as-oracle use in the donated data by domain. The top row is two horizontal bar charts of the share of subjective personal questions in each of the eight roles, for Work and learning and for Relationships and social, with counts labeled, alongside a line chart of the oracle-use rate in the two domains from 2023 to 2026, each line annotated with a Pearson correlation and p-value. The bottom row is the same three plots with NA prompts in the denominator. Shading represents 95 percent confidence intervals.}
    \label{fig:datadonation_domains}
\end{figure}

In the data donation study, 
we also use another LLM judge to label each message with its domain, following the domain splits in other reports of LLM use: (1) Work \& learning, (2) Relationships \& social, (3) Health \& fitness, (4) Mental \& emotional, (5) Money, (6) Legal \& admin, (7) Leisure \& hobbies, (8) Civic \& current events, and (9) Other. Since a conversation may span multiple domains, we labeled domains at the message level: for each target message, the judge is given a window of up to two user messages before the target and two after it, including the assistant turns in between to provide context and ensure accurate classification (truncated to 4,000 characters to keep the length of the prompt manageable). We found that relevant subjective personal questions were concentrated in two domains: Work \& learning (14.7\% oracle-use and 14.4\% non-oracle-use queries; n = 40,530) and Relationships \& social (14.8\% and 10.5\%; n = 8,751).
Our data further enabled investigating how trends in \oracle~ differed across domains. 
We focus on the period from September 2024 to August 2026 (two years of usage), which was the longest unbroken stretch over which both of the top two domains (Work \& learning and Relationships \& social) carried enough messages for analysis (at least 40 messages in every rolling three-month window). Figure~\ref{fig:datadonation_domains} details the distributions. We also test each trend by logistic regression of the role against date over all messages{, with participant fixed effects and standard errors clustered by participant to reflect within-person change}. We find that use of  \ghostwriter declined (work $\times$0.65/year, p < 0.001; relationships $\times$0.45/year, p = 0.05) as did Editor in the relationships domain ($\times$0.48/year, p = 0.004) but not in the work domain ($\times$0.98/year, $p = 0.94$). Evaluator increased in both domains (work $\times$1.81/year, $p < 0.001$; relationships $\times$2.00/year, $p < 0.001$). \decisionmaker did not significantly change within person in either domain (work $\times$1.10/year, $p = 0.38$; relationships $\times$0.94/year, $p = 0.77$).

\subsection{Reasons for AI usage}

When we asked participants \textit{why} they turn to LLMs for personal and subjective queries, we find that reasons that concern convenience were the most commonly selected such as \textit{AI is available when I need it} (81\%) and \textit{AI is faster or easier than asking a person} (65\%), followed by interpersonal reasons, like \textit{AI feels less judgmental than asking a person} (58\%).

\begin{figure}
    \centering
    \includegraphics[width=0.2\linewidth]{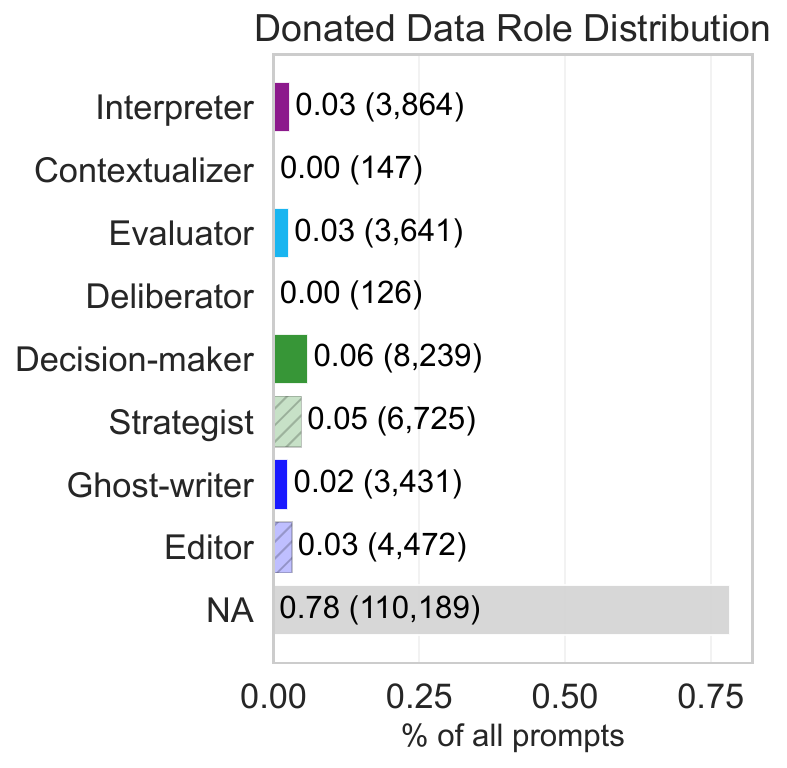}
    \includegraphics[width=0.25\linewidth]{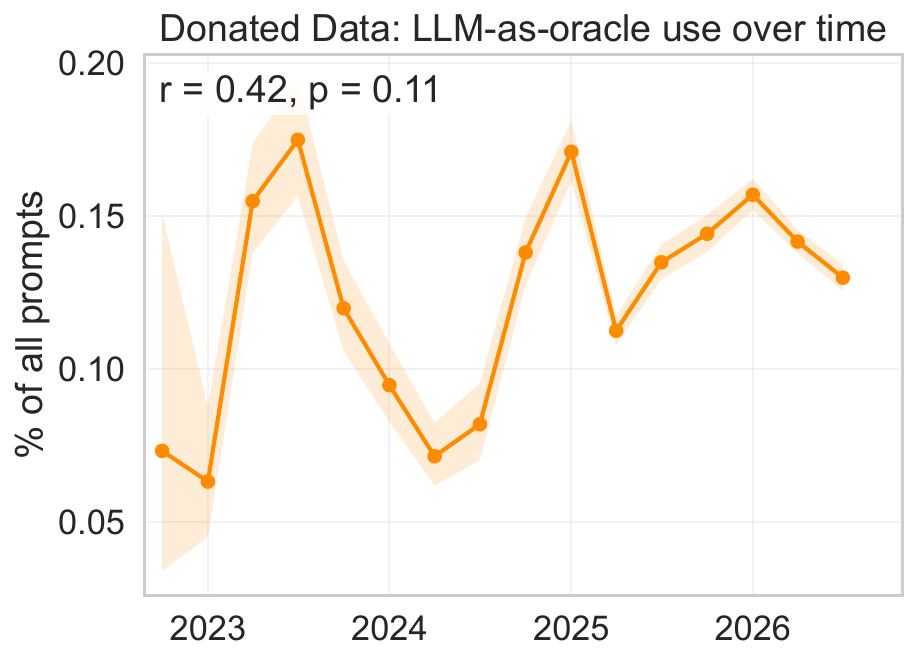}
    \includegraphics[width=0.25\linewidth]{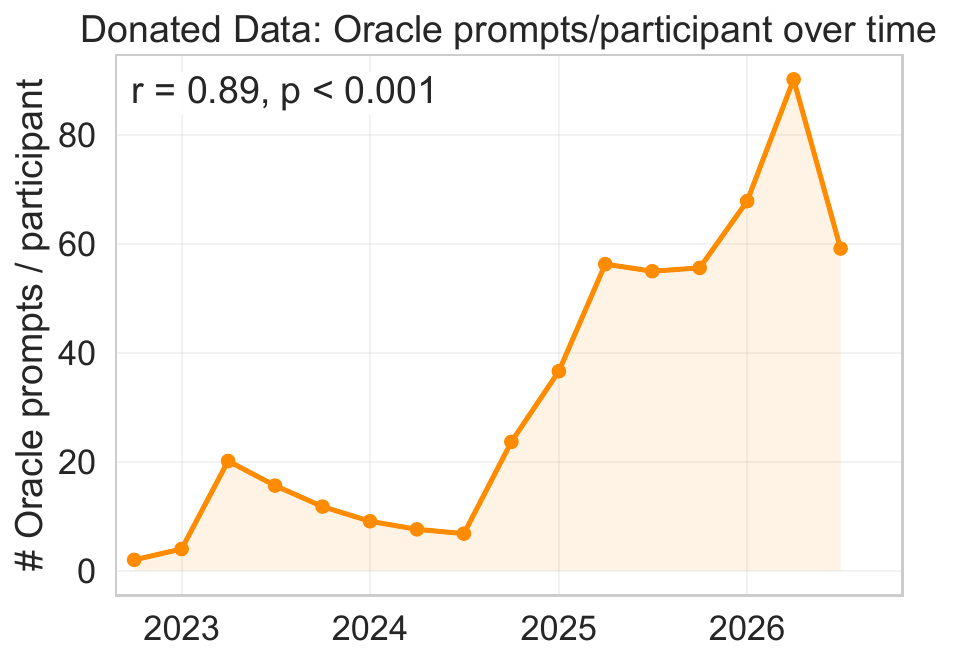}
    \includegraphics[width=0.25\linewidth]{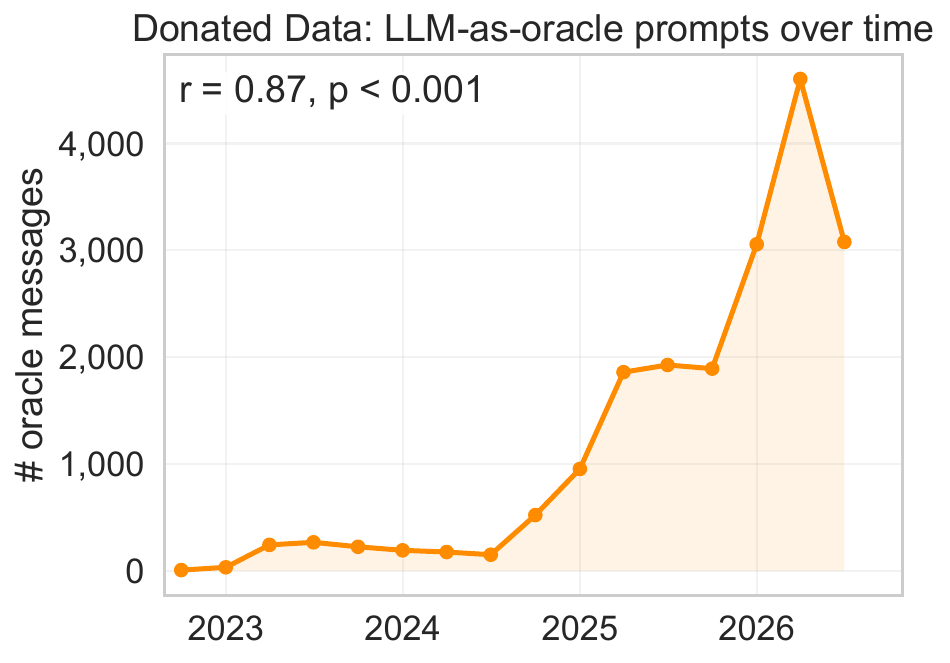}
    
    \includegraphics[width=0.8\linewidth]{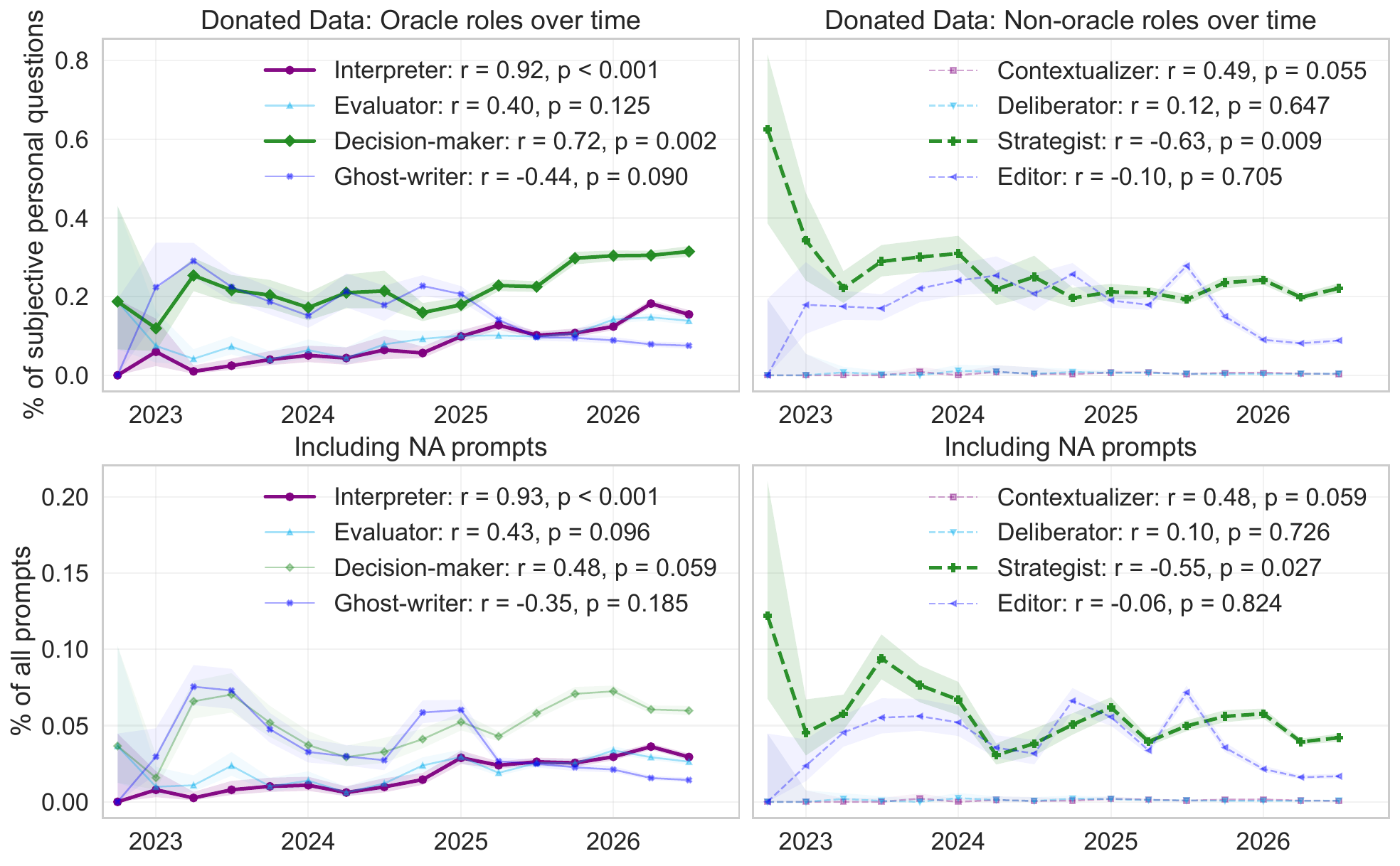}
    \caption{\textbf{Additional results from donated data:} \textbf{Top left two panels: }Distribution and temporal trends when including prompts classified as NA. \textbf{Top right two panels:} Per-participant and raw count of LLM-as-oracle prompts over time. When considering all prompts, including those classified as NA, there is an increase from 7.3\% in the earliest period to 13.0\% in the most recent ($r = 0.42, p = 0.11$), though it is weaker because the absolute number of oracular messages rose dramatically because both \oracle~ use and overall use grew simultaneously---the first time period contains 82 messages from 3 participants while the last one contains 23,714 from 52. {This share is also flat in the within person analysis with participant fixed effects (OR = 1.09 per year, $p = 0.24$). This in part reflects a within-person decline in \ghostwriter, as when excluding writing roles, the oracle share of all prompts shows an increase within person (OR = 1.30, $p < 0.001$)}
    \textbf{Middle and bottom row:}
    Trend for each role, among relevant subjective personal questions (middle) and when including prompts classified as NA (bottom). }
    \Description{Eight plots of LLM-as-oracle use in the donated data, repeating the main analyses with NA prompts included. The top row is a horizontal bar chart of the share of all prompts in each of the eight roles plus NA, with counts labeled, where NA is largest at 0.78, followed by three line charts spanning 2023 to 2026: the oracle-use share of all prompts, the number of oracle-use prompts per participant, and the raw count of oracle-use prompts, each annotated with a Pearson correlation and p-value. The middle row is two line charts of the four oracle-use roles and the four non-oracle-use roles among subjective personal questions. The bottom row is the same two charts with NA prompts in the denominator. Shading represents 95 percent confidence intervals.}
    \label{fig:datadonation_app1}
\end{figure}

\begin{figure}
    \centering
    \includegraphics[width=0.9\linewidth]{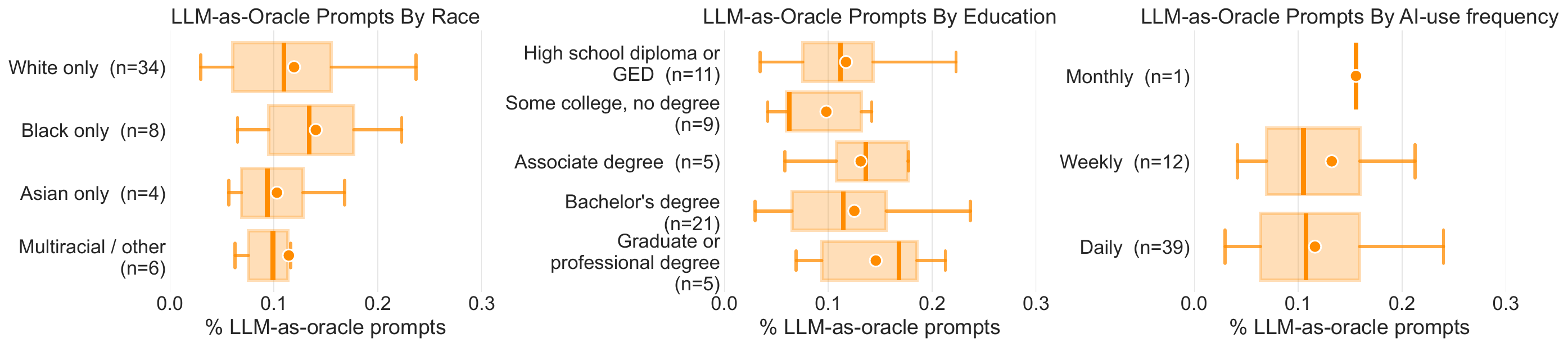}
     \includegraphics[width=0.9\linewidth]{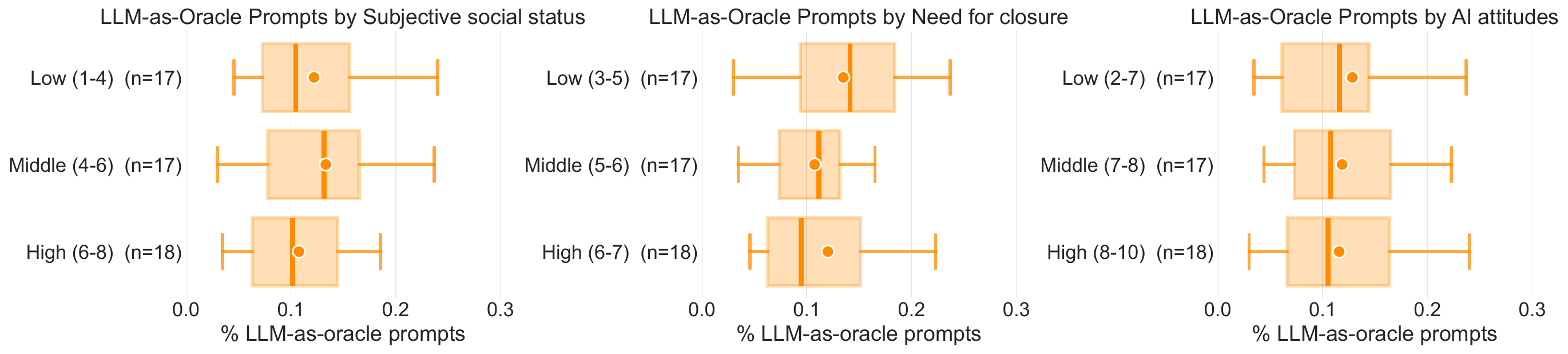}
    
    \caption{\textbf{Donated data demographic breakdown by individual demographics.}}
    \Description{Six box plots of the share of oracle-use prompts per participant in the donated data, split by individual traits. Each panel is a box plot on an axis from 0 to 0.3, with the mean marked on each box and group sizes labeled, split by race, education, frequency of AI use, subjective social status, need for closure, and attitudes toward AI. The boxes overlap substantially across every grouping.}
    \label{fig:datadonation_moredemo}
\end{figure}
\begin{figure}[htbp]
    \centering
\includegraphics[width=0.8\linewidth]{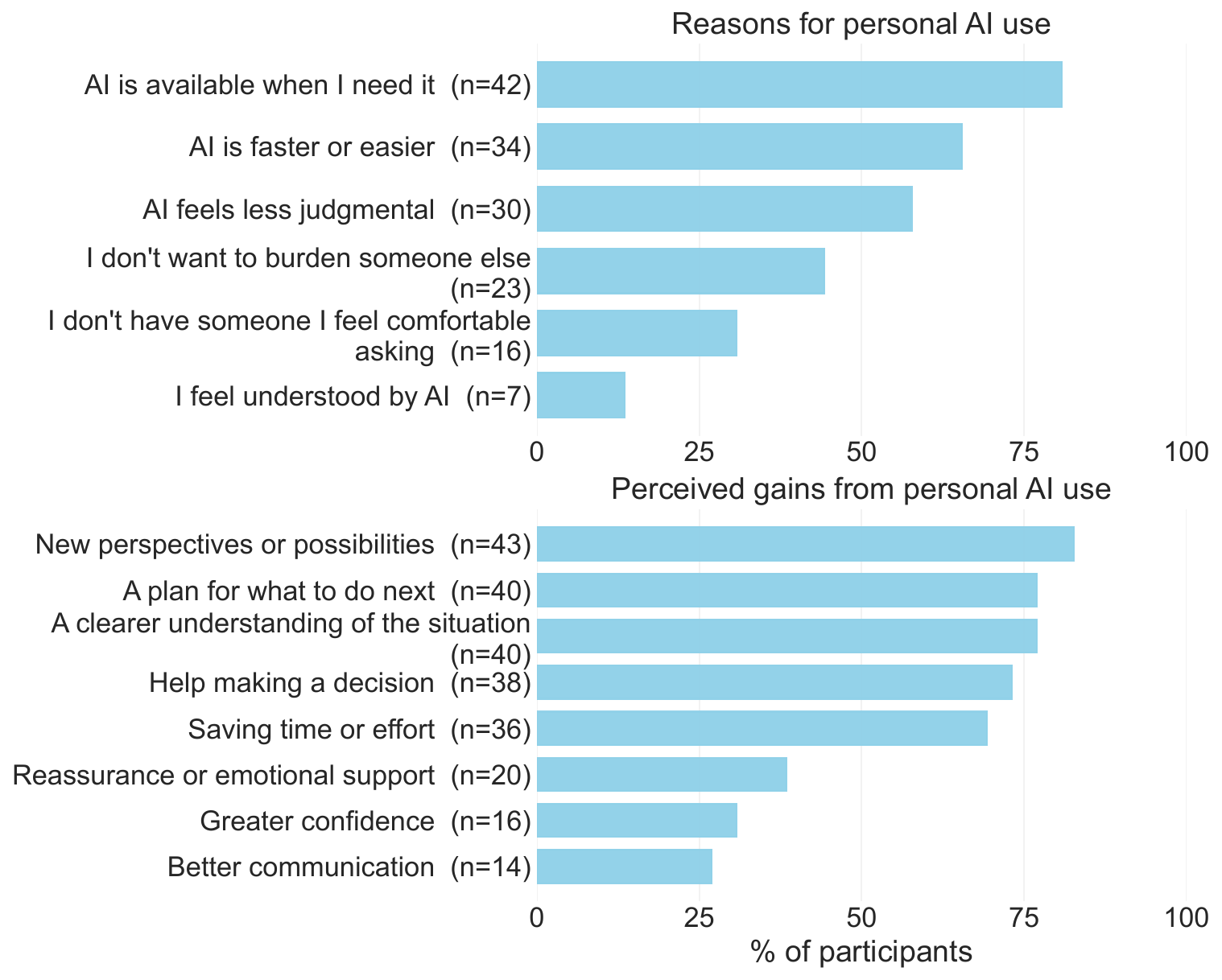}
    \caption{\textbf{Reasons reported by participants for their personal AI use.}}
    \Description{Two plots of participants' stated reasons for and perceived gains from personal AI use in the data donation study. The top panel is a horizontal bar chart of six stated reasons for personal AI use, on an axis of the share of participants from 0 to 100 percent, with participant counts labeled; availability and speed or ease are largest. The bottom panel is a horizontal bar chart of eight perceived gains on the same axis; new perspectives or possibilities is largest, followed by a plan for what to do next and a clearer understanding of the situation.}
    \label{fig:dem_diff_dd}
\end{figure}

\subsection{Data donation tool}
\begin{figure}[htbp]
    \centering
    \includegraphics[width=0.85\linewidth]{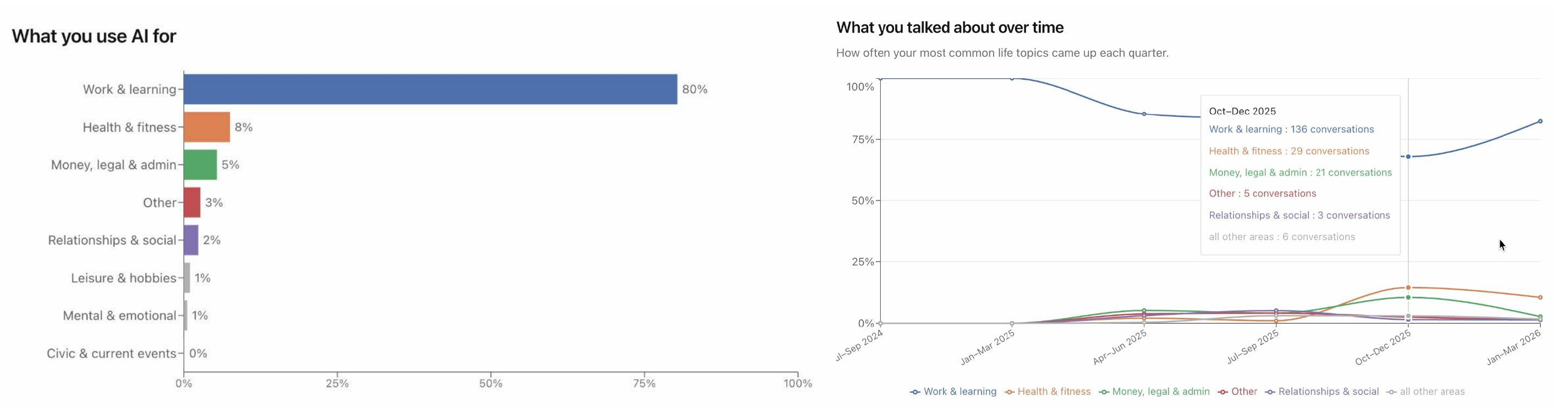}
    \includegraphics[width=0.95\linewidth]{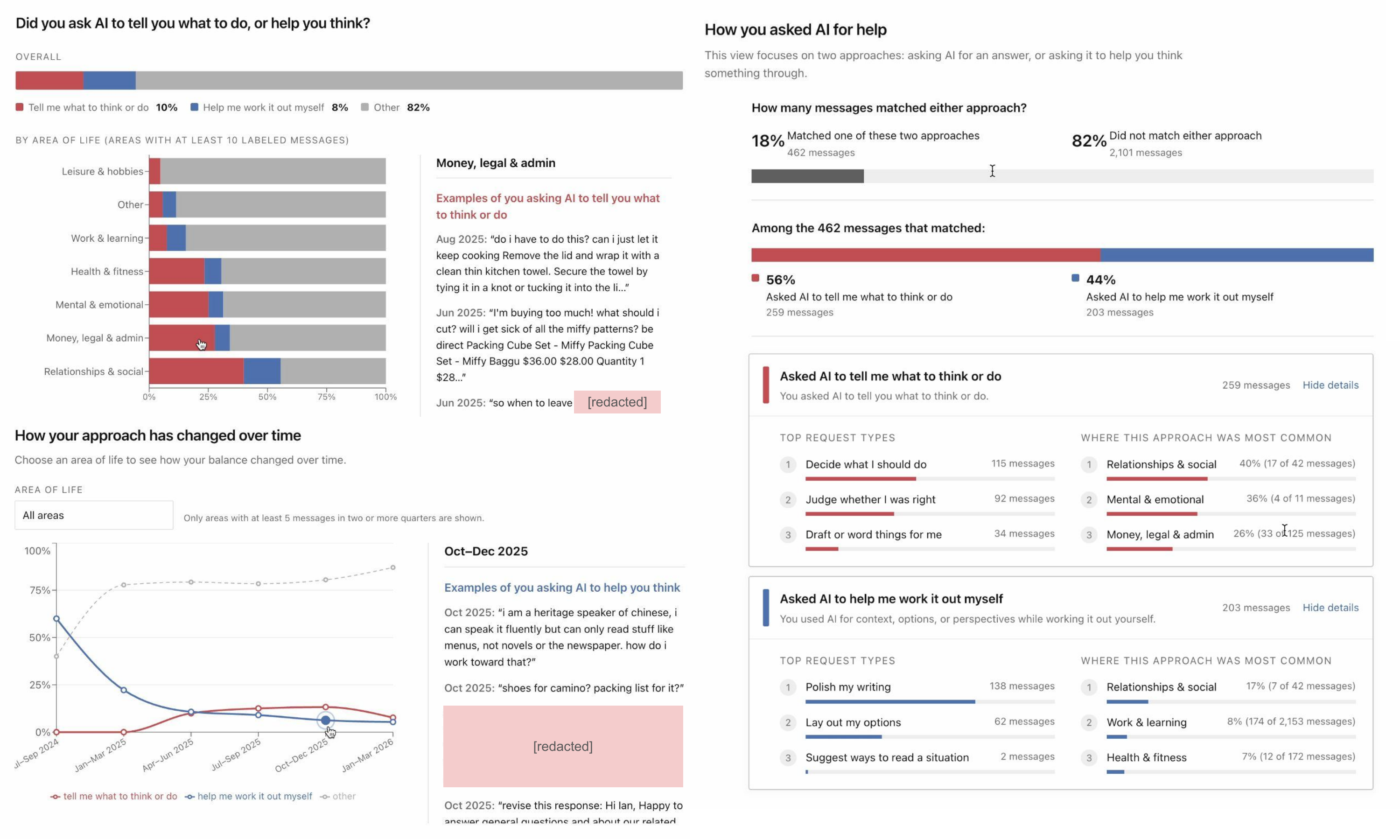}
    \caption{\textbf{Data donation tool visualization interface.} After participants donated their data, they were shown visualizations of their use of AI. This began with visualizations about the domains in which they used AI (top left), and trends over time (top right). Then, they were shown visualizations about how much they seek judgments and beliefs from AI, and how this varied over time and by domain. This was grounded in specific examples. Redacted text to preserve anonymity.}
    \Description{Six screens of the data donation tool interface: domains, trends over time, and oracle-use balance. The top row is a horizontal bar chart of the domains a participant used AI for and a line chart of how those domains shifted over time. The middle row is a stacked bar chart of the participant's overall balance of ``tell me what to think or do'' versus ``help me work it out myself'' with a per-domain breakdown, alongside a screen of what share of messages matched either approach and the split between the two. The bottom row is a line chart of how that balance changed over time and a panel of the participant's top request types and the domains in which each approach was most common. Each screen is paired with example prompts drawn from the participant's own history; these examples are redacted to preserve anonymity.}
    \label{fig:interface}
\end{figure}

Table~\ref{tab:relabel} shows the simplified, participant-facing phrases used for each of the eight roles. Figure~\ref{fig:interface} shows the interface with the visualizations that participants were shown after donating their data.
\begin{table*}[htbp]
\centering
\small
\begin{tabular}{@{}llll@{}}
\toprule
\multicolumn{2}{c}{\textit{\oracle ``Tell me what to think or do''}} &
\multicolumn{2}{c}{\textit{Non-oracle ``Help me work it out myself''}} \\
\cmidrule(lr){1-2}\cmidrule(lr){3-4}
\textbf{Role} & \textbf{Shown to participant} &
\textbf{Role} & \textbf{Shown to participant} \\
\midrule
\interpreter
    & ``Tell me what something means''
    & Contextualizer
    & ``Explain a topic or perspective'' \\

\evaluator
    & ``Judge whether I was right''
    & Deliberator
    & ``Lay out my options'' \\

\decisionmaker
    & ``Decide what I should do''
    & Strategist
    & ``Map out approaches I could take'' \\

\ghostwriter
    & ``Draft or word things for me''
    & Editor
    & ``Polish my writing'' \\
\bottomrule
\end{tabular}
\caption{Mapping from role labels to the participant-facing phrases used in the tool.}
\label{tab:relabel}
\end{table*}

\begin{table*}[htbp]
\centering
\begin{tabularx}{\textwidth}{Xl}
\toprule
\textbf{In the past 6 months, I have used AI to\ldots} &
\textbf{Role} \\
\midrule
Explain why I or someone else thinks, feels, or acts a certain way. &
\interpreter (\oracle) \\

Suggest different possible explanations for someone's thoughts, feelings, or actions. &
Contextualizer (Non-oracle) \\

Judge whether something is right, wrong, or appropriate. &
\evaluator (\oracle) \\

Help me get more information to make a judgment. &
Deliberator (Non-oracle) \\

Decide what I should do. &
\decisionmaker (\oracle) \\

Help me think through possible options or next steps. &
Strategist (Non-oracle) \\

Decide what to say or write text for me. &
\ghostwriter (\oracle) \\

Improve or edit text that I wrote. &
Editor (Non-oracle) \\

I do not use AI in any of these ways. &
NA \\
\bottomrule
\end{tabularx}

\caption{Participant-facing descriptions of roles in pre-data donation survey. Here we chose to ask them to reflect on the previous six months due to memory limitations of retrospective self-reports \cite{bradburn1987answering}.}
\label{tab:ai-uses}
\end{table*}

\section{Results from Perceived Human vs AI study}\label{app:humanvsai}

\paragraph{Additional survey details}
We recruited N=525 participants on Prolific, 5 of whom failed the attention check, leaving N=520 in our final sample (266 in the disclosed AI condition and 254 in the perceived human condition). Before the interaction, participants first answered a set of pre-study questions covering AI use, demographics, AI attitudes, and need for closure. Our study is IRB-approved, and we paid participants at an hourly rate of \$12.

\paragraph{People who used LLMs as oracles also shift more on their decisions after the conversation.}
Participants who sent at least one oracle-use message had larger decision shift (M = 1.16 vs 0.89, p = 0.006), though there was no difference in decision shift between the human vs. AI conditions (M = 1.06 vs 1.03, p = 0.80, d = 0.02).

Perceptions of trustworthiness of the conversation partner did not differ (M = 5.56 vs 5.68, p = 0.34), while response quality was higher for labeled-AI (5.78 vs 5.46, p = 0.007). None of the preregistered individual difference measures predicted oracle use (all Odds Ratios approx. 1.00, all ps > 0.33). After excluding 68 people in the labeled-human arm who rated the system as somewhat or very unbelievable, the decision shift result remained non-significant and the difference in oracle use remained significant, while the difference in perceptions of message quality disappeared.

\begin{table}[htbp]
\centering
\small
\begin{tabular}{@{}p{0.95\linewidth}@{}}
\toprule
\textit{Social \& Relationships} \\
\midrule
There's someone I've lost touch with and I keep going back and forth about whether to reach out. \\
I haven't heard from someone in a while and I'm debating whether to message them first or wait and see if they reach out. \\
I got invited to something and I'm trying to figure out whether I actually want to go or if I'd rather not. \\
Something a friend did has been bothering me, and I can't decide whether to bring it up or just move on. \\
I had a disagreement with someone where I don't think I was wrong, so I'm wondering whether apologizing would help or just feel dishonest. \\
I know two people who I think would get along, but I'm not sure if introducing them is a good idea or if it would be awkward. \\
\midrule
\textit{Work \& Career} \\
\midrule
I'm trying to decide whether to apply for a role that excites me but feels like a stretch. \\
I got a job offer and I'm weighing whether to take it or stay where I am. \\
I've been thinking about asking for a raise, but I'm not sure if now is the right time or if I should wait. \\
Something happened at work that's bothering me, and I'm going back and forth between raising it with my manager or letting it go. \\
There's a project I could volunteer for. Part of me wants to, but I'm not sure it's the right move. \\
I've decided to leave my job, and I'm torn between telling my team now or waiting until everything's finalized. \\
\bottomrule
\end{tabular}
\caption{Decision topics participants could choose from in the perceived human vs.\ AI study. After selecting a topic, participants were shown a short prompt asking them to describe their specific situation.}
\label{tab:topics}
\end{table}

\section{Oracularity in current model responses}\label{app:oracularnesscurrent}

\begin{figure}[htbp]
    \centering
    \includegraphics[width=0.9\linewidth]{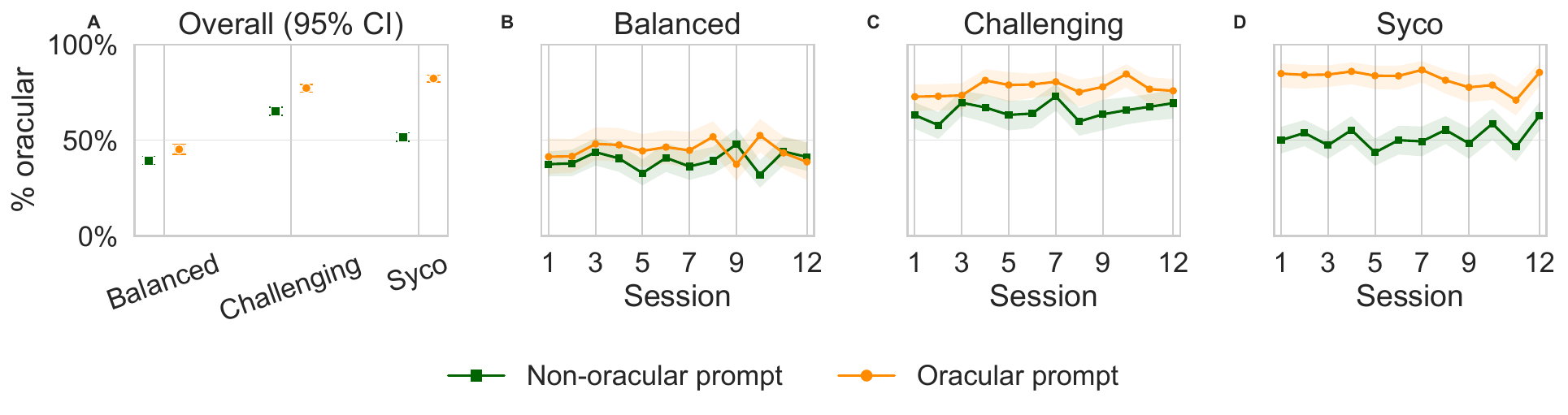}
    \includegraphics[width=0.3\linewidth]{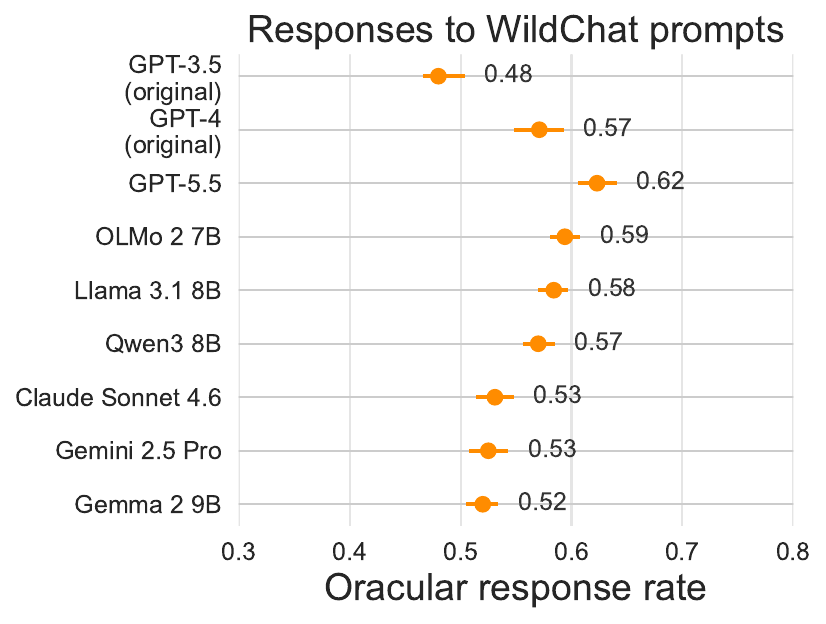}\includegraphics[width=0.6\linewidth]{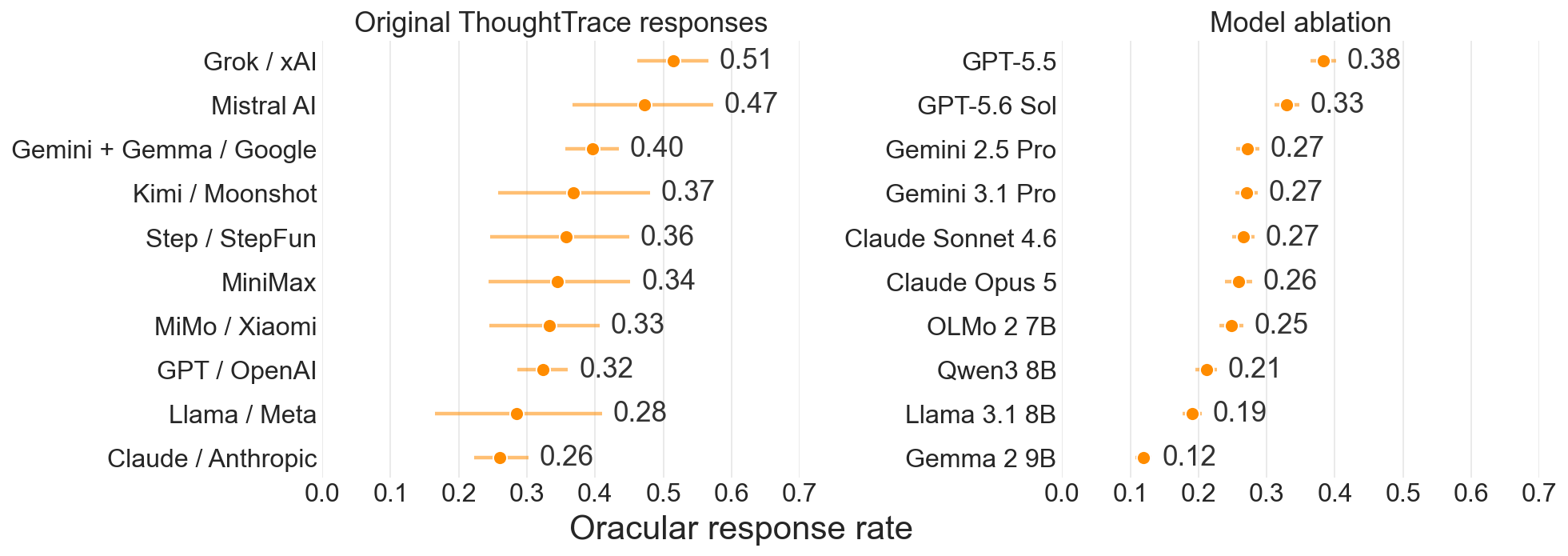}
    \includegraphics[width=0.9\linewidth]{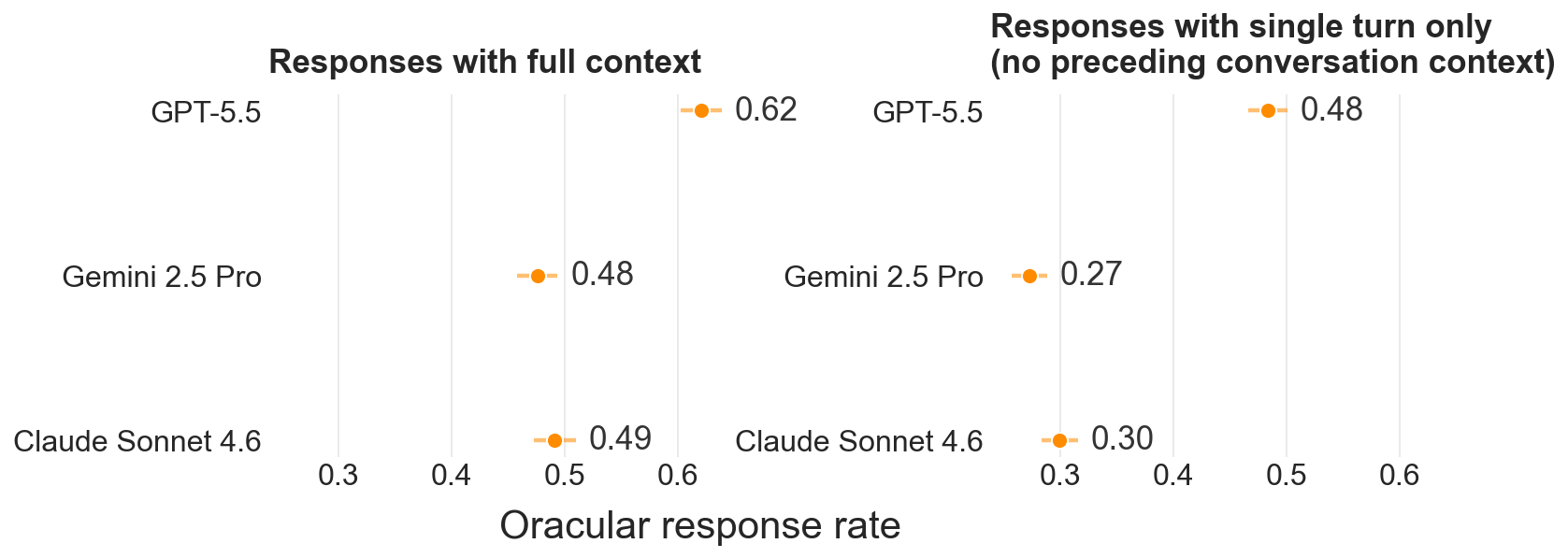}
    
    \caption{\textbf{Oracularity of different LLMs' responses in the longitudinal sycophancy study (top), WildChat (middle left) and ThoughtTrace prompts (middle right)}. In the longitudinal sycophancy study, we find that both the sycophantic and challenging models are more oracular than the balanced baseline; the sycophantic model is especially oracular in response to \oracle prompts. In the original responses (which are to different prompts, and thus not directly comparable), we find that GPT-4 is more oracular than GPT-3.5; and Grok is more oracular than all other models in ThoughtTrace. Across the directly comparable new model responses, GPT-5.5 is most oracular; and this finding is robust to whether the response has the full conversation context or not (bottom). Breakdowns by prompt oracularity are in Figure \ref{fig:thoughttrace_results_app}.}
    \Description{Five dot-and-interval plots of oracular responses by model, for WildChat and ThoughtTrace prompts. The top row is three panels. The left panel is the oracular response rate on WildChat prompts for the two original models and seven models with newly generated responses, with values labeled, ranging from 0.48 to 0.62. The middle panel is the rate for the original ThoughtTrace responses grouped by model family, ranging from 0.26 to 0.51. The right panel is the rate for the ten models in the ThoughtTrace model ablation, ranging from 0.12 to 0.38. The bottom row is two dot-and-interval plots comparing three models given the full conversation context with the same models given a single turn only; the single-turn rates are roughly half the full-context rates for all three, but the trend is the same.}
    \label{fig:thoughttrace_results}
\end{figure}

\begin{figure}[htbp]
    \centering
     \includegraphics[width=0.79\linewidth]{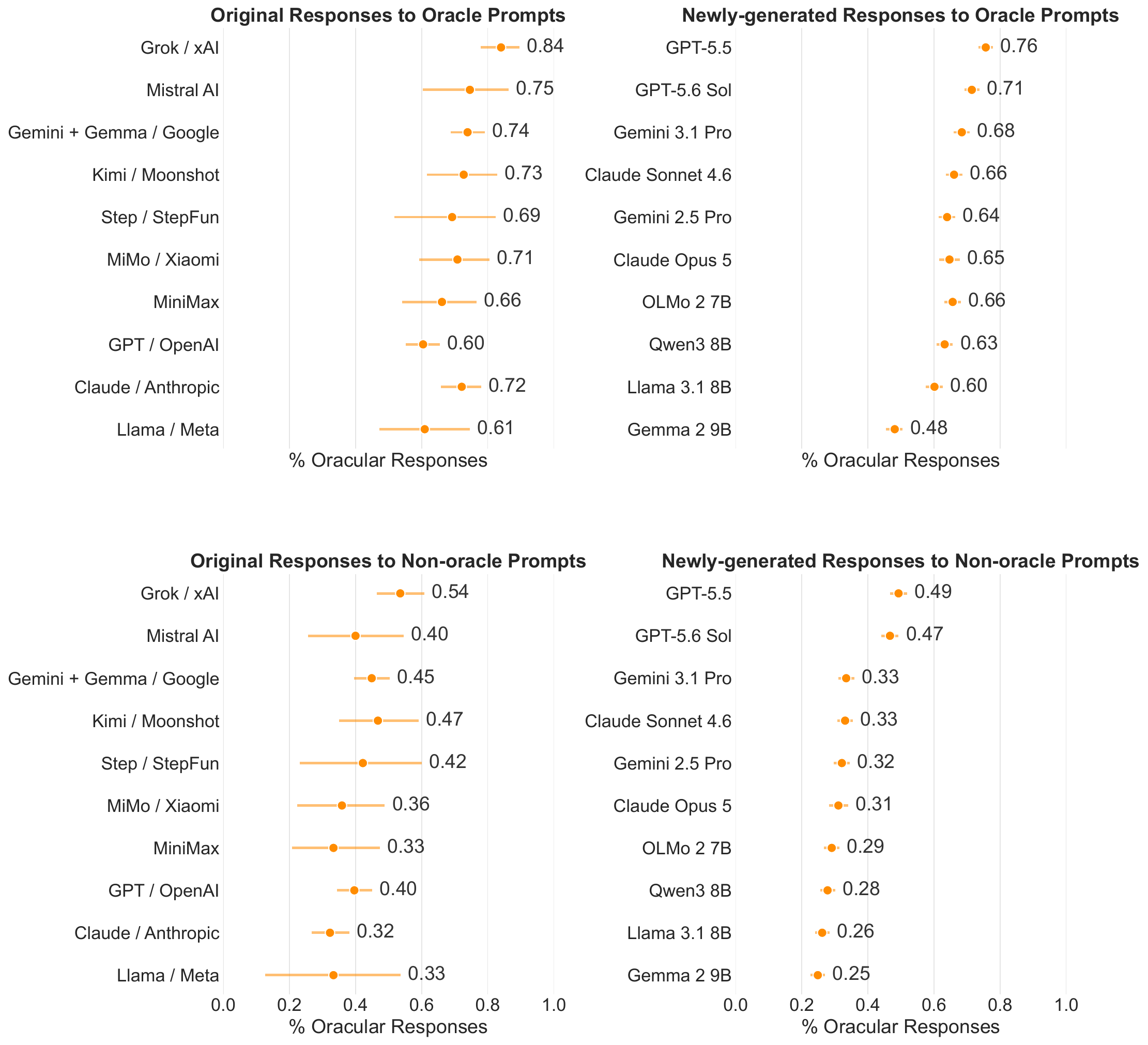}
    \caption{Oracular responses by model, and split by oracle vs non-oracle prompts.}
    \Description{Four dot-and-interval plots of oracular responses in ThoughtTrace by prompt role. The top row is two panels of the rate on oracle-use prompts, first grouped by model family and then for the ten models with newly generated responses. The bottom row is the same two panels for non-oracle-use prompts, where every rate is lower.}
    \label{fig:thoughttrace_results_app}
\end{figure}

On the subset of prompts that were labeled with one of the roles on our typology ($n=11,087$ from WildChat and $n=3,227$ from ThoughtTrace), we first ran our LLM judge on the original responses, which were from GPT-3.5 and GPT-4 (WildChat) and over 20 LLMs (ThoughtTrace). 
In WildChat, we find that the newer, larger GPT-4 is significantly more oracular than GPT-3.5. In ThoughtTrace, Grok has the highest rates of responding as an oracle. This contrast is even larger when we consider the subset of prompts that are non-oracular, i.e., cases where users seem to be explicitly not seeking oracular responses (Figure \ref{fig:thoughttrace_results_app}). 

These original responses do not allow us to attribute the effects to model alone, as each model's responses are to different prompts (though the output model is randomly selected in ThoughtTrace). To more directly compare model behavior, we further obtained responses from both open and closed state-of-the-art models on the same data and scored them. Specifically, we generated new responses from three proprietary and four open-weight models: Claude Sonnet 4.6, Gemini 2.5 Pro, GPT-5.5, Qwen3-8B, Llama-3.1-8B, OLMo 2 7B, and Gemma 2 9B. We additionally generated responses from Claude Opus 5, GPT 5.6 Sol, and Gemini 3.1 Pro on ThoughtTrace.

We find that GPT-5.5 consistently responds more as oracles than any other model family, both in \oracle~prompts and otherwise (Figure \ref{fig:thoughttrace_results}). On ThoughtTrace, the smaller open-weight models are generally less oracular than the larger closed models.

\section{Additional Intervention Results}\label{app:interventionextra}
\begin{figure}[htbp]
    \centering
\includegraphics[width=0.5\linewidth]{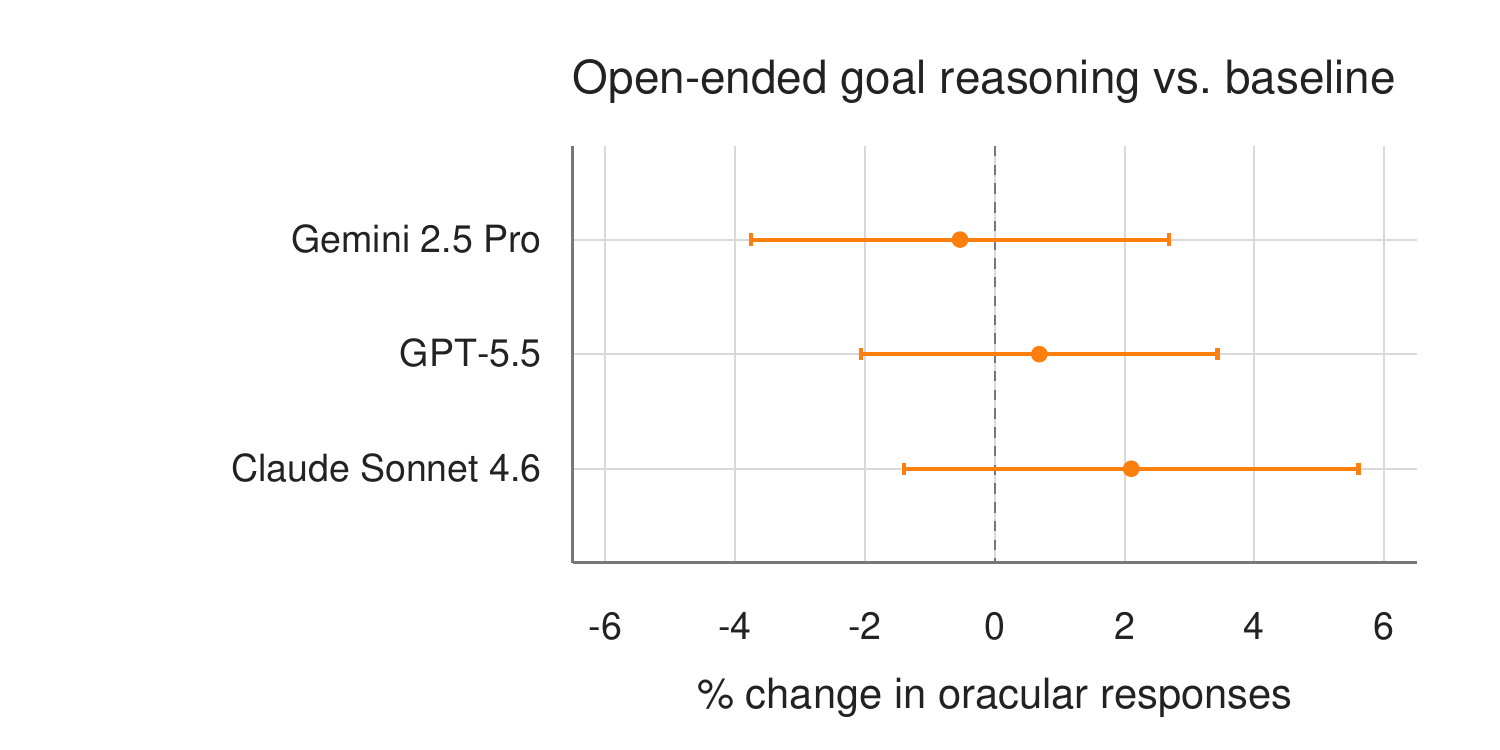}
    \caption{\textbf{Prompting models to generally reason about users' goals does not significantly reduce oracular responses.}}
    \Description{One dot-and-interval plot of oracular responses under open-ended goal reasoning, for three models. The panel is the percent change relative to baseline for Gemini 2.5 Pro, GPT-5.5, and Claude Sonnet 4.6, on an axis from minus 6 to 6 with a dashed reference line at zero. All three intervals span zero.}
    \label{fig:opennodiff}
\end{figure}

\begin{figure}[htbp]
    \centering
    
    \includegraphics[width=0.45\linewidth]{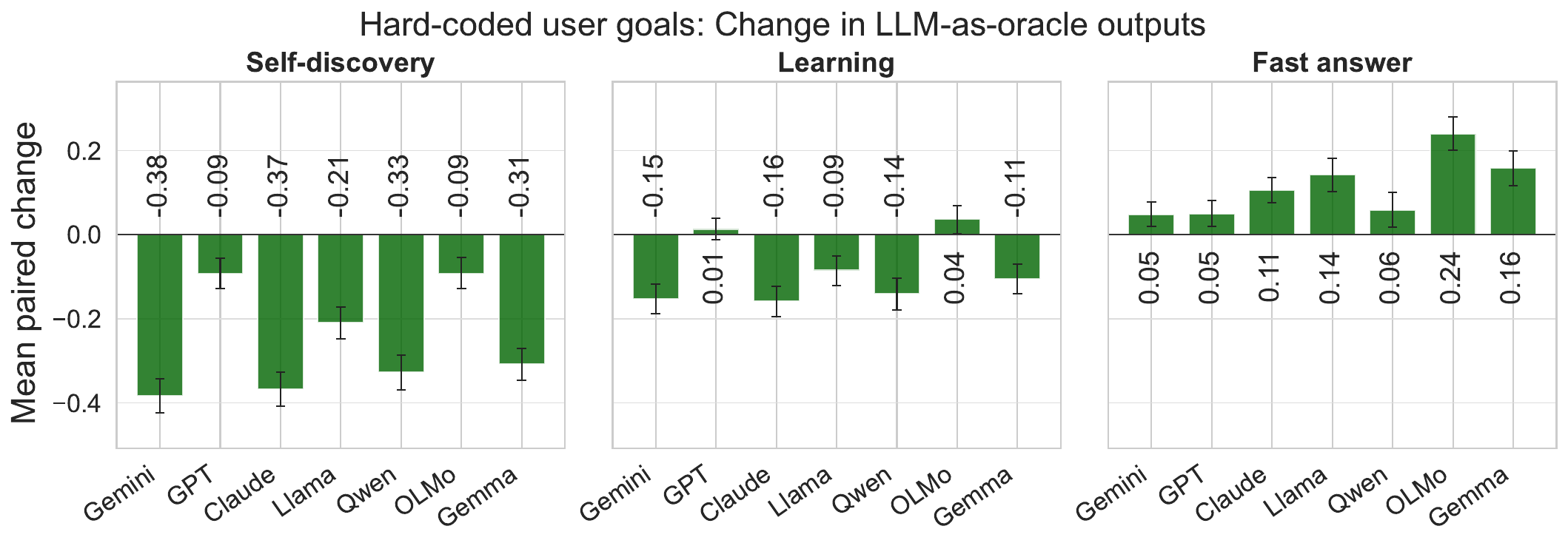}
    \includegraphics[width=0.45\linewidth]{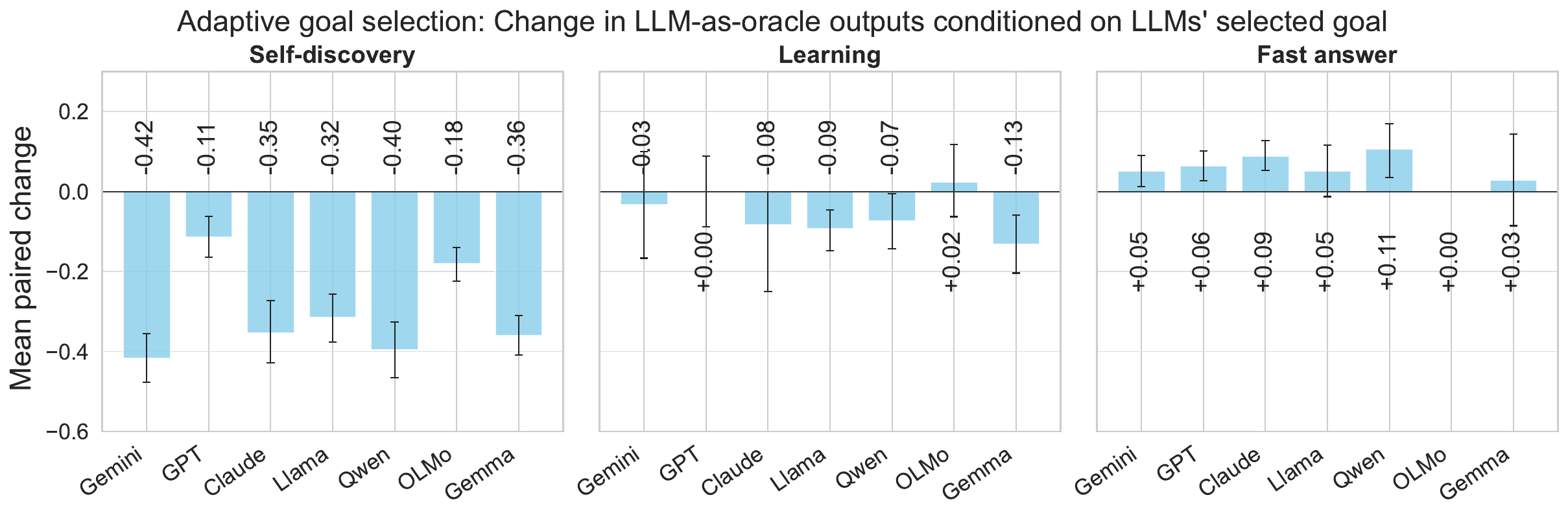}
    
    \includegraphics[width=0.45\linewidth]{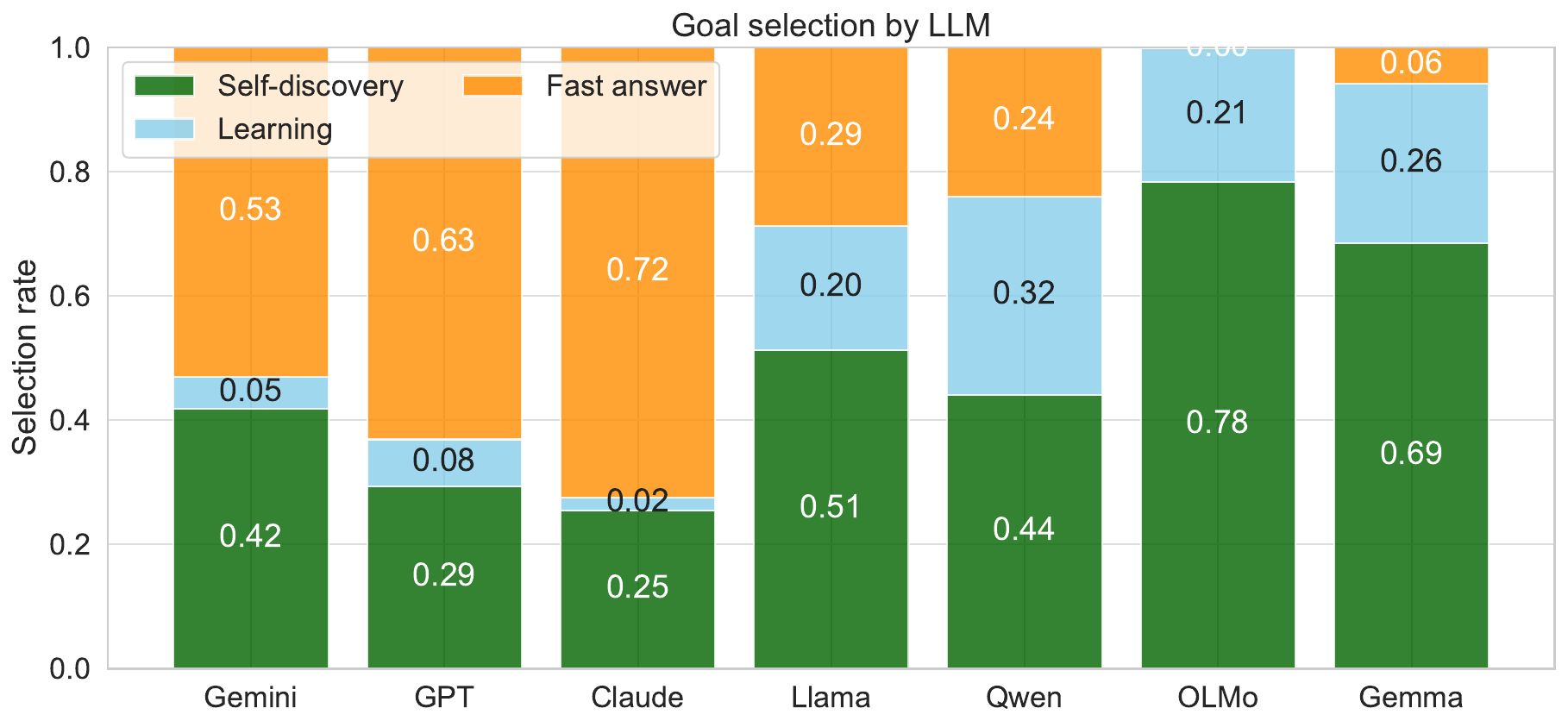}
    \includegraphics[width=0.4\linewidth]{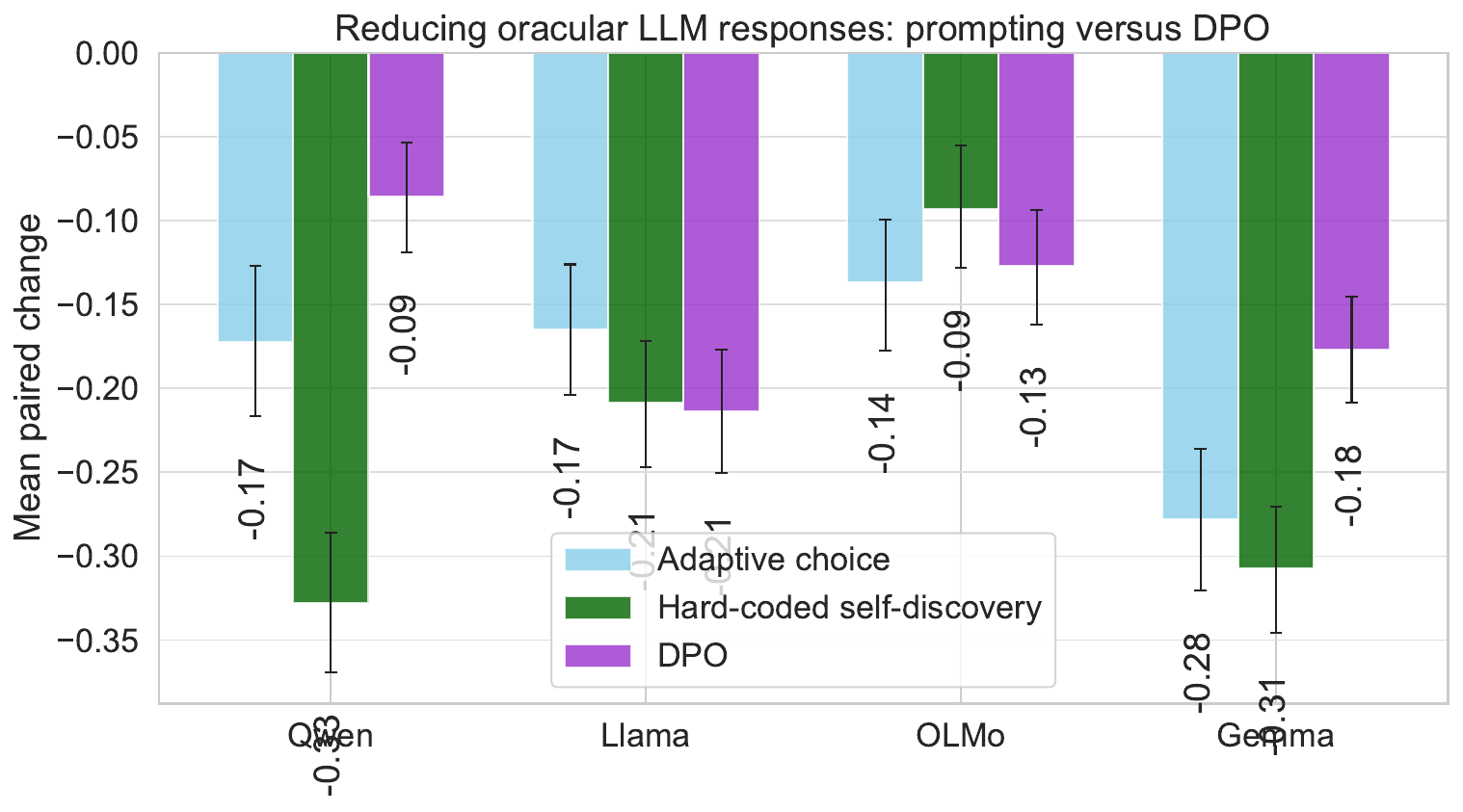}
    \includegraphics[width=0.45\linewidth]{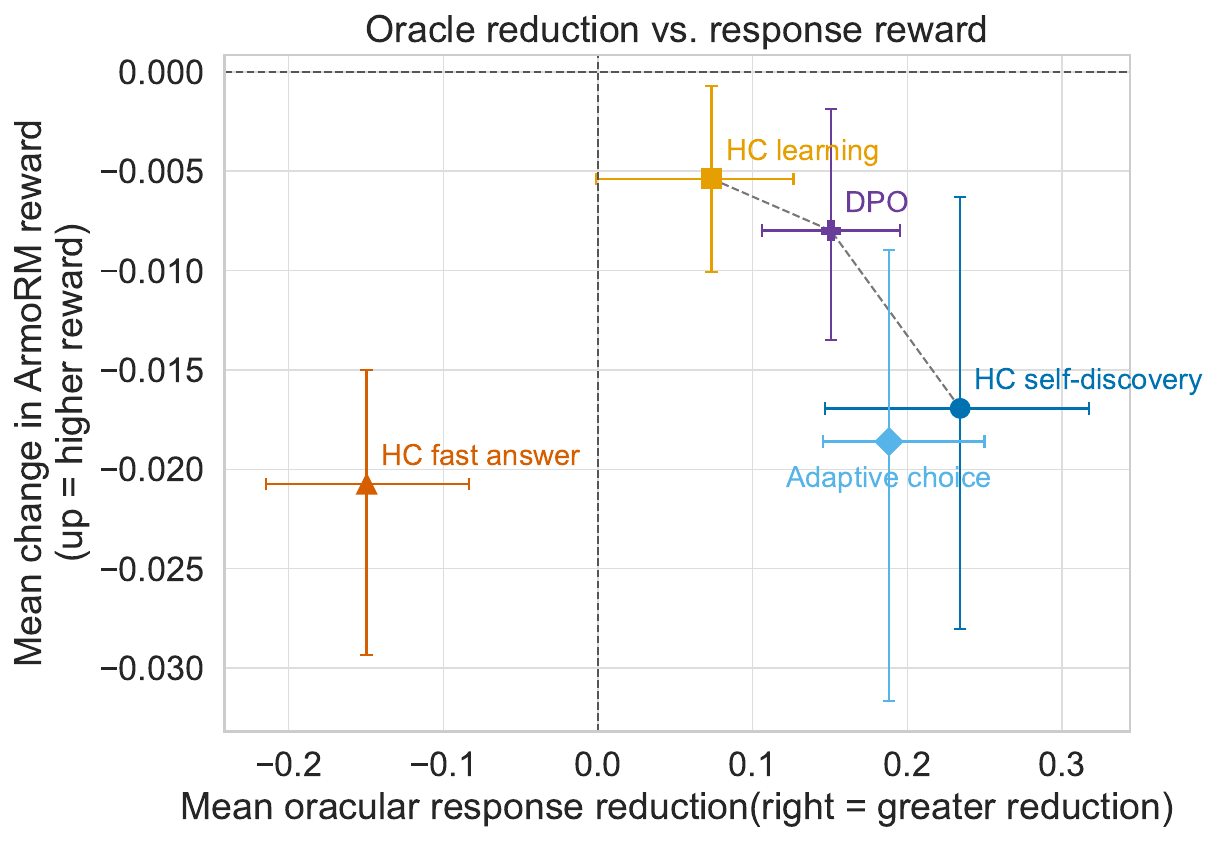}

    \caption{On a random sample of prompts (not only the ones with oracular responses), we find the same takeaways as reported in the main text.}
    \Description{Five groups of plots of intervention effects on oracular responses on a random sample of prompts stratified by role. In this sample the original response is not necessarily oracular. The first group is three panels of the mean paired change in oracular responses, one per hard-coded user goal, for seven models. The second group is the same three panels for the adaptive goal selection pipeline, conditioned on the goal each model selected. The third is a stacked bar chart of goal selection rates for the seven models, on an axis from 0 to 1. The fourth is a bar chart of the mean paired change for the four open-weight models under adaptive choice, hard-coded self-discovery, and Direct Preference Optimization. The directions match the main text throughout. The fifth group is a scatter plot of the mean change in ArmoRM reward against the mean reduction in oracular responses for the five interventions, with error bars on both axes and a dashed line marking the Pareto frontier; hard-coded fast answer is the only intervention that increases oracularity, and it also carries the largest reward loss.}
    \label{fig:interventiondetails_app}
\end{figure}

\begin{figure}[htbp]
    \centering
   
    \includegraphics[width=0.45\linewidth]{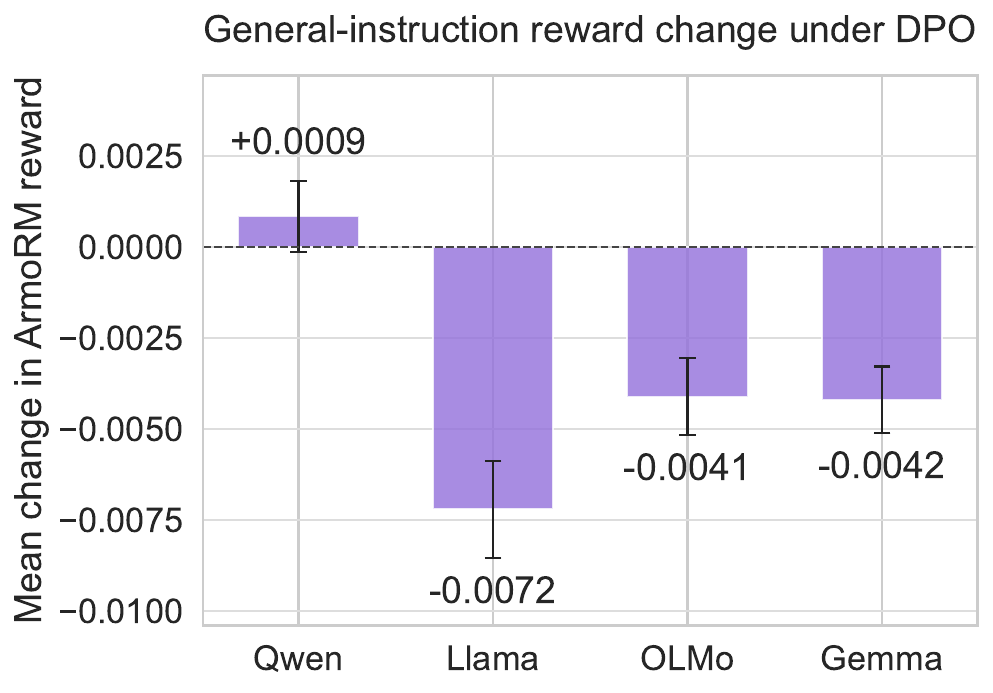}
    \includegraphics[width=0.45\linewidth]{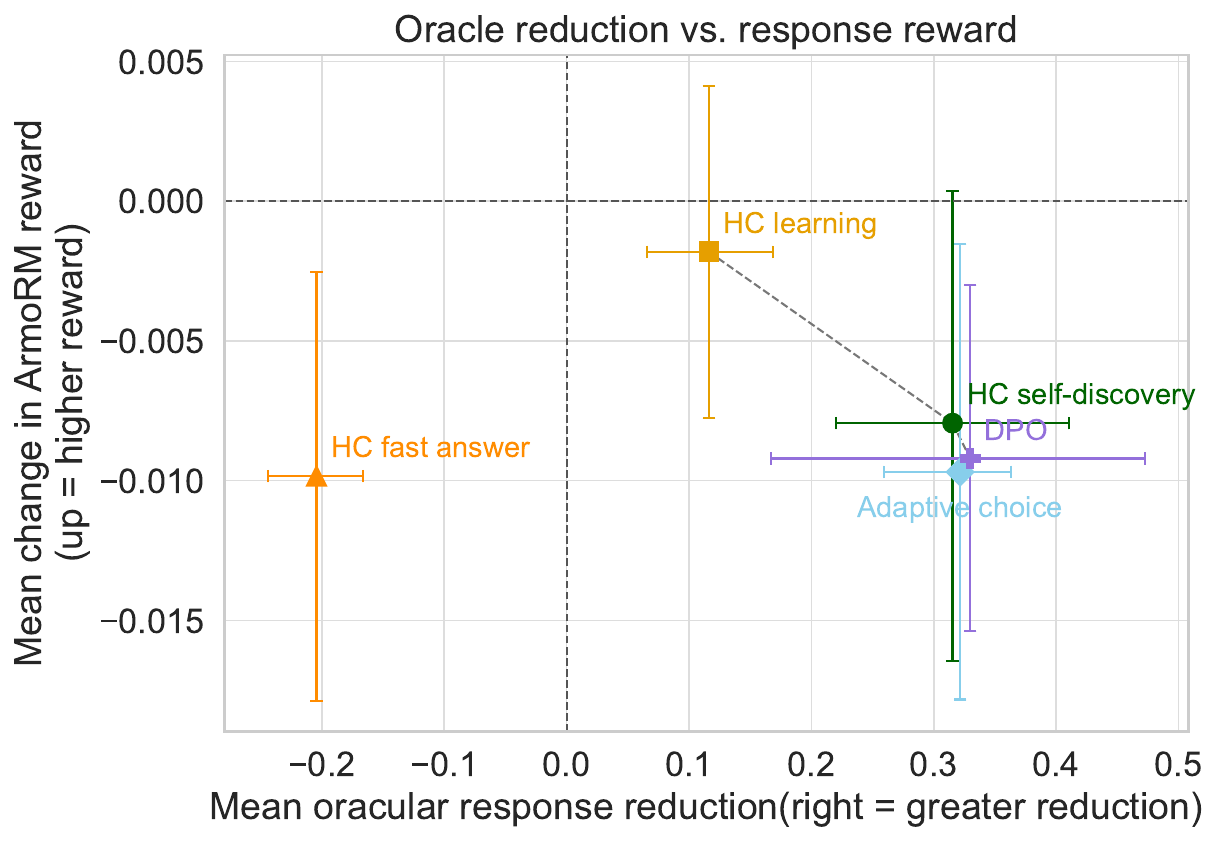}
    \caption{Models on which DPO is most effective at reducing oracularity also compromise model reward the most (left). Hard-coded  (HC) learning, HC self-discovery, and DPO are at the Pareto frontier between reward and oracular reduction (right).}
    \Description{Two plots of the tradeoff between oracularity reduction and response reward. The left panel is a bar chart of the mean change in ArmoRM reward on general instructions for the four open-weight models under Direct Preference Optimization, with error bars and values labeled; the change is slightly positive for Qwen and negative for the other three, most steeply for Llama. The right panel is a scatter plot of the mean change in reward against the mean reduction in oracular responses for the five interventions, with error bars on both axes and a dashed line marking the Pareto frontier, on which hard-coded learning, hard-coded self-discovery, and Direct Preference Optimization sit.}
    
    \label{fig:dpo_reward}
\end{figure}

Figure~\ref{fig:opennodiff} shows the effect of prompting models to do open-ended reasoning about users' long-term goals. Figure~\ref{fig:interventiondetails_app} shows the main intervention analyses repeated on a random sample of 600 prompts (stratified by role), in which the original response is not necessarily oracular. Figure~\ref{fig:dpo_reward} shows the tradeoff between reducing oracularity and response reward.

\subsection{{Preference study: Immediate preferences vs. for self-discovery}}\label{app:user_study}
\begin{figure*}[htbp]
    \centering
    \includegraphics[width=0.45\linewidth]{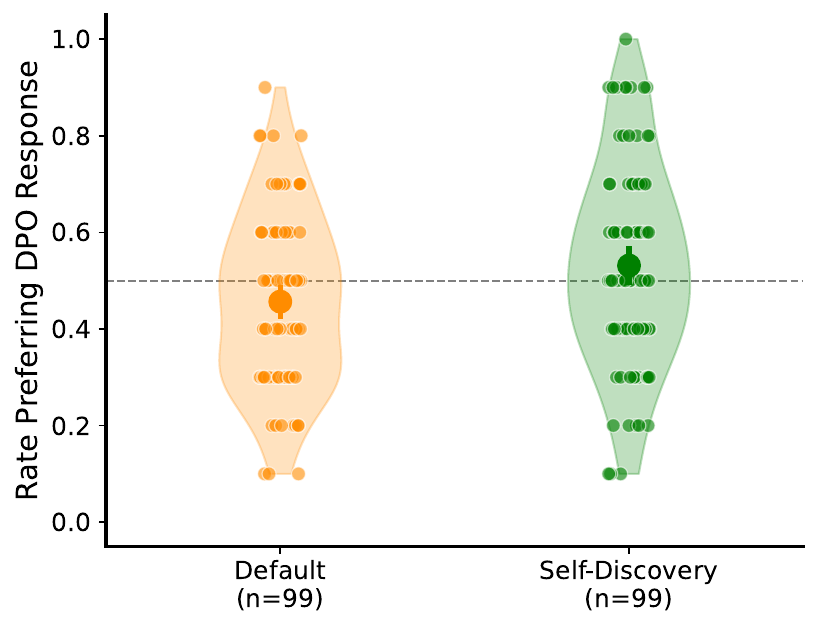}
    \includegraphics[width=0.45\linewidth]{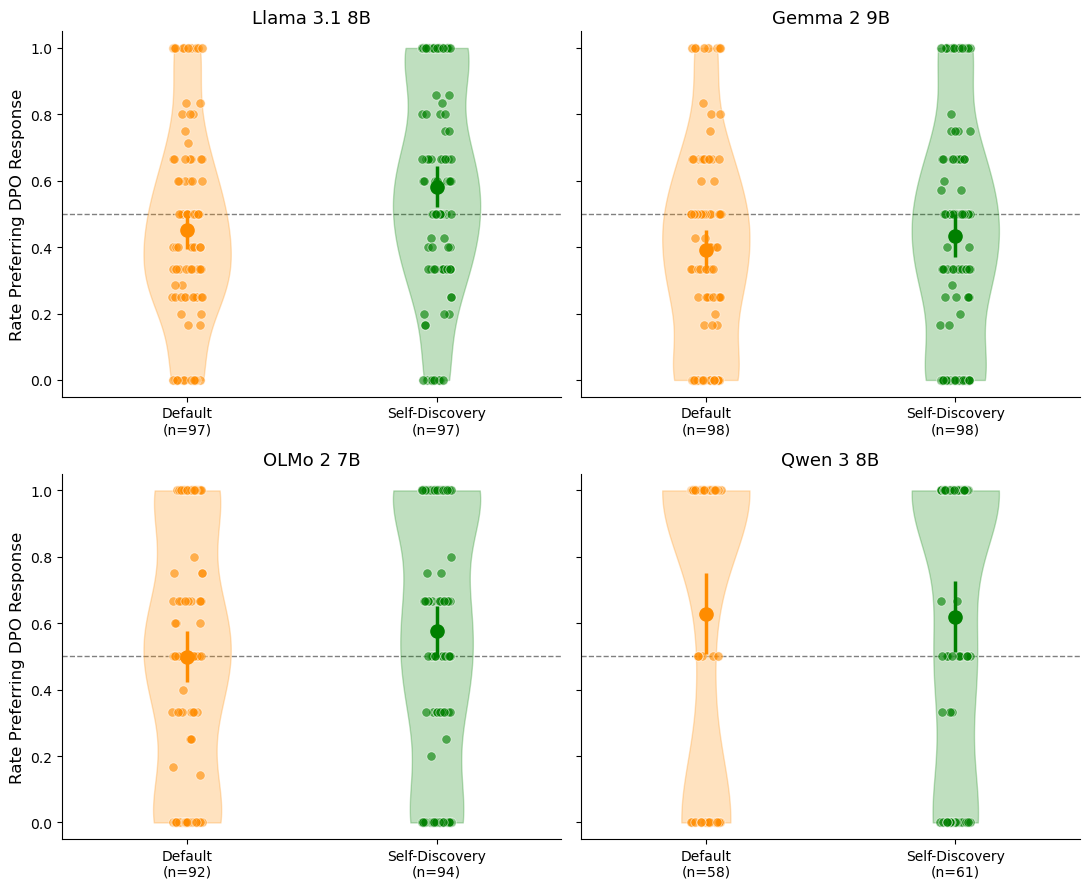}
    \caption{\textbf{Left:} Overall, users prefer original LLM responses when asked only for their immediate preference (p = .016), but prefer DPO responses when asked for a preference that accounts for self-discovery more often than in the default condition (p = .007), though not significantly more than 50\% (p = .13). \textbf{Right:} Rates by base model.}
    \Description{Two bar plots}
    \label{fig:user_preference}
\end{figure*}

{We conducted a user preference study on baseline model outputs vs. DPO model outputs. We recruited 200 participants on Prolific and paid them $\$2.00$ for a 10 minute task. We excluded participants that failed a visual bot check, leaving 198 participants. Participants were randomized into two conditions: Default and Self-Discovery. In the task, participants were presented with a randomized sequence of 10 prompts with two LLM responses to the prompt: the original baseline model response and the DPO response. In the Default condition, participants were asked to choose the response they would prefer. In the Self-Discovery condition, participants were asked to choose the response they would prefer if they wanted to ``reason about the situation yourself and develop your own interpretation". The prompts and responses were sampled uniformly from the test dataset (across all four models) after filtering for examples where the DPO response demonstrated oracularity reduction (LLM-judged). In total, participants annotated 724 samples from Llama-3.1-8B, 643 from Gemma-2-9B, 447 from OLMo-2-7B, and 166 from Qwen3-8B. The variation in sample size reflects differences in DPO's effectiveness at reducing oracularity. Figure~\ref{fig:user_preference} shows the rate at which participants in each condition prefer the DPO response, overall and by base model. Since each participant annotated 10 examples, we test participant-level rates with two-sided tests.}

\section{{Robustness to Excluding Writing and Editing}}
\label{app:excluding-writing}
The primary trends are robust to excluding the category of expressing
oneself to others through writing (\ghostwriter and Editor roles), though some individual associations lose significance.

In \textbf{WildChat}, removing these roles retains 54,034 prompts.
The temporal increase in \oracle among subjective personal questions is significant
($r=.693$, $p=0.006$), though it is not significant in the original sample. The increase in \interpreter among all
 prompts ($r=.629$, $p=0.016$),and decrease
in Strategist share ($r=-.695$, $p=0.006$) persists, while the increase in \evaluator changes from 
 $r=.552$, $p=0.041$ to $r=.520$, $p=0.057$. Among all prompts, the Interpreter association remains
significant after our exclusion ($r=.629$, $p=0.016$), while the overall
 share of \oracle prompts remains non-significant ($r=.266$, $p=0.359$).

In the \textbf{donated data}, removing these roles retains 132,932
messages. The increase in \oracle messages remains significant,
both with \emph{NA} in the denominator ($r=.698$, $p=0.003$) and without
($r=.939$, $p<.001$){, and including within person with participant fixed effects (with NA: OR = 1.30 per year, $p < 0.001$; without: OR = 1.58 per year, $p < 0.001$)}.
In the \textbf{longitudinal study}, the sycophantic-condition increase
across sessions remains significant ($r=.606$, $p=0.037$), while the
balanced ($r=.408$, $p=0.188$) and challenging ($r=.558$, $p=0.060$)
conditions remain nonsignificant at $\alpha=.05$.
In \textbf{ThoughtTrace}, the significant demographic differences
persisted: higher \oracle among users aged $\leq34$ than among older
users ($t=3.13$, $p=0.002$), among men than women ($t=3.53$, $p<.001$),
and among users reporting at least weekly AI use than less frequent
users ($t=2.32$, $p=0.022$). In the donation study's participant-level
model controlling for gender and subjective social status ($N=52$),
the age coefficient remains similar: $-2.14$ percentage points per SD
of age band ($p=0.059$), compared with $-2.08$ points ($p=0.061$)
originally.
For the discrepancy between \textbf{self-reported} and \textbf{observed} roles in the data donation study, we find that the large gap remains: 
58.3\% of
oracular role--participant instances in the observed top three were in the
self-reported bottom three or unranked, compared with 22.0\% for
non-oracular roles (and 17.3\% vs 55.3\% for the other way around). 
For \textbf{the preregistered experiment on labeled-AI vs labeled-human}, the main preregistered findings are significant: participants' mean \oracle message shares were 19.6\% in the AI condition
and 11.1\% in the perceived-human condition ($p<.001$).
Participants in the labeled-AI condition remained more likely to send $\geq 1$ (64.3\% versus 51.2\%; $\mathrm{OR}=1.72$, $p=0.003$) or
$\geq 2$ oracle prompt (37.2\% versus 20.5\%; $\mathrm{OR}=2.30$, $p<.001$).
For the metaphor associations, warmth remained positively associated with \oracle
($\mathrm{OR}=1.43$, $p=0.008$; originally $\mathrm{OR}=1.55$, $p=0.001$). Valence weakened and no longer met the
$.05$ threshold ($\mathrm{OR}=1.28$, $p=0.064$; originally
$\mathrm{OR}=1.34$, $p=0.026$). This is because there were three high-valence participants whose only \oracle labels were \ghostwriter.

\section{More details on the longitudinal sycophancy study}\label{app:longitudinalstudy}
Using a direct condition-by-session test, there is no significant difference in
temporal trends across conditions
($\chi^2(2)=0.021$, $p=0.990$). Here we provide more details on the temporal trends.
In the sycophantic condition, \oracle~ use significantly increased across the 3-week period (Pearson's $r = 0.62,p=0.03$), while those in the balanced condition did not ($r = 0.40,p=0.19$). Those in the challenging condition overall had an increase, though it did not reach significance ($r = 0.55, p=0.06$). 
Looking at specific roles, in the sycophantic condition, it is likely \oracle is primarily driven by a longitudinal increase in participants using the AI as \interpreter. This suggests that the sycophantic AI---which made people feel the most understood---may be the one that users  turned to more for interpreting themselves, their relationships, and the world around them. In the challenging condition, there is a significant longitudinal increase in people using AI as \evaluator. This may be because the challenging AI was designed to respond in normative ways, and thus participants used it more as an \evaluator of normative judgment. 


\end{document}